\documentclass[12pt]{article}

\pdfoutput=1

\usepackage{graphicx}
\usepackage{amssymb}
\usepackage{amsmath}
\usepackage{multirow}

\newcommand{\eq}[1]{\begin{equation} #1 \end{equation}}

\newcommand{\eqa}[1]{\begin{eqnarray} #1 \end{eqnarray}}
\newcommand{\al}[1]{\begin{align} #1 \end{align}}

\newcommand{\av}[1]{\langle #1 \rangle}

\newcommand{\ds}{\displaystyle}

\newcommand{\op}{\mathcal{O}}
\newcommand{\nn}{\nonumber}

\newcommand{\n}{\noindent}

\newcommand{\afb}{A_{\rm FB}}
\newcommand{\aim}{A_{\rm im}}
\newcommand{\bin}{{\rm bin}}

\newcommand{\C}[1]{{\cal C}_{#1}}
\newcommand{\Cp}[1]{{\cal C}^{\prime}_{#1}}

\newcommand{\intbin}[1]{\av{#1}_{\rm bin}}

\begin{document}


\begin{flushright}
{\small
UAB-FT-710\\
LPT-ORSAY/12-81
}
\end{flushright}
$\ $
\vspace{1.5cm}
\begin{center}
\Large\bf
Implications from clean observables for the binned analysis of $B\to K^* \mu^+\mu^-$ at large recoil
\end{center}

\vspace{3mm}
\begin{center}
{\sc S\'ebastien Descotes-Genon}\\[2mm]
{\em \small
Laboratoire de Physique Th\'eorique, CNRS/Univ. Paris-Sud 11 (UMR 8627)\\ 91405 Orsay Cedex, France
}
\end{center}

\begin{center}
{\sc  Joaquim Matias, Marc Ramon, Javier Virto}\\[2mm]
{\em 
Universitat Aut\`onoma de Barcelona, 08193 Bellaterra, Barcelona, Spain
}
\end{center}

\vspace{1mm}
\begin{abstract}\noindent
We perform a frequentist analysis of $q^2$-dependent $B \to K^*(\to K\pi) \ell^+ \ell^-$ angular observables at large recoil, aiming at bridging the gap between current theoretical analyses and the actual experimental measurements. We focus on the most appropriate set of observables to measure and on the role of the $q^2$-binning.
We highlight the importance of the observables $P_i$ exhibiting a limited sensitivity to soft form factors for the search for New Physics contributions. We compute predictions for these binned observables in the Standard Model, and we compare them with their experimental determination extracted from recent LHCb data. 
Analysing $b\to s$ and $b\to s\ell^+\ell^-$ transitions within four different New Physics scenarios, we identify several New Physics benchmark points which can be discriminated through the measurement of $P_i$ observables with a fine $q^2$-binning. We emphasise the importance (and risks) of using observables with (un)suppressed
dependence on soft form factors for the search of New Physics, which we illustrate
by the different size of hadronic uncertainties attached to two related observables ($P_1$ and $S_3$).
We illustrate how the $q^2$-dependent angular observables measured in several bins can help to unravel New Physics contributions to $B \to K^*(\to K\pi) \ell^+ \ell^-$, and show the extraordinary constraining power that the clean observables  will have in the near future. We provide semi-numerical expressions for these observables as functions of the relevant Wilson coefficients at the low scale.
\end{abstract}


\thispagestyle{empty}

\newpage

\section{Introduction}

The set of rare $B$-meson decays mediated by the $b\to s$ transition has been thoroughly studied for many years both from the theoretical and experimental sides. Recently, this set has been complemented by increasingly precise experimental measurements of $q^2$-dependent angular observables in the decay $B \to K^*(\to K\pi) \ell^+ \ell^-$ \cite{0904.0770,1108.0695,1204.3933,LHCbinned,1205.3422}. This has triggered extensive theoretical work studying the constraining power of radiative and semileptonic $B$ decays on New Physics in the framework of the weak effective Hamiltonian~\cite{DescotesGenon:2011yn,1202.2172,bobeth2,Altmannshofer:2011gn,bobeth3,bobeth4,straub,nejc,hurth}.  These constraints apply mostly to the Wilson coefficients $\C7^{(')}$, $\C9^{(')}$ and $\C{10}^{(')}$ related to the magnetic and semileptonic operators $\op_7$, $\op_9$, $\op_{10}$ and the corresponding chirality-flipped operators (which are highly suppressed in the Standard Model). In addition, the tight experimental bounds set in the last few months on the differential decay rate of $B_s\to \mu^+\mu^-$ \cite{1107.2304,1203.4493,1203.3976,1204.0735}, pushed it close to its small Standard Model (SM) prediction~\cite{1106.4041,1106.0998} -- the theoretical and experimental values are actually brought even closer by the $B_s$ mixing correction $O(\Delta\Gamma_s)$ to $B_s$ branching ratios measured at LHCb, discussed in Refs.~\cite{1111.4882,1204.1735} and applied to $B_s\to \mu^+\mu^-$ in Ref.~\cite{1204.1737}.
This puts strong constraints on $\C{10}^{(')}$, as well as on the coefficients $\C{S}^{(')}$ and $\C{P}^{(')}$ of scalar and pseudoscalar operators, specially when combined with $B\to K\ell^+\ell^-$ data \cite{1205.1845,1205.5811}.

Small experimental errors and a good control over hadronic uncertainties on the theory side are the key ingredients for these constraints to be efficient. At present, the branching ratio of the inclusive radiative decay $B\to X_s\gamma$, and the CP asymmetry of $B\to K^*\gamma$ constitute the strongest constraints on the $\C7$-$\Cp7$ plane. However, the complementarity of constraints among different observables can be exploited to reduce considerably the parameter space. In this respect, the inclusion of the isospin asymmetry of $B\to K^*\gamma$, together with the forward-backward asymmetry $\afb$ and the longitudinal polarization fraction $F_L$ in $B \to K^*\ell^+ \ell^-$ integrated over the dilepton invariant mass $q^2$ between 1 and 6 GeV$^2$ already impose additional nontrivial constraints on the $\C7$-$\Cp7$ plane, as well as on $\C9^{(')}$ and $\C{10}^{(')}$ \cite{DescotesGenon:2011yn,1202.2172}.

This complementarity can be exploited further by considering the $q^2$-dependence of angular $B \to K^* \ell^+ \ell^-$ observables. Indeed, increasingly precise measurements of these observables integrated in smaller bins are being presented, due in part to the important statistics obtained from the large data sets collected at the LHCb experiment. This in turn allows for more complete angular analyses providing more observables \cite{LHCbinned}. The prospects for the near future are very good, aiming towards a complete angular analysis with a fine $q^2$-binning.

At this point, it becomes crucial to handle theoretical uncertainties as accurately as possible. 
The lack of huge deviations in $B$ physics up to now forces us to be precise and conservative in our theoretical predictions. Before claiming any discrepancy, indication or discovery of New Physics, one must be sure that a description in terms of observables with little dependence on the specific choice of hadronic parameters has been used.
While the uncertainties related to the form factors constitute a dominant part of the theoretical error, there is a wide spread of quoted uncertainties for $B\to K^*$ form factors in the recent literature, ranging from a $\sim 10\%$ to a $\sim 40\%$ error for the same form factor~\cite{buras,1006.4945}. The size of this error does not depend only on the particular theoretical method used to compute the form factor, but also on the delicate estimation of errors associated to the assumptions built in each procedure. For example,  the values $A_0(0)=0.33\pm 0.03$ and $V(0)=0.31\pm 0.04$  given in Ref.~\cite{buras} should be compared to the values $A_0(0)=0.29\pm 0.10$ and $V(0)=0.36\pm 0.17$ as quoted in Ref.~\cite{1006.4945}. Even central values have shifted significantly, see for instance the value $V(0)=0.41\pm0.05$ from Ref.~\cite{0412079} before its update of Ref.~\cite{buras} (also consistent with Ref.~\cite{1004.3249}). Without attempting to discuss the related conceptual issues in any further depth, it is clear that the impact of such discrepancies between different groups concerning hadronic uncertainties can be reduced greatly if one selects quantities that show a suppressed dependence on the soft form factors, such as the observables $P_i$ or $A_T^i$ defined in Refs.~\cite{kruger,matias1,primary}. These observables can be considered as being theoretically clean in the kinematic range of interest here.

The construction of theoretically clean observables in $B \to K^* \ell^+ \ell^-$ has been the subject of theoretical work for some time \cite{kruger,matias1,matias2,becirevic,primary,bobeth}. Based on the symmetries of the $B \to K^* \ell^+ \ell^-$ angular distribution discussed in Ref.~\cite{matias2}, a complete characterization of the full distribution in terms of a minimal basis of clean observables has recently been proposed \cite{primary}. The conclusion is that,
a complete description of the differential decay rate in the limit of massless leptons and in the absence of scalar contributions can be achieved through
a set of 6 clean observables $P_{1,2,3,4,5,6}$ complemented by two observables with a significant sensitivity to form factors, e.g.,
the differential decay rate and the forward-backward asymmetry $A_{FB}$ (or equivalently $F_L$)~\footnote{Mass and scalar effects can be taken into account by including 4 extra clean observables (see Ref.~\cite{primary}). The explicit form of the symmetries in the presence of mass terms and scalar operators can be found in the same paper.}.  In a short term, these clean observables ($P_i$) should play a leading role in detecting deviations from the SM in a safe way, relegating less clean observables such as $\av{\afb}$, $\av{F_L}$ or $\av{S_3}$ to a secondary role of useful cross-checks.

Measurements of the transverse asymmetry $A_T^{(2)}$ have been already provided by the CDF collaboration \cite{1108.0695}. However, there has recently been some reluctance from the experimental side to extract such clean observables because their theoretical predictions have been mostly presented as functions of $q^2$, while the experimental results are obtained integrated in $q^2$-bins. This issue becomes relevant when a coefficient in the angular distribution is expressed as a product of various observables. For example, in Ref.~\cite{LHCbinned}, the authors prefer to fit for $F_L$ and $S_3$ instead of $F_L$ and $A_T^{(2)}$ (where $S_3\sim (1-F_L) A_T^{(2)}$), arguing that a rapid variation of both $F_L$ and $A_T^{(2)}$ with $q^2$ could result in a biased estimate of $A_T^{(2)}$ when averaging over large $q^2$-bins. This is a perfectly correct statement if one wishes to compare the experimental measurements with (differential) $q^2$-dependent observables such as $A_T^{(2)}(q^2)$. 
Actually, in the specific case of $A_T^{(2)}$ (for the SM case as well as for particular NP values of the Wilson coefficients involved), we observed a very good agreement
between the observable integrated naively over the low-$q^2$ bins (i.e., $\int_{\bin} dq^2 A_T^{(2)}$)  and its
value derived from the binned observables as they are being measured (denoted $\intbin{A_T^{(2)}}$ and defined in the next section), showing that the bias is small in this particular case. Though encouraging, this
remark will not prevent us from trying to simplify the comparison between theory and experiment by providing theoretical predictions for the \emph{exact} integrated quantities as measured in experiments. These kind of integrated observables have been already discussed in the context of other transverse asymmetries in Refs.~\cite{bobeth,bobeth4}. 

In the present paper we address precisely how to analyse efficiently the LHCb measurements on $B\to K^*\ell^+\ell^-$ at large recoil by choosing a set of clean observables 
integrated over $q^2$-bins. We begin by building a set of integrated observables that correspond in the limit of small binning to the observables in Ref.~\cite{primary}, with the aim of making contact with experimentally measurable quantities. In Sections \ref{sec:2} and \ref{sec:sm}, we present Standard Model predictions for these integrated observables, providing the results for different choices of the $q^2$-binning. In Section \ref{sec:const} we perform a model-independent analysis setting constraints on the Wilson coefficients $\C7^{(')}$, $\C9^{(')}$ and $\C{10}^{(')}$, using data from radiative $B$ decays and including the forward-backward asymmetry and the longitudinal polarization fraction of $B \to K^* \ell^+ \ell^-$, both integrated in the full low-$q^2$ bin $[1,6]$ GeV$^2$. This updates the analysis of Refs.~\cite{DescotesGenon:2011yn,1202.2172} (and related analyses) with several improvements. We use these results to identify a number of New Physics ``benchmark points'' that are allowed at the 95.5\% confidence level by all the constraints considered. In Section \ref{sec:comp} we discuss the potential complementarity of $q^2$-dependent observables in $B \to K^* \ell^+ \ell^-$ by analyzing our set of clean observables within the NP scenarios specified by the benchmark points. This analysis indicates the scenarios that are more likely to be affected predominantly by the binned observables in $B \to K^* \ell^+ \ell^-$.

In Section \ref{sec:S3vsP1} we open up a parenthesis to discuss the impact of hadronic uncertainties on the different observables, and the resulting model-independent constraints that follow from different choices of observables. We demonstrate the advantages of using a complete set of clean observables as the one introduced in Section \ref{sec:2} and Ref.~\cite{primary}. 

We then address the model-independent constraints from $q^2$-dependent observables in Section \ref{sec:exp}. We extract the experimental values for the three clean observables ($P_1,P_2$ and $P_3$) that can be related to the measurements provided in Ref.~\cite{LHCbinned}.  
Our determinations are affected by uncertainties that could be considerably improved, since we lack the experimental information concerning the correlation between the LHCb measurements, which is essential to assess uncertainties in a proper and accurate way. The constraints from $q^2$-dependent observables are studied in Section \ref{sec:const2}. First we consider the constraints from the measured $\afb$ and $F_L$ in the two $q^2$-bins $[2,4.3]$, $[4.3,8.68]$ GeV$^2$, then we turn to the constraints imposed by the clean observables $P_{1,2,3}$ in the same bins. In Section \ref{sec:compar}, we compare briefly our study with other similar works.

After concluding in Section \ref{sec:con}, we include in Appendix \ref{appA} the expression of the coefficient $J_8$ in term of observables and in Appendix \ref{appB} a collection of the relevant formulas used to derive the New Physics constraints. In particular we provide the explicit expressions of the clean integrated observables used throughout the paper, for the different choices of $q^2$-binning. In Appendix \ref{appC} we describe the statistical approach used in the fits.

As a summary of the most important findings, we anticipate the following conclusions of this work:
\begin{itemize}

\item Three ($P_{1,2,3}$) out of the six clean observables describing the massless distribution can already be extracted from current measurements, as shown in Section \ref{sec:exp}. While $P_3$ sets no relevant constraints yet, $P_1$ and $P_2$ are complementary to other radiative and semileptonic observables. Combining the measurements of $P_2$ in two different bins leads to a mild tension with respect to the SM, compatible at 95.5\%C.L. but pointing towards a negative contribution to the Wilson coefficient $\C7$.

\item The explicit form of the coefficients of the massless angular distribution in terms of the basis of observables is given in Eq.~(\ref{Jsintermsobs}). The expressions turn out to be very simple and exhibit two important features:

\begin{itemize}
\item A more natural basis devised to extract information from the distribution in a clean way emerges in the massless case, with a slight redefinition of the observables $P_{4,5,6} \to P_{4,5,6}^\prime$, which are also clean and defined in Eqs.~(\ref{P4'})-(\ref{P6'}) (in the SM $P_{4,5,6} \sim P_{4,5,6}^\prime$ to a very good approximation).

\item The clean observables in the natural basis ($P_{1,2,3} , P'_{4,5,6}$) can be related to form-factor sensitive observables $S_i$ through the following simple rule:
\eq{
\frac{S_{3,6,9}}{F_T} \to P_{1,2,3}   \quad,\quad \frac{S_{4,5,7}}{\sqrt{F_T F_L}} \to  P_{4,5,6}^\prime\ ,
}
where the exact relationships are given in Eqs.~(\ref{Pis}) and (\ref{P456}).

\end{itemize}
 
\item The ``flipped-sign solution" for $\C7$ is in general disfavoured by present data at the 95.5\% confidence level depending on the NP scenario considered. The isospin asymmetry in $B\to K^* \gamma$ plays an important role (independent of Wilson coefficients other than $\C7^{(')}$),  as well as the forward-backward asymmetry in $B\to K^*\ell\ell$. This confirms the result of Refs.~\cite{DescotesGenon:2011yn,1202.2172}. 
 
 \item We show explicitly the strong impact of the different computations available in the literature for the soft form factors on the theoretical uncertainties for observables like $A_{FB}$, $F_L$ and $S_3$, and the robustness of the clean observables $P_i$. While the impact on the theoretical error in $F_L$ is evident, the problem for observables like $S_3$ is more subtle. In the case of $S_3$ the theoretical uncertainty in the SM is protected by its tiny central value, but away from the SM point the impact can be substantial, preventing this observable from discriminating NP scenarios. None of these problems affect the clean observables $P_i$ or $A_T^i$.  
 
\end{itemize}

\section{Integrated observables in $q^2$-bins}
\label{sec:2}

The differential decay rate of the process $\bar B_d \to \bar K^*(\to K\pi) \ell^+ \ell^-$ can be written as:
\eqa{\label{dist}
\frac{d^4\Gamma}{dq^2\,d\!\cos\theta_K\,d\!\cos\theta_l\,d\phi}&=&\frac9{32\pi} \bigg[
J_{1s} \sin^2\theta_K + J_{1c} \cos^2\theta_K + (J_{2s} \sin^2\theta_K + J_{2c} \cos^2\theta_K) \cos 2\theta_l\nn\\[1.5mm]
&&\hspace{-2.7cm}+ J_3 \sin^2\theta_K \sin^2\theta_l \cos 2\phi + J_4 \sin 2\theta_K \sin 2\theta_l \cos\phi  + J_5 \sin 2\theta_K \sin\theta_l \cos\phi \nn\\[1.5mm]
&&\hspace{-2.7cm}+ (J_{6s} \sin^2\theta_K +  {J_{6c} \cos^2\theta_K})  \cos\theta_l    
+ J_7 \sin 2\theta_K \sin\theta_l \sin\phi  + J_8 \sin 2\theta_K \sin 2\theta_l \sin\phi \nn\\[1.5mm]
&&\hspace{-2.7cm}+ J_9 \sin^2\theta_K \sin^2\theta_l \sin 2\phi \bigg]\,,
}
where the kinematical variables $\phi$, $\theta_\ell$, $\theta_K$, $q^2$ are defined as in Refs.~\cite{bobeth,buras,primary}. The decay rate $\bar \Gamma$ of the CP-conjugated process $B_d \to K^*(\to K\pi) \ell^+ \ell^-$ is obtained from Eq.~(\ref{dist}) by replacing $J_{1,2,3,4,7}\to \bar J_{1,2,3,4,7}$ and $J_{5,6,8,9}\to -\bar J_{5,6,8,9}$, where $\bar J$ is equal to $J$ with all weak phases conjugated. This corresponds to the same definition of $\theta_\ell$ for both $B$ and $\bar B$ (see for example \cite{0805.2525,buras}). In this paper we assume that all the observables are CP-averaged, and so are always functions of $J_i + \bar J_i$. Therefore, $J_i\to J_i + \bar J_i$ and $\Gamma\to \Gamma+\bar\Gamma$ should be  understood in all the formulas below, and in particular all the observables $O(J)$ are assumed to be $O(J+\bar J)$.

In order to cope with limited statistics, one can write down integrated distributions, such as the uniangular distributions, which depend on a subset of coefficients $J_i$. This is the way observables such as $F_L$, $\afb$ or $A_T^{(2)}$ have been measured traditionally. A more recent approach deals with ``folded'' distributions, with the double advantage of increasing the statistics and focusing on a restricted set of angular coefficients. For example, in Ref.~\cite{LHCbinned}, the identification of $\phi \leftrightarrow \phi + \pi$ has been used to produce a ``folded'' angle $\hat \phi \in [0,\pi]$ in terms of which a (folded) differential rate $d\hat\Gamma(\hat \phi) = d\Gamma(\phi)+d\Gamma(\phi-\pi)$ becomes
\eqa{
\frac{d^4\hat\Gamma}{dq^2\,d\!\cos\theta_K\,d\!\cos\theta_l\,d\hat\phi} &=&
\frac{9}{16\pi}
\bigg[ J_{1c} \cos^2{\theta_K} + J_{1s} (1 - \cos^2{\theta_K})
+ J_{2c} \cos^2{\theta_K} (2\cos^2{\theta_\ell}-1)\nn\\[2mm]
&&\hspace{-1.2cm} + J_{2s} (1 - \cos^2{\theta_K}) (2\cos^2{\theta_\ell} - 1)
+ J_3 (1 - \cos^2{\theta_K})(1 - \cos^2{\theta_\ell}) \cos{2\hat \phi}\nn\\[2mm]
&&\hspace{-1.2cm} + J_{6s} (1 - \cos^2{\theta_K}) \cos{\theta_\ell}
+ J_9 (1 - \cos^2{\theta_K})(1 - \cos^2{\theta_\ell}) \sin{2\hat \phi}
\bigg]\ .
}
In the following we will neglect scalar and lepton mass effects. A detailed analysis of the impact of neglecting lepton masses can be found in Ref.~\cite{1209.1525}. Concerning scalar contributions, two observables called $S_1$ and $S_2$ were designed in Ref.~\cite{primary} to explore the measurable impact of scalar effects. However, the strong constraint from the $B_s\to \mu^+\mu^-$ branching ratio already makes these effects negligible. Still it will be interesting to include these corrections once enough statistics is collected.

In this approximation, this distribution can be written as a function of the observables in Ref.~\cite{primary} as follows:
\eqa{
\frac{d^4\Gamma}{dq^2\,d\!\cos\theta_K\,d\!\cos\theta_l\,d\hat\phi} &=&
\frac{9}{16\pi}
\bigg[ F_L \cos^2{\theta_K} + \frac34F_T (1 - \cos^2{\theta_K})
- F_L \cos^2{\theta_K} (2\cos^2{\theta_\ell}-1)\nn\\[2mm]
&&\hspace{-2.5cm} + \frac14F_T (1 - \cos^2{\theta_K}) (2\cos^2{\theta_\ell} - 1)
+ \frac12 P_1 F_T (1 - \cos^2{\theta_K})(1 - \cos^2{\theta_\ell}) \cos{2\hat \phi}\nn\\[2mm]
&&\hspace{-2.5cm} + 2 P_2 F_T (1 - \cos^2{\theta_K}) \cos{\theta_\ell}
- P_3 F_T (1 - \cos^2{\theta_K})(1 - \cos^2{\theta_\ell}) \sin{2\hat \phi}
\bigg]\, \frac{d\Gamma}{dq^2}\ ,
\label{fd}}
where $P_1$, $P_2$ and $P_3$ are theoretically clean observables \cite{primary} that in terms of form factor dependent observables\footnote{The observables $S_i$ are defined as $S_i=(J_i+\bar J_i)/(\bar\Gamma+\Gamma)$ \cite{buras}, while $\aim=S_9$ \cite{matias1}.} are given as
\eq{P_1 F_T = 2 S_3\ ,\quad P_2 F_T=S_{6s}/2 \ , \quad P_3 F_T= -S_9\ ,
\label{Pis}}
or alternatively\footnote{Note that the $\cos{\theta_\ell}$ term in Eq.~(\ref{fd}) has opposite sign with respect to Ref.~\cite{LHCbinned} because of the different definition of the angle $\theta_\ell$: $\theta_\ell^{\rm us}=\pi-\theta_\ell^{\rm LHCb}$ for the $\bar B$ decay.},
\eq{
P_2 F_T=-2 \afb/3\ ,\quad P_3  F_T = -\aim\ .\label{Pis2}}
The quantity $F_T$ is defined as $F_T\equiv 1-F_L$. The observable $P_1$ is better know by its original name, $A_T^{(2)}$ \cite{kruger}.

Experimentally, one can fit separately each of the following five independent coefficients that appear in the folded distribution in Eq.~(\ref{fd}):
\eq{
\begin{array}{lll}
\ds c_0(q^2)= \frac{d\Gamma}{dq^2}\ ,\quad
&
\ds c_1(q^2)= P_1 F_T\frac{d\Gamma}{dq^2}\ ,\quad
&
\ds c_2(q^2)= P_2 F_T\frac{d\Gamma}{dq^2}\ ,\\[8mm]
\ds c_3(q^2)= P_3 F_T \frac{d\Gamma}{dq^2} \ ,\quad
&
\ds c_4(q^2)= F_T \frac{d\Gamma}{dq^2}\ . & 
\end{array}
\label{cis}}
For each $q^2$ one can then in principle extract the theoretically clean observables $P_1\equiv A_T^{(2)}$, $P_2$ and $P_3$, as well as the transverse polarization fraction and the differential decay rate. However, in practice the $q^2$-dependence is discretised in a number of bins, and the coefficients $c_i(q^2)$ thus extracted are quantities integrated over particular $q^2$-intervals:
\eq{ \av{c_i}_{\rm bin}=\frac{\int_{{\rm bin}} dq^2 c_i(q^2)}{\int_{{\rm bin}} dq^2}\ .
\label{intci}}
One could hope to be able to extract also corresponding integrated theoretically clean observables such as
\eq{ \av{A_T^{(2)}}_{\rm bin}^{\rm ({\bf naive})}=\frac{\int_{{\rm bin}} dq^2 A_T^{(2)}(q^2) }{\int_{{\rm bin}} dq^2 }\ ,}
but due to the experimental procedure used, such a determination is achievable asymptotically only, as the bin size goes to zero. The actual theoretically clean quantities that can be extracted from experiment and on which we will focus from now on, must be composed of the integrated quantities in Eq.~(\ref{intci}):
\eqa{
\av{P_1}_{\rm bin}\equiv \av{A_T^{(2)}}_{\rm bin} &=& \frac{\int_{{\rm bin}} dq^2 c_1(q^2)}{\int_{{\rm bin}} dq^2 c_4(q^2)}=\frac{\av{c_1}_{\rm bin}}{\av{c_4}_{\rm bin}}\ ,\label{P1bar}\\
\av{P_2}_{\rm bin} &=& \frac{\int_{{\rm bin}} dq^2 c_2(q^2)}{\int_{{\rm bin}} dq^2 c_4(q^2)}
=\frac{\av{c_2}_{\rm bin}}{\av{c_4}_{\rm bin}}\ ,\label{P2bar}\\
\av{P_3}_{\rm bin} &=& \frac{\int_{{\rm bin}} dq^2 c_3(q^2)}{\int_{{\rm bin}} dq^2 c_4(q^2)}
=\frac{\av{c_3}_{\rm bin}}{\av{c_4}_{\rm bin}}\ .\label{P3bar}
}
Other observables are accessible to the current LHCb data set by means of similar partial angular analyses. Three observables related to $P_4$, $P_5$ and $P_6$ (see Ref.~\cite{primary}) could be extracted by the LHCb collaboration in the near future. This means that, without actually performing a full angular analysis, the LHCb collaboration could be able to provide measurements of the complete set of 8 observables that describe the full distribution in the massless approximation (six of them being theoretically clean \cite{primary}). Therefore, we will work here under the assumption that integrated versions (exactly as in Eq.~(\ref{intci})) of the observables $c_{0,1,2,3,4}$ and $J_{4,5,7}$ are available experimentally. Any measurable observable must be a combination of the $\av{c_i}_{\rm bin}$ in Eq.~(\ref{intci}), and of the observables:
\eq{ \av{J_{4,5,7}}_{\rm bin}=\frac{\int_{{\rm bin}} dq^2 J_{4,5,7}(q^2)}{\int_{{\rm bin}} dq^2}\ .
\label{intJi}}

The coefficients of the angular distribution can be written in terms of the basis of observables (see Ref.~\cite{primary}), and in terms of the coefficients $c_i$ of Eq.~(\ref{cis}) as follows \footnote{A generalization of this parameterization including scalars and lepton masses can be found in Ref.~\cite{primary}. An alternative parametrization including lepton masses is given in Ref.~\cite{1209.1525}.}:

 \al{\label{Jsintermsobs}
J_{1s} &=   \dfrac{3}{4} F_T \frac{d\Gamma}{dq^2} =  \dfrac{3}{4} c_4\ , &
& J_{2s} = \dfrac{1}{4} F_T \frac{d\Gamma}{dq^2} = \dfrac{1}{4} c_4   \nn \\
J_{1c} &= F_L \frac{d\Gamma}{dq^2} =  c_0-c_4\ ,&
&J_{2c} = - F_L \frac{d\Gamma}{dq^2} = c_4-c_0\ ,\nn\\
J_3  &= \dfrac{1}{2} P_1 F_T \frac{d\Gamma}{dq^2} = \dfrac{1}{2} c_1\ ,&
& J_{6s} = 2 P_2 F_T \frac{d\Gamma}{dq^2} = 2 c_2\ ,\nn\\
J_4 &= \dfrac1{2} {P_4'} \sqrt{F_T F_L}\ \frac{d\Gamma}{dq^2} 
= \dfrac1{2} {P_4'} \sqrt{c_4 (c_0-c_4)}\ ,&
& J_9 = - P_3 F_T \frac{d\Gamma}{dq^2} = - c_3\ ,\nn\\
 J_5 &=  P_5' \sqrt{F_T F_L}\ \frac{d\Gamma}{dq^2} = P_5' \sqrt{c_4 (c_0-c_4)} \ ,\nn\\
J_7 &= - P_6' \sqrt{F_T F_L }\ \frac{d\Gamma}{dq^2} = - P_6' \sqrt{c_4 (c_0-c_4)}\ ,   }
 where the primed observables are defined as:
\eqa{
P_4' \equiv  P_4 \sqrt{1-P_1 } &=& \frac{J_4}{\sqrt{-J_{2c} J_{2s}}} \label{P4'}\\
P_5' \equiv  P_5 \sqrt{1+P_1 } &=& \frac{J_5}{2 \sqrt{-J_{2c} J_{2s}}} \\
P_6' \equiv  P_6 \sqrt{1-P_1 } &=& -\frac{J_7}{2 \sqrt{-J_{2c} J_{2s}}} 
\label{P6'}}
The case of the coefficient $J_8$ is discussed separately in detail in Appendix \ref{appA}.
These observables $P_{4,5,6}'$ are clean and coincide to a good approximation with $P_{4,5,6}$ in the SM (due to the fact that $P_1 \simeq 0$ in the SM). The whole analysis can be performed directly in terms of the observables $P_{4,5,6}\,$; however, from the experimental point of view, fitting the primed observables is simpler and more efficient.

These observables can be related to the observables $S_{4,5,7}$ of Ref.~\cite{buras}:
\eq{
P_4' = 2 \frac{S_4}{\sqrt{F_T F_L}}\ ,\quad
P_5' = \frac{S_5}{\sqrt{F_T F_L}} \ ,\quad
P_6' = -\frac{S_7}{\sqrt{F_T F_L}}.
\label{P456}}
There is therefore no particular advantage for the experimental extraction of the observables $S_i$ instead of the $P_i'$, while from the theory point of view the $P_i'$ are under better control and suffer from smaller uncertainties. 

We want now to construct the theoretically clean  integrated observables that correspond to those in Ref.~\cite{primary} (or variations thereof). For $P_{1,2,3}$ the answer is precisely $\av{P_1}_{{\rm bin}}$, $\av{P_2}_{{\rm bin}}$ and $\av{P_3}_{{\rm bin}}$ defined in Eqs.~(\ref{P1bar}-\ref{P3bar}).
In analogy with Eqs.~(\ref{P1bar}-\ref{P3bar}), integrated versions of the observables $P'_{4,5,6}$ can be defined:
\eqa{
\av{P_4'}_\bin &=& \frac{2 \int_{{\rm bin}} dq^2 J_4(q^2)}{\sqrt{\int_{{\rm bin}} dq^2 c_4(q^2)  \int_{{\rm bin}} dq^2 (c_0(q^2)-c_4(q^2))}} = \frac{2 \intbin{J_4}}{\sqrt{\intbin{c_4}\big(\intbin{c_0}-\intbin{c_4}\big)}}\ ,
\label{P4bar}\\
\av{P_5'}_\bin &=& \frac{\int_{{\rm bin}} dq^2 J_5(q^2)}{\sqrt{\int_{{\rm bin}} dq^2 c_4(q^2)  \int_{{\rm bin}} dq^2 (c_0(q^2)-c_4(q^2))}}
= \frac{\intbin{J_5}}{\sqrt{\intbin{c_4}\big(\intbin{c_0}-\intbin{c_4}\big)}}\ ,
\label{P5bar}\\
\av{P_6'}_\bin &=&\frac{-\int_{{\rm bin}} dq^2 J_7(q^2)}{\sqrt{\int_{{\rm bin}} dq^2 c_4(q^2)  \int_{{\rm bin}} dq^2 (c_0(q^2)-c_4(q^2))}}
= \frac{-\intbin{J_7}}{\sqrt{\intbin{c_4}\big(\intbin{c_0}-\intbin{c_4}\big)}}\ .\label{P6bar}\hspace{0.8cm}
}
Finally, integrated versions of the longitudinal polarization fraction $F_L$ and the forward-backward asymmetry $A_{\rm FB}$ can be defined in terms of the coefficients $c_i$ in the following way:
\eqa{
\av{A_{\rm FB}}_\bin &=& -\frac32 \frac{\int_{{\rm bin}} dq^2 c_2(q^2)}{\int_{{\rm bin}} dq^2 c_0(q^2)}=
-\frac32 \frac{\av{c_2}_{\rm bin}}{\av{c_0}_{\rm bin}}\ ,\label{AFBbar}\\[2mm]
\av{F_L}_\bin &=& \frac{\int_{{\rm bin}} dq^2 (c_0(q^2)-c_4(q^2))}{\int_{{\rm bin}} dq^2 c_0(q^2)}
=\frac{\av{c_0}_{\rm bin} - \av{c_4}_{\rm bin}}{\av{c_0}_{\rm bin}}\ .\label{FLbar}
}\\
In the following sections we will study these integrated observables in detail.

\section{SM predictions for integrated observables}
\label{sec:sm}

\begin{figure}\begin{center}
\includegraphics[height=15.5cm]{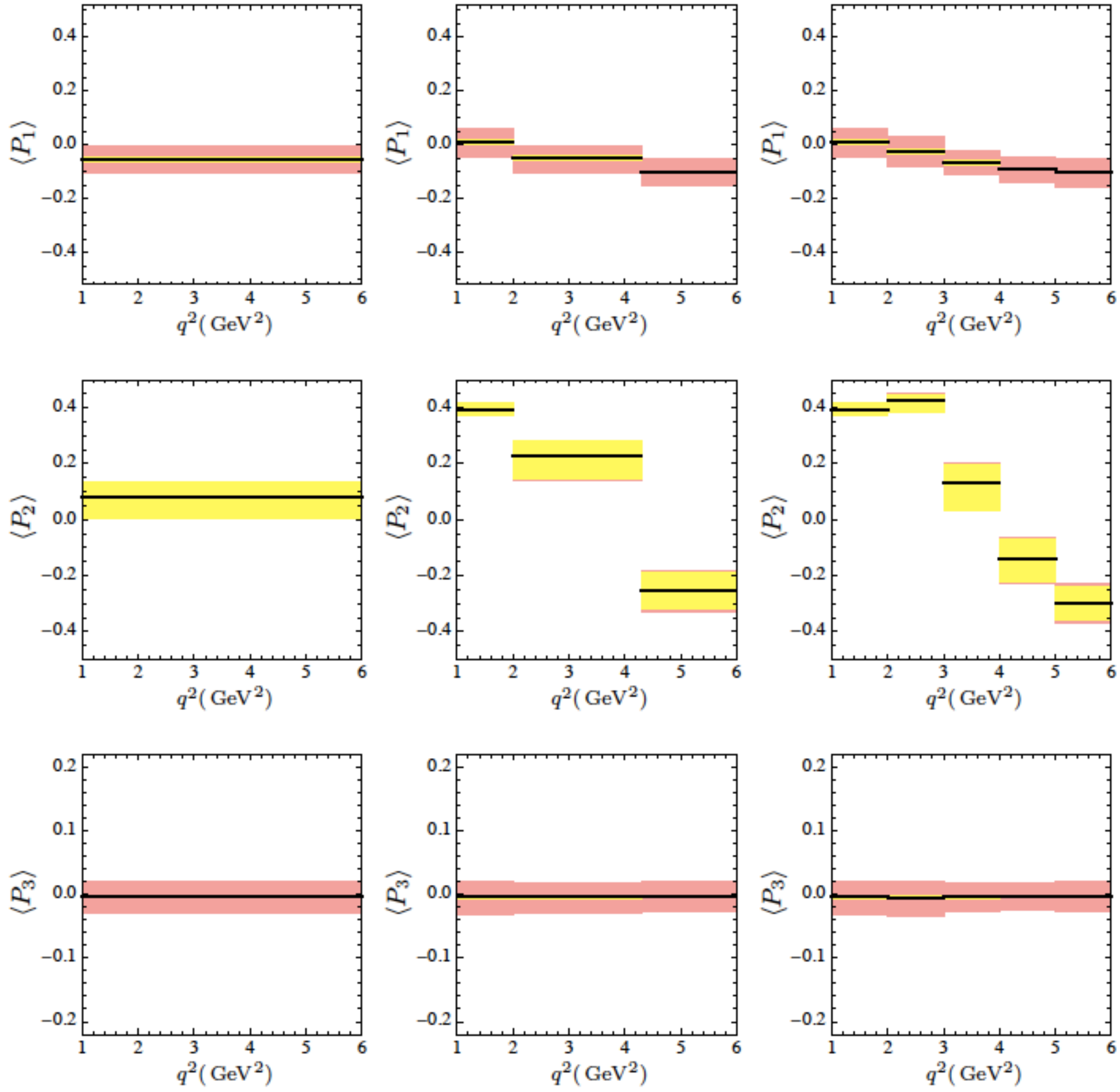}
\end{center}
\vspace{-0.4cm}
\caption{Binned Standard Model predictions for the clean observables $\av{P_{1,2,3}}$, for a single bin $[1,6]$ GeV$^2$ (left column), three bins $[1,2]$, $[2,4.3]$, $[4.3,6]$ GeV$^2$ (central column), and five bins of width 1 GeV$^2$ each (right column). The red (dark gray) error bar correspond to the $\Lambda/m_b$ corrections, the yellow one (light gray) to the other sources of uncertainties. If one of the two bands is missing, it means the associated uncertainty is negligible compared to the dominant one.
}
\label{SMplotsPs1}
\end{figure}

\begin{figure}\begin{center}
\includegraphics[height=15.5cm]{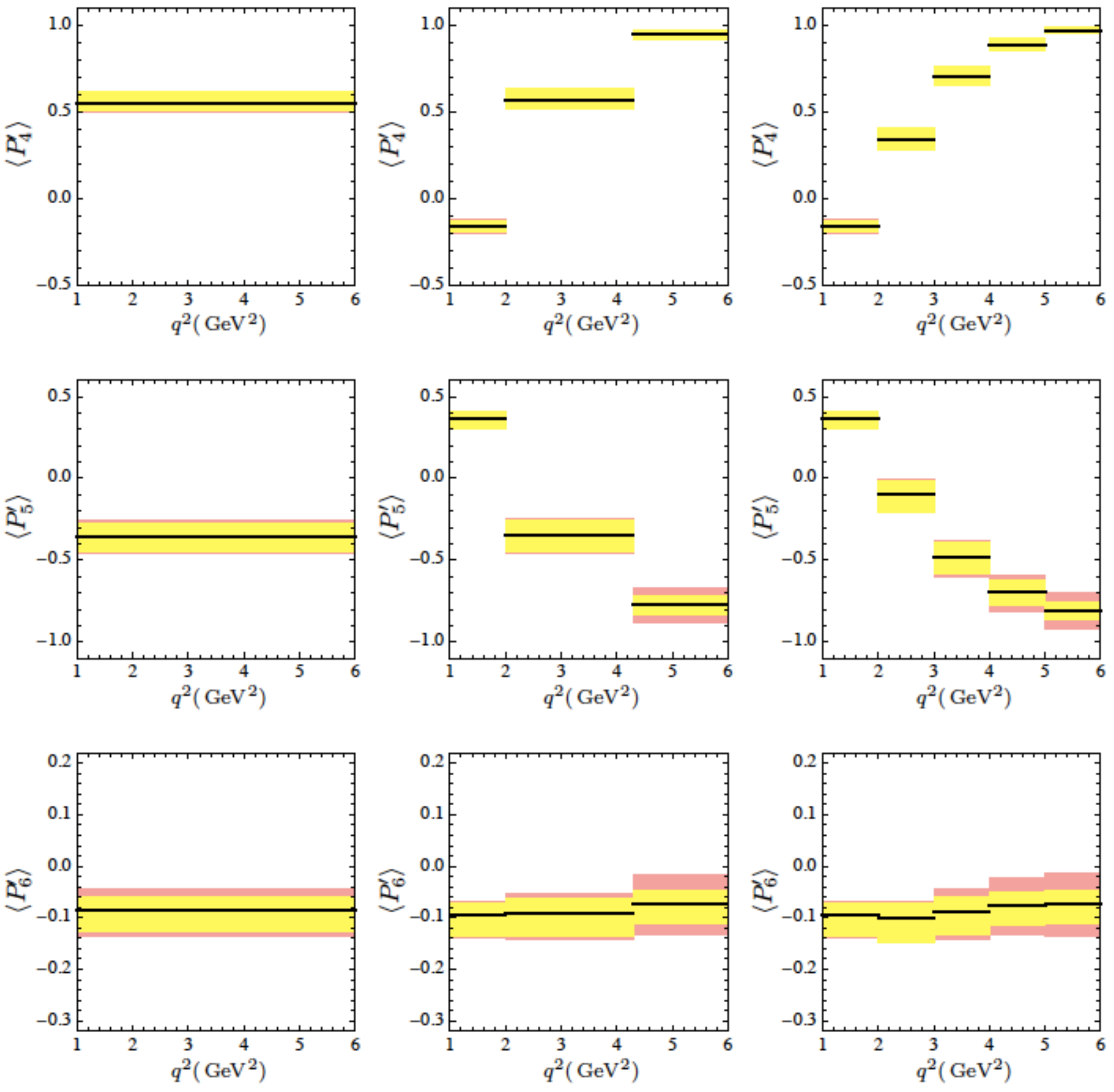}
\end{center}
\vspace{-0.4cm}
\caption{Binned Standard Model predictions for the clean observables $\av{P'_{4,5,6}}$,  with the same conventions as in Fig.~\ref{SMplotsPs1}.}
\label{SMplotsPs2}
\end{figure}

\begin{figure}\centering
\includegraphics[height=9cm,width=15cm]{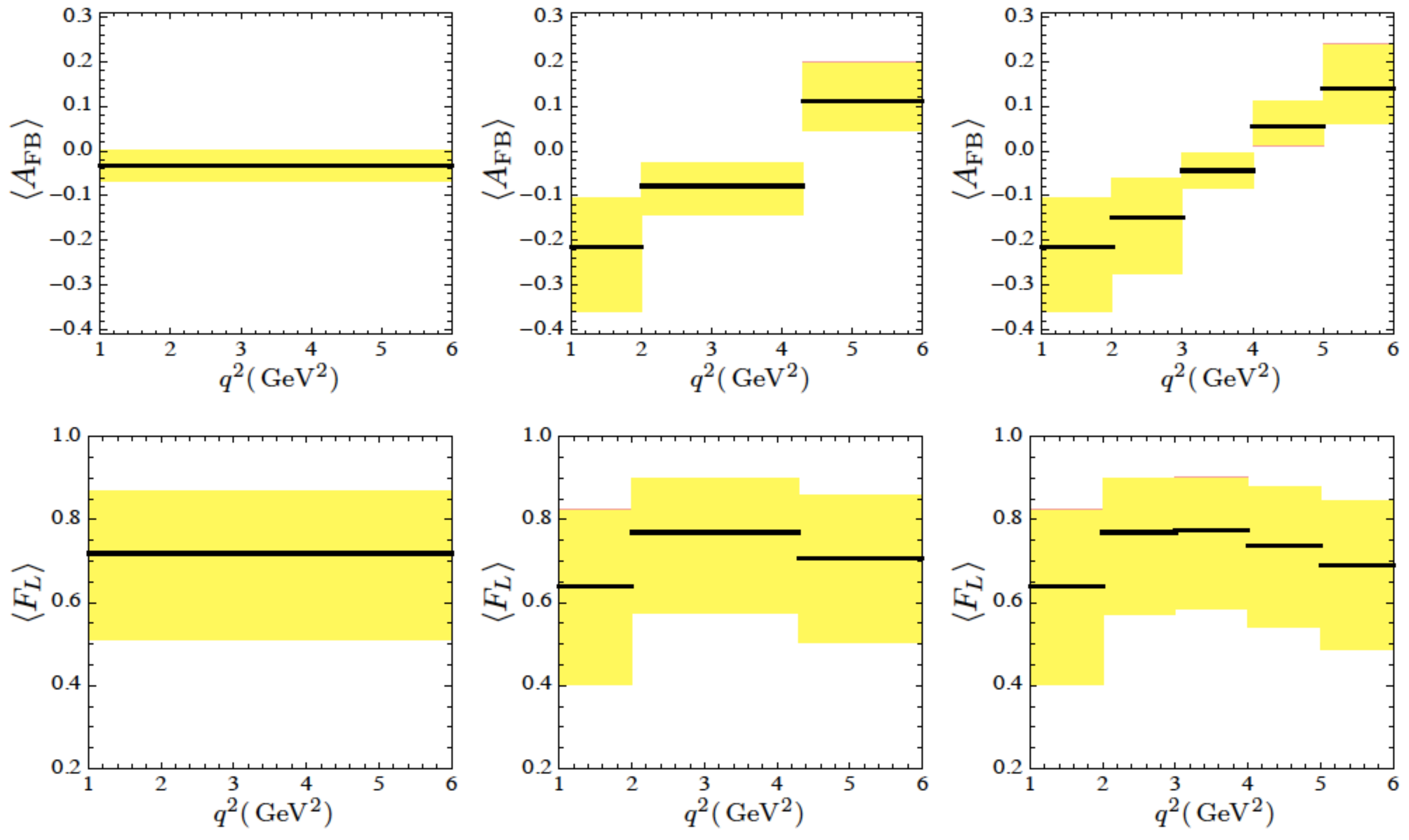}
\caption{Binned Standard Model predictions for the observables $\av{A_{\rm FB}}$ and $\av{F_L}$,  with the same conventions as in Fig.~\ref{SMplotsPs1}.}
\label{SMplotsAFBFL}
\end{figure}

We can provide SM predictions for the set of integrated observables $\av{P_i}$ as well as $\av{\afb}$ and $\av{F_L}$. In Tables \ref{TableSMPs} and \ref{TableSMAFBFL} we show the predictions in the $q^2$-bins [1,2], [2,4.3], and [4.3,6] (GeV$^2$) -- following the binning  used by the experimental collaborations up to now (except for the first bin) -- as well as the predictions for the integrated low-$q^2$ observables, in the region [1,6] GeV$^2$. The first error accounts for all parametric uncertainties, while the second error corresponds to an estimate of $\Lambda/m_b$ corrections, as described below. In Figs.~\ref{SMplotsPs1}, \ref{SMplotsPs2} and \ref{SMplotsAFBFL} we show the corresponding SM predictions for the observables in the case of one and three bins (corresponding to the predictions in Tables \ref{TableSMPs} and \ref{TableSMAFBFL}), as well as for five bins with a width  of 1 GeV$^2$.
\begin{table}
\small
\begin{center}
\begin{tabular}{||c||r|r|r||r||}
\multicolumn{1}{}{}   \\
\hline\hline
\begin{minipage}{1.7cm}
\centering
\vspace{4mm}
$q^2$~(GeV$^2$)
\vspace{0.5mm}
\end{minipage}
& $[\,1\,,\,2\,]$\hspace{1cm}  & $[\,2\,,\,4.3\,]$\hspace{1cm} & $[\,4.3\,,\,6\,]$\hspace{1cm} & $[\,1\,,\,6\,]$\hspace{1cm} \\ 
\hline\hline
%
\begin{minipage}{1cm}
\centering
\vspace{3mm}
$\av{P_1}$
\vspace{3mm}
\end{minipage}  & 
$0.008^{+0.009 + 0.051}_{- 0.005 - 0.053}$ &
$-0.051^{+0.010 + 0.048}_{- 0.009 - 0.050}$ &
$-0.100^{+0.001 + 0.049}_{- 0.001 - 0.053}$ & 
$-0.055^{+0.009 + 0.049}_{- 0.008 - 0.052}$\\
\hline
\begin{minipage}{1cm}
\centering
\vspace{3mm}
$\av{P_2}$
\vspace{3mm}
\end{minipage} & 
$0.395^{+0.020 + 0.011}_{- 0.021 - 0.012}$ &
$0.227^{+0.055 + 0.014}_{- 0.083 - 0.016}$ &
$-0.254^{+0.063 + 0.034}_{- 0.068 - 0.035}$ & 
$0.080^{+0.054 + 0.020}_{- 0.073 - 0.021}$\\
\hline
\begin{minipage}{1cm}
\centering
\vspace{3mm}
$\av{P_3}$
\vspace{3mm}
\end{minipage}  & 
$-0.003^{+0.001 + 0.025}_{- 0.002 - 0.028}$ &
$-0.004^{+0.001 + 0.023}_{- 0.003 - 0.025}$ &
$-0.002^{+0.001 + 0.022}_{- 0.002 - 0.024}$ & 
$-0.003^{+0.001 + 0.023}_{- 0.002 - 0.025}$\\
\hline
\begin{minipage}{1cm}
\centering
\vspace{3mm}
$\av{P_4'}$
\vspace{3mm}
\end{minipage}  & 
$-0.160^{+0.036 + 0.024}_{- 0.027 - 0.025}$ &
$0.570^{+0.067 + 0.000}_{- 0.054 - 0.002}$ &
$0.944^{+0.025 + 0.000}_{- 0.025 - 0.004}$ & 
$0.553^{+0.060 + 0.004}_{- 0.050 - 0.008}$\\
\hline
\begin{minipage}{1cm}
\centering
\vspace{3mm}
$\av{P_5'}$
\vspace{3mm}
\end{minipage} & 
$0.369^{+0.044 + 0.000}_{- 0.061 - 0.002}$ &
$-0.343^{+0.089 + 0.043}_{- 0.108 - 0.046}$ &
$-0.774^{+0.061 + 0.087}_{- 0.059 - 0.093}$ & 
$-0.353^{+0.081 + 0.050}_{- 0.095 - 0.053}$\\
\hline
\begin{minipage}{1cm}
\centering
\vspace{3mm}
$\av{P_6'}$
\vspace{3mm}
\end{minipage}  & 
$-0.095^{+0.025 + 0.012}_{- 0.042 - 0.011}$ &
$-0.092^{+0.029 + 0.026}_{- 0.045 - 0.024}$ &
$-0.074^{+0.027 + 0.051}_{- 0.038 - 0.046}$ & 
$-0.085^{+0.027 + 0.033}_{- 0.041 - 0.029}$\\
%
\hline\hline 
\end{tabular}
\caption{SM predictions for the clean observables $\av{P_i}$.
}
\label{TableSMPs}
\end{center}
\end{table}
\begin{table}
\begin{center}
\begin{tabular}{||c||r|r|r||r||}
\multicolumn{1}{}{}   \\
\hline\hline
\begin{minipage}{1.7cm}
\centering
\vspace{4mm}
$q^2$~(GeV$^2$)
\vspace{0.5mm}
\end{minipage}
& $[\,1\,,\,2\,]$\hspace{1cm}  & $[\,2\,,\,4.3\,]$\hspace{1cm} & $[\,4.3\,,\,6\,]$\hspace{1cm} & $[\,1\,,\,6\,]$\hspace{1cm} \\ 
\hline\hline
%
\begin{minipage}{1cm}
\centering
\vspace{3mm}
$\av{A_{\rm FB}}$
\vspace{3mm}
\end{minipage}  & 
$-0.214^{+0.111 + 0.003}_{- 0.144 - 0.002}$ &
$-0.079^{+0.053 + 0.004}_{- 0.065 - 0.003}$ &
$0.112^{+0.086 + 0.017}_{- 0.065 - 0.016}$ & 
$-0.034^{+0.035 + 0.009}_{- 0.033 - 0.008}$\\
\hline
\begin{minipage}{1cm}
\centering
\vspace{3mm}
$\av{F_L}$
\vspace{3mm}
\end{minipage} & 
$0.638^{+0.185 + 0.007}_{- 0.236 - 0.006}$ &
$0.769^{+0.129 + 0.006}_{- 0.194 - 0.006}$ &
$0.706^{+0.151 + 0.004}_{- 0.201 - 0.004}$ & 
$0.719^{+0.149 + 0.006}_{- 0.208 - 0.006}$\\
%
\hline\hline 
\end{tabular}
\caption{SM predictions for $\av{\afb}$ and $\av{F_L}$.
}
\label{TableSMAFBFL}
\end{center}
\end{table}

The SM predictions are obtained as follows. The observables integrated over each bin are defined in terms of the coefficients $c_i(q^2)$ in Eqs.~(\ref{P1bar})-(\ref{P3bar}), (\ref{P4bar})-(\ref{P6bar}), (\ref{AFBbar}) and (\ref{FLbar}). The coefficients $c_i(q^2)$ are simple functions of transversity amplitudes (see for example Ref.~\cite{primary}). The transversity amplitudes can be written in terms of Wilson coefficients and $B\to K^*$ form factors following Refs.~\cite{0106067,0412400}. Concerning the Wilson coefficients, the form factors, and the treatment of uncertainties, we proceed as in Refs.~\cite{primary,DescotesGenon:2011yn,1202.2172}, with a slight revision in the treatment of form factors:\\

\n {\bf Wilson coefficients:} The SM Wilson coefficients are evaluated at the matching scale $\mu_0=2 M_W$, and evolved down to the hadronic scale $\mu_b=4.8\,{\rm GeV}$ following Refs.~\cite{0512066,0306079,0411071,0312090,0609241}. The running of quark masses and couplings proceeds analogously. The SM Wilson coefficients at the scale $\mu_b$ are shown in Table~\ref{TabInputs}.\\

\begin{table}
\begin{center}
\begin{tabular}{||lr|lr||}
\hline\hline
$\mu_b=4.8 \ {\rm GeV}$ &   & $\mu_0=2M_W$ &  \cite{Misiak:2006zs} \\
\hline
$m_B=5.27950   \ {\rm GeV}$ & \cite{pdg} & $m_{K^*}=0.89594   \ {\rm GeV}$  & \cite{pdg} \\
\hline
$m_{B_s} = 5.3663  \ {\rm{GeV}}$ & \cite{pdg} & $m_{\mu} = 0.105658367  \ {\rm GeV}$ & \cite{pdg} \\
\hline
$ \sin^2 \theta_W                      = 0.2313$ & \cite{pdg}         &  & \\
$M_W=80.399\pm 0.023  \ {\rm GeV}$ &  \cite{pdg} &
$M_Z=91.1876 \ {\rm GeV}$ &  \cite{pdg} \\
\hline
$ \alpha_{em}(M_Z)                     =1/128.940 $ & \cite{Misiak:2006zs}&
$ \alpha_s(M_Z)                        = 0.1184 \pm 0.0007  $ & \cite{pdg} \\
\hline
$ m_t^{\rm pole} = 173.3\pm 1.1  \ {\rm GeV}      $ & \cite{Alcaraz:2009jr}
 & $ m_b^{1S}                   = 4.68 \pm 0.03   \ {\rm GeV}   $  &\cite{Bauer:2004ve} \\
$ m_c^{\overline{MS}}(m_c)                   = 1.27 \pm 0.09  \ {\rm GeV}    $ & \cite{pdg} &
$ m_s^{\overline{MS}}(2\ {\rm GeV})=0.101 \pm 0.029 \ {\rm GeV}$ &   \cite{pdg} \\
\hline
$\lambda_{CKM}=0.22543\pm 0.0008$ &  \cite{ckmfitter} &
$A_{CKM}=0.805\pm 0.020$ & \cite{ckmfitter}  \\
$\bar\rho=0.144\pm 0.025$ & \cite{ckmfitter} &
$\bar\eta=0.342\pm 0.016$ & \cite{ckmfitter}  \\
\hline
${\cal B}(B \to X_c e \bar\nu) = 0.1061 \pm 0.00017$ 
\hspace{0.15cm}
& \cite{Misiak:2006zs} & $ C                                  = 0.58 \pm 0.016     $  & \cite{Misiak:2006zs} \\
$ \lambda_2                            = 0.12      \ {\rm GeV}^2         $ & \cite{Misiak:2006zs} & &\\
\hline
$ \Lambda_h=0.5\ {\rm GeV}$ & \cite{Kagan:2001zk} & $ f_B = 0.190 \pm 0.004\ {\rm GeV}    $ & \cite{1203.3862}\\
$f_{K^*,||}=0.220 \pm 0.005$\ {\rm GeV} & \cite{buras}
 & $f_{K^*,\perp}(2\ {\rm GeV})=0.163(8)\ {\rm GeV}$
 \hspace{0.15cm}
 & \cite{buras}\\
$ V(0)=    0.36^{+0.23}_{-0.12}$ &  \cite{1006.4945} &$  A_0(0)=    0.29^{+0.10}_{-0.07}  $ & \cite{1006.4945}\\
$ a_{1,||,\perp}(2\ {\rm GeV})=0.03\pm 0.03$ & \cite{buras}& $ a_{2,||,\perp}(2\ {\rm GeV})=0.08\pm 0.06$ & \cite{buras}\\
$\lambda_B(\mu_h)=0.51\pm 0.12\ {\rm GeV}$& \cite{buras}&&\\
\hline
$ f_{B_s} = 0.227 \pm 0.004\ {\rm GeV}$ & \cite{1203.3862} & $ \tau_{B_s} = 1.497 \pm 0.015\ {\rm ps}    $ & \cite{pdg} \\[1mm]
\end{tabular}
{\small
\begin{tabular}{||c|c|c|c|c|c|c|c|c|c||}
\hline\hline
$\!\C1(\mu_b)\!$ &   $\!\C2(\mu_b)\!$ &  $\!\C3(\mu_b)\!$ &  $\!\C4(\mu_b)\!$
& $\!\C5(\mu_b)\!$ & $\!\C6(\mu_b)\!$ & $\!\C7^{\rm eff}(\mu_b)\!$ & $\!\C8^{\rm eff}(\mu_b)\!$
& $\!\C9(\mu_b)\!$ & $\! \C{10}(\mu_b)\!$ \\
\hline
-0.2632 & 1.0111 & -0.0055 & -0.0806 & 0.0004 &
0.0009 &  -0.2923 & -0.1663 & 4.0749 & -4.3085\\
\hline\hline
\end{tabular}
}
\end{center}
\caption{Input parameters used in the analysis and Wilson coefficients at $\mu_b$.}
\label{TabInputs}
\end{table}

\n {\bf Form factors:} There are seven $B\to K^*$ form factors: $V(q^2)$, $A_{0,1,2}(q^2)$ and $T_{1,2,3}(q^2)$. Their determination involves the computation at $q^2=0$ and the parameterization of the $q^2$-dependence. At $q^2=0$, these form factors can be obtained from light-cone sum rules with B-meson distribution amplitudes (see Ref.~\cite{1006.4945}). Concerning their dependence on $q^2$, Ref.~\cite{1006.4945} provides a conservative and convenient parameterization (the prospects for these form factors from lattice QCD have been discussed in detail in Ref.~\cite{becirevic}). The soft form factors $\xi_{\|,\bot}$ are defined in terms of the full form factors following Ref.~\cite{0412400}. The soft form factor $\xi_\bot(0)$ at $q^2=0$ is obtained directly from $V(0)$ as given in Ref.~\cite{1006.4945}. The form factor $\xi_\|(0)$ is defined as a combination of the form factors $A_1(0)$ and $A_2(0)$, and in the large-recoil limit and at leading order in $\alpha_s$ it is proportional to $A_0(0)$. We use $A_0(0)$ to fix $\xi_\|(q^2)$ at $q^2=0$ to a good accuracy, and set its $q^2$-dependence to reproduce its exact expression in terms of $A_1(q^2)$ and $A_2(q^2)$ using the parameterization of Ref.~\cite{1006.4945}.   The numerical inputs used are collected in Table~\ref{TabInputs}. 

In the present paper we take the form factors of Ref.~\cite{1006.4945} for two reasons. The first one is to be consistent with the analysis of Ref.~\cite{primary}. The second is to be conservative in the treatment of hadronic uncertainties, showing at the same time that clean observables are mostly insensitive to this choice.
The use of the value for $V(0)$ from Ref.~\cite{buras} would shift the central values of $\afb$ and $F_L$ in the whole low-$q^2$ region, while only a mild effect
around $q^2 \simeq$ 6 GeV$^2$ would be seen in some of the $P_i$ observables. 
These form factors have much larger uncertainties than those of  Refs.~\cite{buras,0412079}, and translate into large error bars in $\afb$, $F_L$ and other form factor dependent observables.

This difference in the size of the uncertainties can be partly explained by the approaches taken to apply light-cone sum rules in Refs.~\cite{buras,0412079} and \cite{1006.4945}. In Refs.~\cite{buras,0412079}, the sum rules are written using the light-meson distribution amplitudes up to twist 4 and including ${\cal O}(\alpha_s)$ corrections. In Ref.~\cite{1006.4945}, the sum rules are written for the $B$-meson distribution amplitudes up to twist 3 -- they include the (significant) soft-gluon emission from charm loops not considered in Refs.~\cite{buras,0412079}, but neglect the radiative corrections included in these references, as well as $1/m_b$ HQET corrections. Therefore the two analyses are only partially comparable, which explains why the quoted uncertainties differ in size. But one should also emphasise that beyond the approach taken, the hadronic inputs, i.e., the models used for the distribution amplitudes, play a crucial role concerning the uncertainties quoted for the form factors: in the case of light mesons \cite{buras,0412079}, the shapes are constrained by results coming from other light-cone sum rules, whereas in the case of the $B$ meson \cite{1006.4945}, a large range of variation for the shape models is allowed. At any rate, the discrepancy between Refs.~\cite{buras,0412079} and Ref.~\cite{1006.4945}
is a clear indication that the theoretical uncertainties attached to these observables (i.e., $\afb$, $F_L$) in the literature should be considered with a healthy dose of skepticism, and are strongly dependent on the choice of the $B\to K^*$ form factors.
Fortunately, it will be seen in Section \ref{sec:S3vsP1} that the error bars for the clean observables $P_i$ \emph{are not} affected by this variation of the form factor uncertainties and remain under good theoretical control.\\

\n {\bf Uncertainties:} We consider five main sources of uncertainties: the renormalization scale $\mu_b$, the quark mass ratio $\hat m_c\equiv m_c/m_b$, the form factors, the factor that determines the relative size of the hard-scattering term with respect to the form factor contribution to the amplitude (defined in Eq.~(55) of Ref.~\cite{0106067}) and the $\Lambda/m_b$ power corrections.

We follow the usual procedure consisting in varying the renormalization scale from $\mu_b/2$ to $2\mu_b$. For $\hat m_c$ we take $\hat m_c=0.29\pm 0.02$ (see Refs.~\cite{0106067,0103087}). Concerning the form factors: we express all the observables as a function of $A_0(q^2)$ and $V(q^2)$. The $q^2$-parameterization of these form factors (that of Ref.~\cite{1006.4945}) depends on the normalisation parameters $A_0(0)$, $V(0)$ as well as $b_{1}^V$ and $b_{1}^{A_0}$ (encoding the $q^2$-dependence of the form factor in the so-called $z$-parametrisation). We vary simultaneously $V(0)$ and $b_{1}^V$, and independently, $A_0(0)$ and $b_{1}^{A_0}$, within the errors quoted in Table \ref{TabInputs} and Ref.~\cite{1006.4945}. The two errors are added in quadrature. Concerning the parameter describing the relative weight of the hard-scattering contribution compared to the form-factor one in Ref.~\cite{0106067}, its error is estimated at the level of a 25\%, where its reduction with respect to Ref.~\cite{primary} (where it was 30\%) is due to the updated value of $f_B$ \cite{1203.3862}. Finally, for $\Lambda/m_b$ corrections, we follow the statistical procedure outlined in Ref.~\cite{matias2} to produce upper and lower 1$\sigma$ ranges consistent with a generic 10\% contribution of power corrections to the amplitudes. All individual uncertainties are considered separately and their impact on each observable is monitored to produce asymmetric upper and lower errors. All upper and lower uncertainties, excluding $\Lambda/m_b$ corrections, are added separately in quadrature to produce the first asymmetric error bars shown in Tables~\ref{TableSMPs} and \ref{TableSMAFBFL}, and the yellow bands in Figs.~\ref{SMplotsPs1}, \ref{SMplotsPs2} and \ref{SMplotsAFBFL}. The second error bars in Tables~\ref{TableSMPs} and \ref{TableSMAFBFL} correspond to the $\Lambda/m_b$ corrections. In Figs.~\ref{SMplotsPs1}, \ref{SMplotsPs2} and \ref{SMplotsAFBFL}, both uncertainties are added linearly to give the larger red error bands.

As can be seen from the plots, some observables appear to be almost insensitive to power corrections. This merely indicates that the hadronic uncertainties are much more important than power corrections for these observables.

While the observables $P_i$ are almost insensitive to the choice of form factors, the uncertainties of other observables vary substantially if the form factors of Refs.~\cite{buras,0412079} are used instead. For example, in Table~\ref{TableSMAFBFL} we quote the following prediction for $F_L$: $\av{F_L}_{[1,6]}=0.719^{+0.149 + 0.006}_{- 0.208 - 0.006}$, while if we take the form factors of Ref.~\cite{buras} we obtain $\av{F_L}_{[1,6]}=0.809^{+0.045 + 0.008}_{- 0.032 - 0.002}$ (see Figure 1 of Ref.~\cite{1209.0262}).

It could be wrongly concluded that $F_L$ has a small error in comparison with the predictions for the $P_i$, if one compares the percentage over the central value of the observable. However, the percentage with respect to the central value is not a sensible measure to compare the size of the errors. In the case of $F_L$, defined in the range [0,1], the size of the error should be compared with 1. Otherwise, taking instead the observable $F_T$=1-$F_L$ (which has a central value 5 times smaller) one would judge the error as 5 times larger, while the two observables are effectively the same. For the observables $P_i$, defined in the range [-1,1] (this is also true approximately for $P'_{4,5,6}$), the same argument applies, and therefore the error percentage should be evaluated over 2 and not the central value. In practice one is concerned with the discriminating power between SM and NP points, which gives further support to this argument: since one expects general (unconstrained) NP to give contributions in the whole range of the observable, the size of the error (as a measure of uncertainty of the position of the SM point) should be compared to the full range.

If this prescription is adopted, the relative errors of the $P_i$ are 2.5\%, 3.5\%, 1.2\%, 3\%, 5\%, 2.5\% for $i=1...6$, while for $F_L$ is 18\%. These are the results obtained using the form factors in Ref.~\cite{1006.4945}. If instead Ref.~\cite{buras} is used for the form factors, the errors of the $P_i$ barely change, while the relative error for $F_L$ goes down to 5\%. This can be seen in Figure 1 of Ref.~\cite{1209.0262}. This is one of the benefits of the clean observables: there is no need to rely on complicated estimations of errors in the light-cone sum-rules procedure, as explained above.

The sensitivity to form factor uncertainties of observables such as $S_3$ is more subtle and will be discussed in Section~\ref{sec:S3vsP1}.


\begin{table}
\centering
\begin{tabular}{||c|cr|cr||}
\hline\hline
Observable & \multicolumn{2}{|c|}{Experiment} & \multicolumn{2}{|c||}{SM prediction}\\
\hline\hline
$BR(B\to X_s\gamma)$ & $(3.55 \pm 0.26)\cdot 10^{-4}$ &\cite{hfag}& $(3.15 \pm 0.23)\cdot 10^{-4}$ &\cite{Misiak:2006zs}\\
\hline
$S_{K^*\gamma}$& $-0.16 \pm 0.22$ &\cite{hfag}& $-0.03\pm 0.01$ &\cite{DescotesGenon:2011yn}\\
\hline
$A_I(B\to K^*\gamma)$ & $0.052\pm 0.026$ & \cite{hfag} & $0.041 \pm 0.025$ &\cite{DescotesGenon:2011yn}\\
\hline\hline
$BR(B\to X_s\mu^+\mu^-)_{[1,6]}$& $(1.60\pm 0.50)\cdot 10^{-6}$ &\cite{0712.3009}& $(1.59\pm 0.11)\cdot 10^{-6}$ &\cite{0712.3009}\\
\hline
$\av{A_{\rm FB}}_{[1,6]}$& $-0.13^{+0.068}_{-0.078}$ & $\star$ & $-0.034\pm 0.035$ & $\dagger$\\
\hline
$\av{F_L}_{[1,6]}$& $0.622^{+0.059}_{-0.057}$ & $\star$ & $0.719\pm 0.179$ & $\dagger$\\
\hline\hline
$BR(B_s\to\mu^+\mu^-)$& $<4.5\cdot 10^{-9}$ (at 95.5\% C.L.) &\cite{1203.4493}& $(3.32\pm 0.17)\cdot 10^{-9}$ & \cite{straub}\\
\hline\hline
\end{tabular}
\caption{Experimental numbers and Standard Model predictions of the observables used in the analysis of Section \ref{sec:const}. $\star$ indicates our own average of the data. $\dagger$ indicates our SM prediction. (See also Ref.~\cite{1106.4041} concerning the $B_s\to\mu^+\mu^-$ branching ratio.)}
\label{TabExp}
\end{table}

\section{Model-independent constraints without $q^2$-binned observables}
\label{sec:const}

In this section we revisit the model-independent constraints on the Wilson coefficients $\C7$, $\Cp7$, $\C9^{(')}$ and $\C{10}^{(')}$ from well controlled observables, excluding all $B\to K^*\ell^+\ell^-$ observables except for $\av{A_{\rm FB}}_{[1,6]}$ and $\av{F_L}_{[1,6]}$. This analysis follows closely the study of Ref.~\cite{DescotesGenon:2011yn,1202.2172} (see also Refs.~\cite{buras,bobeth4,Altmannshofer:2011gn,straub,Damir3}). The aim is to reevaluate the constraints on the Wilson coefficients taking into account the following updates:
\begin{itemize}
%
%
\item Updated averages for $\av{A_{\rm FB}}_{[1,6]}^{exp}$ and $\av{F_L}_{[1,6]}^{exp}$ including the new measurements in Ref.~\cite{LHCbinned} by the LHCb collaboration:
\eqa{
\av{A_{\rm FB}}_{[1,6]}^{\rm LHCb} &=& -0.18^{+0.06+0.01}_{-0.06-0.02}\\
\av{F_{L}}_{[1,6]}^{\rm LHCb} &= & 0.66^{+0.06+0.04}_{-0.06-0.03}
}
The experimental averages for these two observables are collected in Table \ref{TabExp}.
\item Updated theoretical predictions for $\av{A_{\rm FB}}_{[1,6]}$ and $\av{F_L}_{[1,6]}$ including subleading corrections of order $|V_{ub}V_{us}|/|V_{tb}V_{ts}|$, as well as a recent update for $f_B$ \cite{1203.3862}:
\eq{f_B=190\pm 4\ {\rm MeV}}
\item Analysis of the constraints using a consistent (frequentist) statistical approach detailed in App.~\ref{appC}.

\end{itemize}
These results will be used in the following sections to study the impact of the inclusion of binned observables, in view of (a) recent measurements at LHCb and (b) the impressive prospects for the near future measurements of $q^2$-dependent $B\to K^*\ell^+\ell^-$ observables by the same collaboration.

We consider the following observables: $BR(B\to X_s\gamma)$, $S_{K^*\gamma}$, $A_I(B\to K^*\gamma)$, $BR(B\to X_s\mu^+\mu^-)$, $\av{A_{\rm FB}}_{[1,6]}$ and $\av{F_L}_{[1,6]}$. The experimental situation is summarised in Table~\ref{TabExp}, together with the SM predictions. As discussed in Refs.~\cite{DescotesGenon:2011yn,1202.2172}, these observables can be classified as class-I (dependence only on $\C7^{(')}$), class-II (dependence only on $\C7^{(')},\C9^{(')},\C{10}^{(')})$ and class-III (depending on all these plus other operators, e.g., scalar operators). The analysis is divided into four NP scenarios:
\begin{itemize}
\item {\bf Scenario A:} New Physics in $\C7$ and $\Cp7$ only, real values only.
\item {\bf Scenario B: } New Physics in $\C7$, $\Cp7$, $\C9$, $\C{10}$ only, real values only.
\item {\bf Scenario B': } New Physics in $\C7$, $\Cp7$, $\Cp9$, $\Cp{10}$ only, real values only.
\item {\bf Scenario C: } New Physics in $\C7$, $\Cp7$, $\C9^{(')}$, $\C{10}^{(')}$, real values only.
\end{itemize}

\begin{figure}\centering
\includegraphics[height=7.5cm]{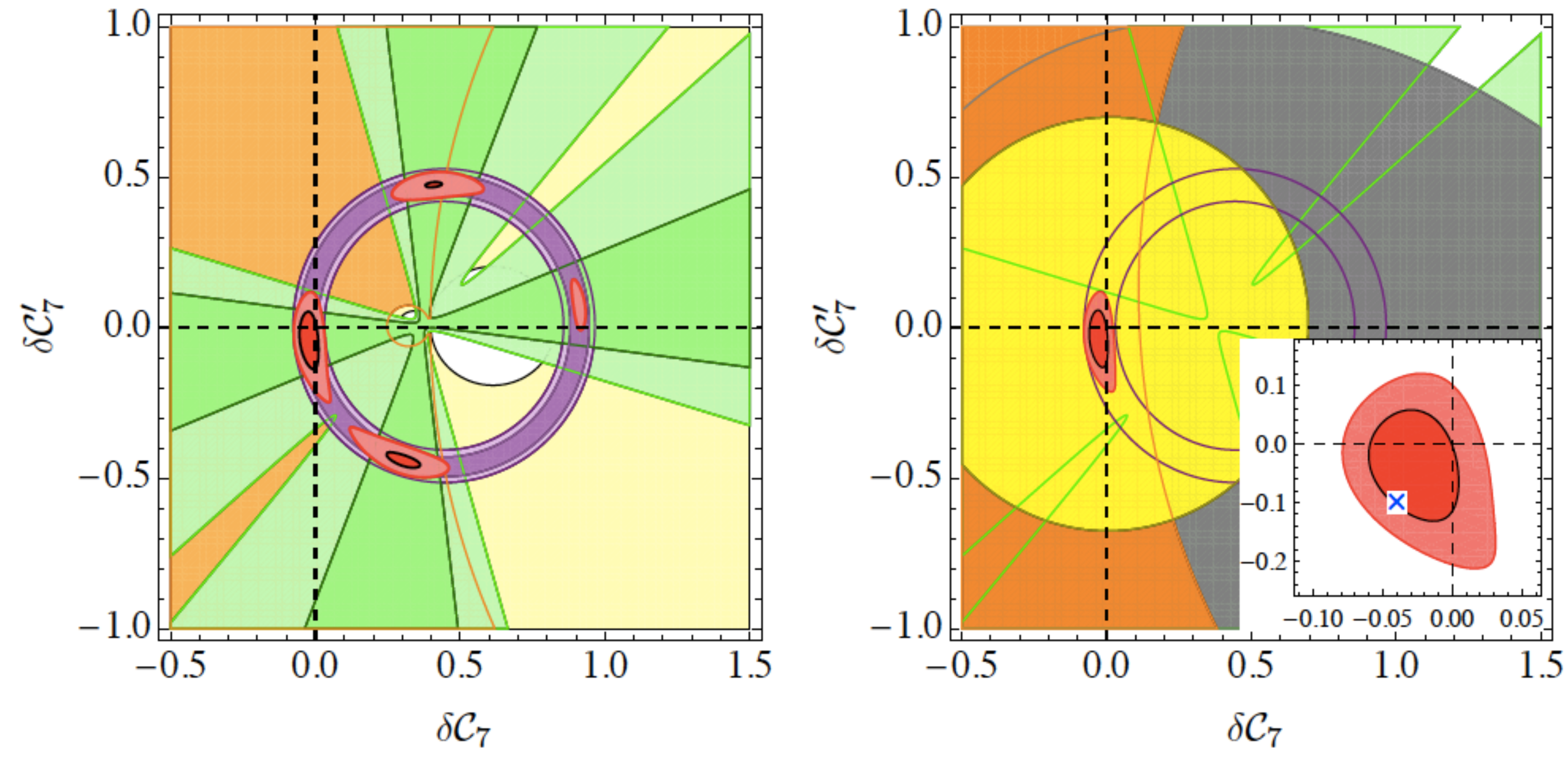}\\[-3mm]
\caption{68.3\% (dark) and 95.5\% (light) CL constraints on $\delta\C7(\mu_b)$, $\delta \Cp7(\mu_b)$. Left: Class-I observables -- $BR(B\to X_s\gamma)$ (purple), $S_{K^*\gamma}$ (green) and  $A_I(B\to K^*\gamma)$ (yellow/orange).
Right: Scenario A (class-I and class-III) --  $BR(B\to X_s\mu^+\mu^-)$ (yellow), $\av{A_{\rm FB}}_{[1,6]}$ (orange) and $\av{F_L}_{[1,6]}$ (gray).
The combined constraints are shown in red.
The cross indicates the position of the benchmark point $a$. The origin $(0,0)$ corresponds to the SM point.}
\label{ConstraintsScA}
\end{figure}

\begin{figure}\centering
\includegraphics[height=9cm]{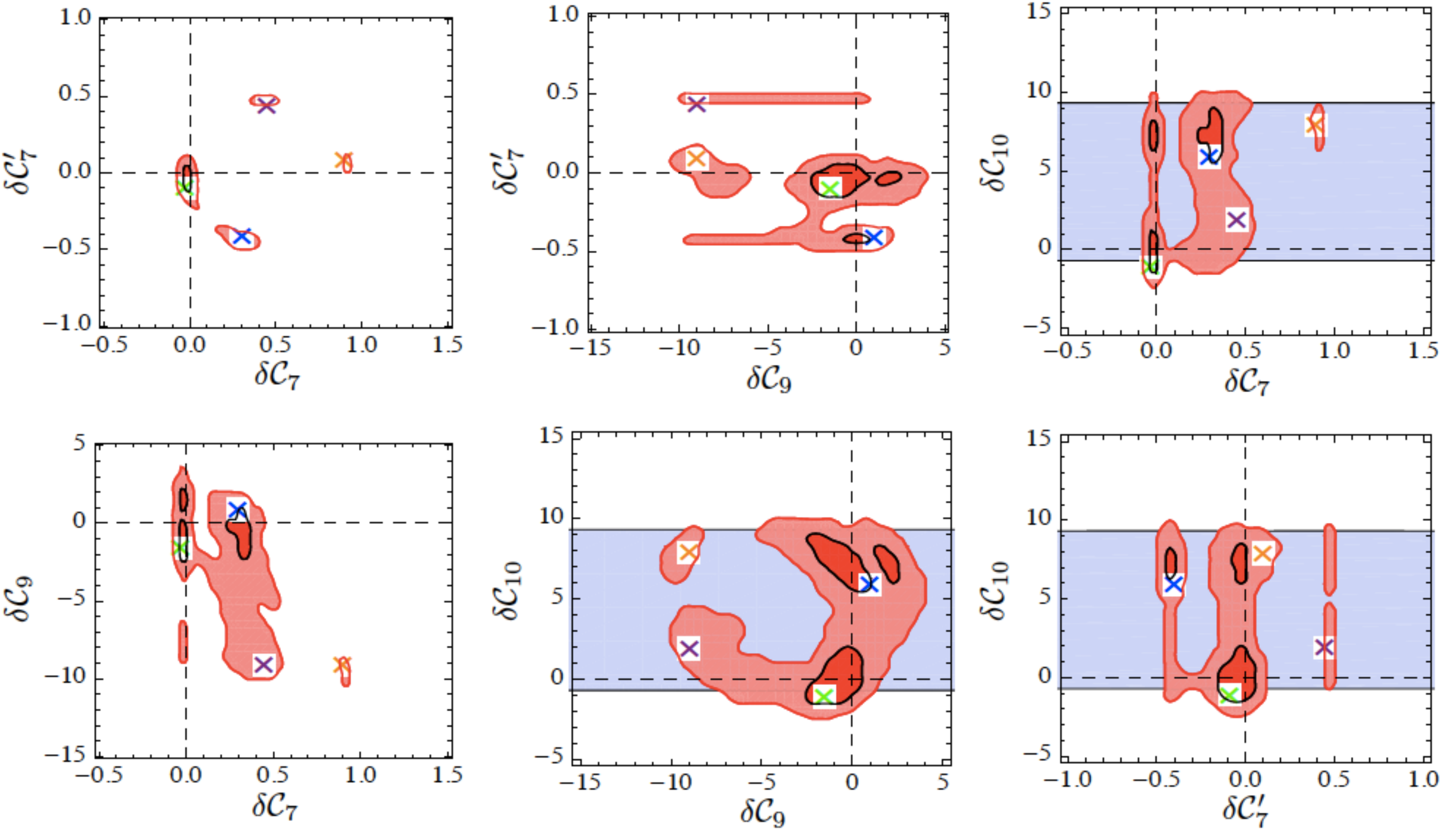}\\[-3mm]
\caption{68.3\% (light red) and 95.5\% (dark red) CL contours for $\delta\C7^{(\prime)}(\mu_b)$, $\delta\C9(\mu_b)$, $\delta\C{10}(\mu_b)$ in Scenario B. The crosses indicate benchmark points $b1$ (green), $b2$ (blue), $b3$ (purple) and $b4$ (orange). The blue band corresponds to the $B_s\to \mu^+\mu^-$ constraint.}
\label{ConstraintsScB}
\end{figure}

\begin{figure}\centering
\includegraphics[height=9cm,width=15.5cm]{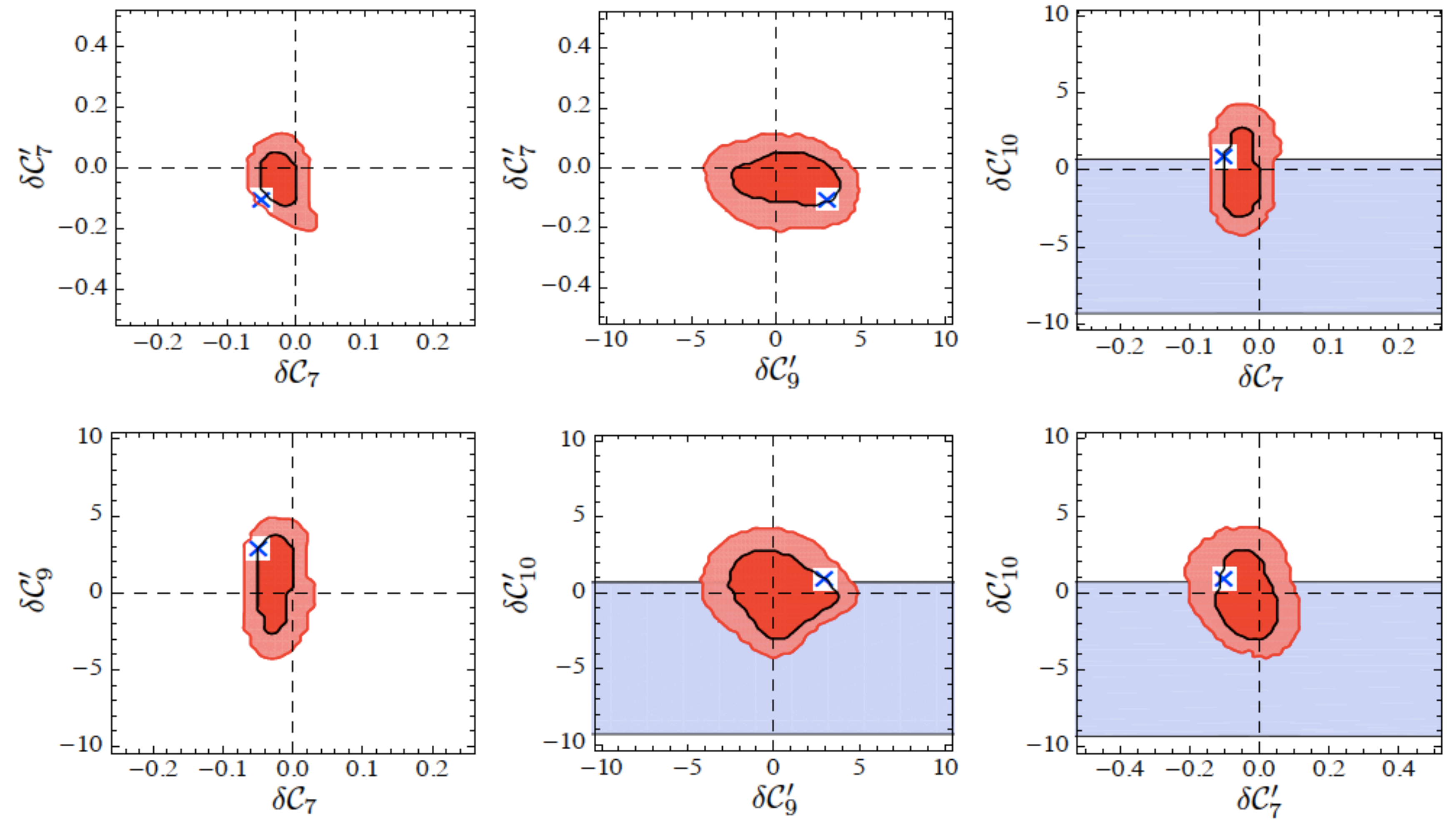}
\caption{68.3\% (light red) and 95.5\% (dark red) CL contours for $\delta\C7^{(\prime)}(\mu_b)$, $\delta\Cp9(\mu_b)$, $\delta\Cp{10}(\mu_b)$ in Scenario B'. The cross is benchmark point $b'$.  The $B_s\to \mu^+\mu^-$ constraint is indicated as a blue band.}
\label{ConstraintsScBp}
\end{figure}

\begin{figure}\centering
\includegraphics[height=8cm]{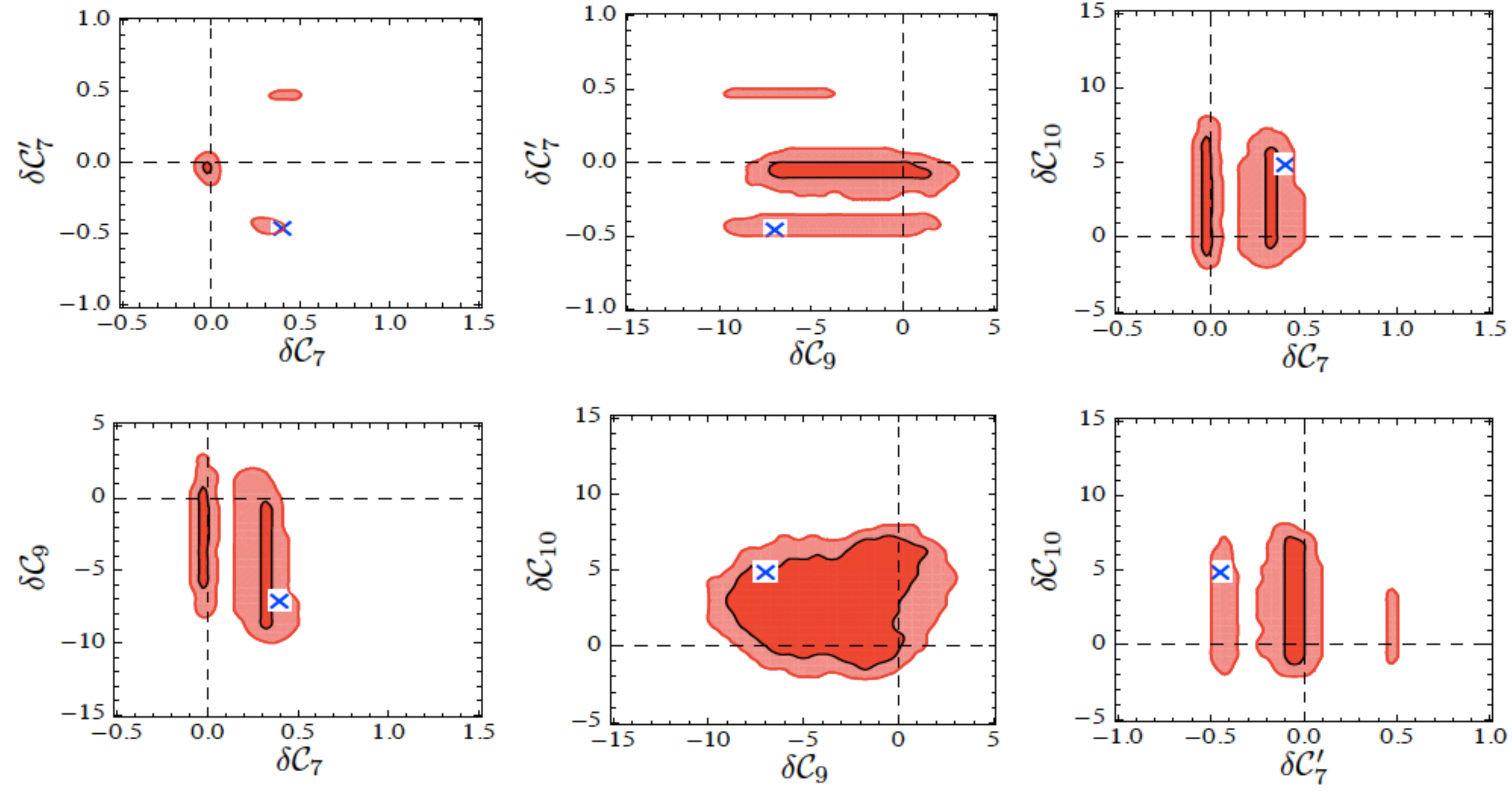}\\
\includegraphics[height=8.1cm]{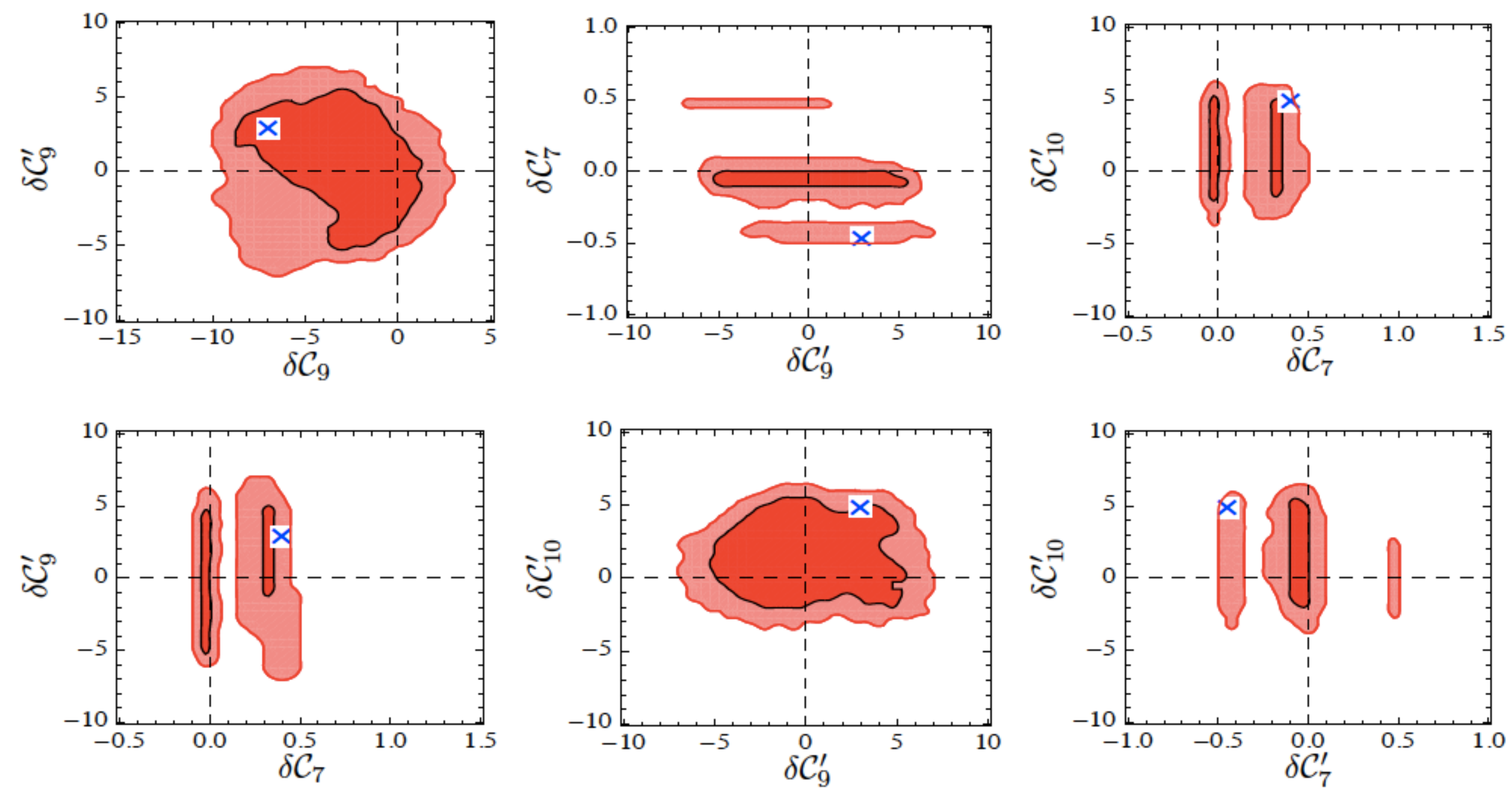}\\
\includegraphics[height=4cm]{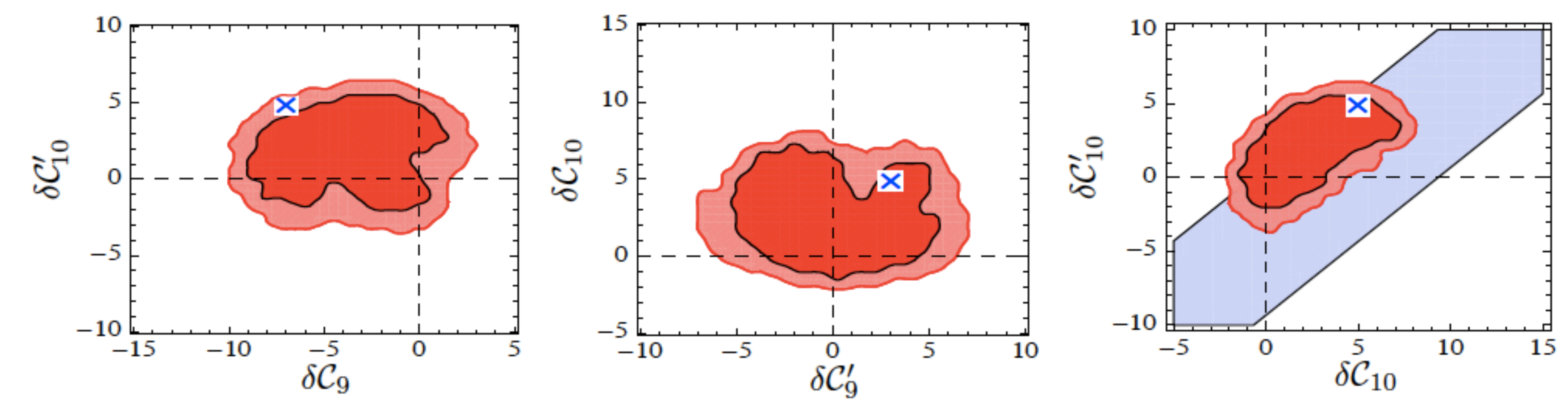}\\
\caption{68.3\% (light red) and 95.5\% (dark red) CL contours for  $\delta\C7(\mu_b)$, $\delta\Cp7(\mu_b)$, $\delta\C9^{(')}(\mu_b)$, $\delta\C{10}^{(')}(\mu_b)$ in Scenario C. The cross is benchmark point $c$. The $B_s\to \mu^+\mu^-$ constraint is indicated as a blue band.}
\label{ConstraintsScC}
\end{figure}

The constraints from $BR(B\to X_s\gamma)$, $S_{K^*\gamma}$, $A_I(B\to K^*\gamma)$ and $BR(B\to X_s\mu^+\mu^-)$ are implemented using the formulas presented in Sections 2.3 and 2.5 of Ref.~\cite{DescotesGenon:2011yn}. The constraints are set on the shift of the Wilson coefficients with respect to their SM value at the hadronic scale $\mu_b$.
Concerning $\av{A_{\rm FB}}_{[1,6]}$ and $\av{F_L}_{[1,6]}$, 
we provide the corresponding coefficients for the semi-numerical expressions of the integrated observables in Appendix \ref{appB}, with an update of
$\av{A_{\rm FB}}_{[1,6]}$ and $\av{F_L}_{[1,6]}$ following the definition of the binned quantities described in Section~\ref{sec:2}.

Inside the framework defined in Ref.~\cite{DescotesGenon:2011yn}, the three observables $BR(B\to X_s\gamma)$, $S_{K^*\gamma}$ and $A_I(B\to K^*\gamma)$ (class-I observables) are insensitive  to New Physics contributions to Wilson coefficients other than the electromagnetic operators $\C7^{(')}$. Therefore, the constraints from these (class-I) observables are common to all NP scenarios and affect only $\C7$ and $\Cp7$.
The joint 68.3\% and 95.5\% CL constraints on $\C7$ and $\Cp7$ are shown in the left panel of Fig.~\ref{ConstraintsScA}, together with the individual constraint from each observable. We find that the isospin asymmetry in $K^*\gamma$ disfavours the ``flipped-sign'' solution for $\C7$, and helps reducing the significance of the regions where $|\C7'|\simeq \C7^{SM},\C7\simeq 0$. We recall that our scenarios assume the coefficients of the chromomagnetic operators $\op_8^{(')}$ to be equal to their SM values.

We then consider  the impact of adding the class-III observables ($BR(B\to X_s\mu^+\mu^-)$, $\av{A_{\rm FB}}_{[1,6]}$ and $\av{F_L}_{[1,6]}$) in the different NP scenarios in turn. Within Scenario A, we obtain the joint 68.3\% and 95.5\% CL constraints in the $\C7$-$\Cp7$ plane shown in the right panel of Fig.~\ref{ConstraintsScA}. In this case, the $B\to K^*\mu^+\mu^-$ forward-backward asymmetry disfavours strongly the two regions with large $\delta\Cp7$ allowed at the 95.5\% CL by class-I observables. This plot features also the benchmark point  $a$, defined in Table~\ref{TabPoints} and  used in the next section to study the power of $q^2$-dependent $B\to K^*\ell^+\ell^-$ observables to discriminate among NP scenarios.

The joint constraints within Scenario B are shown in Fig.~\ref{ConstraintsScB}. Also shown is the constraint from $B_s\to\mu^+\mu^-$, which is a direct constraint on $\C{10}$.
We see that at the 95.5\% CL, there are four allowed regions in space of Wilson coefficients. The four benchmark points $b1$-$b4$ are also indicated, one in each of the four allowed regions. Finally, in Figs.~\ref{ConstraintsScBp} and \ref{ConstraintsScC} we show the constraints within Scenarios B' and C, together with the corresponding benchmark points $b'$ and $c'$. The $B_s\to\mu^+\mu^-$ branching ratio constrains $\Cp{10}$ directly in Scenario B', but only $|\C{10}-\Cp{10}|$ in Scenario C. We plot the constraint based on Ref.~\cite{straub}, including the $O(\Delta\Gamma_s)$ correction needed to connect theory and experiment \cite{1111.4882,1204.1735,1204.1737}. (Using the value from Ref.~\cite{1106.4041} would result in a slightly tighter constraint. Since there is only an experimental bound on this branching ratio, and since the status of the theoretical prediction is unclear, we refrain from including this piece of information in our combined constraints on the Wilson coefficients.)

As a final comment, we note that the ``flipped-sign'' solution for $\C7$ is in general disfavoured, but not very significantly. In fact, in Scenario B, due to the loosening of the constraint from $\av{\afb}_{[1,6]}$ when $\C9$, $\C{10}$ are allowed not to vanish, this flipped-sign solution reappears inside the 95.5\% C.L. region. More precise constraints from $A_I(B\to K^*\gamma)$ and $\av{\afb}_{[1,6]}$ should help to settle this question.


\section{Complementarity of observables for NP studies}
\label{sec:comp}

The complementarity of the different angular observables in the identification of possible NP effects is manifest even if we deal exclusively with observables integrated over the whole $[1,6]$ GeV$^2$ ranges. For example, New Physics contributing predominantly to $\C{10}$ will most likely push substantially $\av{P_4}_{[1,6]}$ below its SM value\footnote{To be specific, since this discussion is for illustrative purposes only, we focus on the New Physics scenarios and the observables $P_{4,5,6}$ of Ref.~\cite{primary}. Below we will study specific benchmark points and the primed observables $P'_{4,5,6}$.} (see Ref.~\cite{primary}). However, this will be essentially indistinguishable from a NP contribution to $\Cp{10}$, which has a very similar effect on $P_4$. This ambiguity can be resolved looking at the measured value of $\av{P_1}_{[1,6]}$, because a New Physics contribution to $\Cp{10}$ can enhance this observable considerably above its SM prediction, while $\C{10}$ has a negligible effect.

Another example would correspond to a moderate enhancement of $\av{P_1}_{[1,6]}$. Assuming no significant deviation is seen in $\av{P_4}_{[1,6]}$, this could signal a non-SM value of $\Cp9$ \emph{or} $\C7,\Cp7$ (in the island around the SM point in the $\C7$-$\Cp7$ plane). The former case will not give any substantial deviation in $\av{P_5}_{[1,6]}$, while the latter tends to increase $\av{P_5}_{[1,6]}$ above its SM value. Comparing $\av{P_1}_{[1,6]}$, $\av{P_4}_{[1,6]}$ and $\av{P_5}_{[1,6]}$ would thus help to distinguish among these scenarios.

The discriminating power of the observables is  substantially increased when we consider, not only different angular observables, but also their $q^2$-dependence. A New Physics contribution to $\Cp{10}$ (in the scenario considered above) will generally increase $\av{P_1}_{[1,6]}$, but its effect on $\av{P_4}_{[1,6]}$ and $\av{P_5}_{[1,6]}$ could be washed out by simultaneous contributions to $\C9$ and $\C{10}$. However, if this enhancement of $\av{P_1}_{[1,6]}$ is also present in the region $\sim 4-6\ {\rm GeV}^2$ (for instance with $\av{P_1}_{[4.3,6]}$ shifted upwards with respect to the SM), the NP effect cannot be misinterpreted as a non-SM value of $\C7^{(\prime)}$ or $\Cp9$, because these would only enhance $P_1$ in the region $\sim 1-3\ {\rm GeV}^2$.

\begin{table}
\centering
\begin{tabular}{||c||c|c|c|c|c|c||}
\hline\hline
Point & $\delta \C7(\mu_b) $ & $\delta \Cp7(\mu_b) $ & $\delta \C9(\mu_b) $ & $\delta \C{10}(\mu_b) $ & $\delta \Cp9(\mu_b) $ & $\delta \Cp{10}(\mu_b) $\\
\hline\hline
$a$ & $-0.04$ & $-0.1$ & 0 & 0 & 0 & 0 \\
\hline
$b1$ & $-0.03$ & $-0.09$ & $-1.5$ & $-1$ & 0 & 0 \\
\hline
$b2$ & $0.3$ & $-0.4$ & $1$ & $6$ & 0 & 0 \\
\hline
$b3$ & $0.45$ & $0.45$ & $-9$ & $2$ & 0 & 0 \\
\hline
$b4$ & $0.9$ & $0.1$ & $-9$ & $8$ & 0 & 0 \\
\hline
$b'$ & $-0.05$ & $-0.15$ & $0$ & $0$ & $3$ & $1.5$ \\
\hline
$c$ & $0.4$ & $-0.45$ & $-7$ & $5$ & $3$ & $5$ \\
\hline\hline
\end{tabular}
\caption{NP benchmark points used in the analysis of Section \ref{sec:comp}.}
\label{TabPoints}
\end{table}

\begin{figure}\centering
\includegraphics[height=9.4cm]{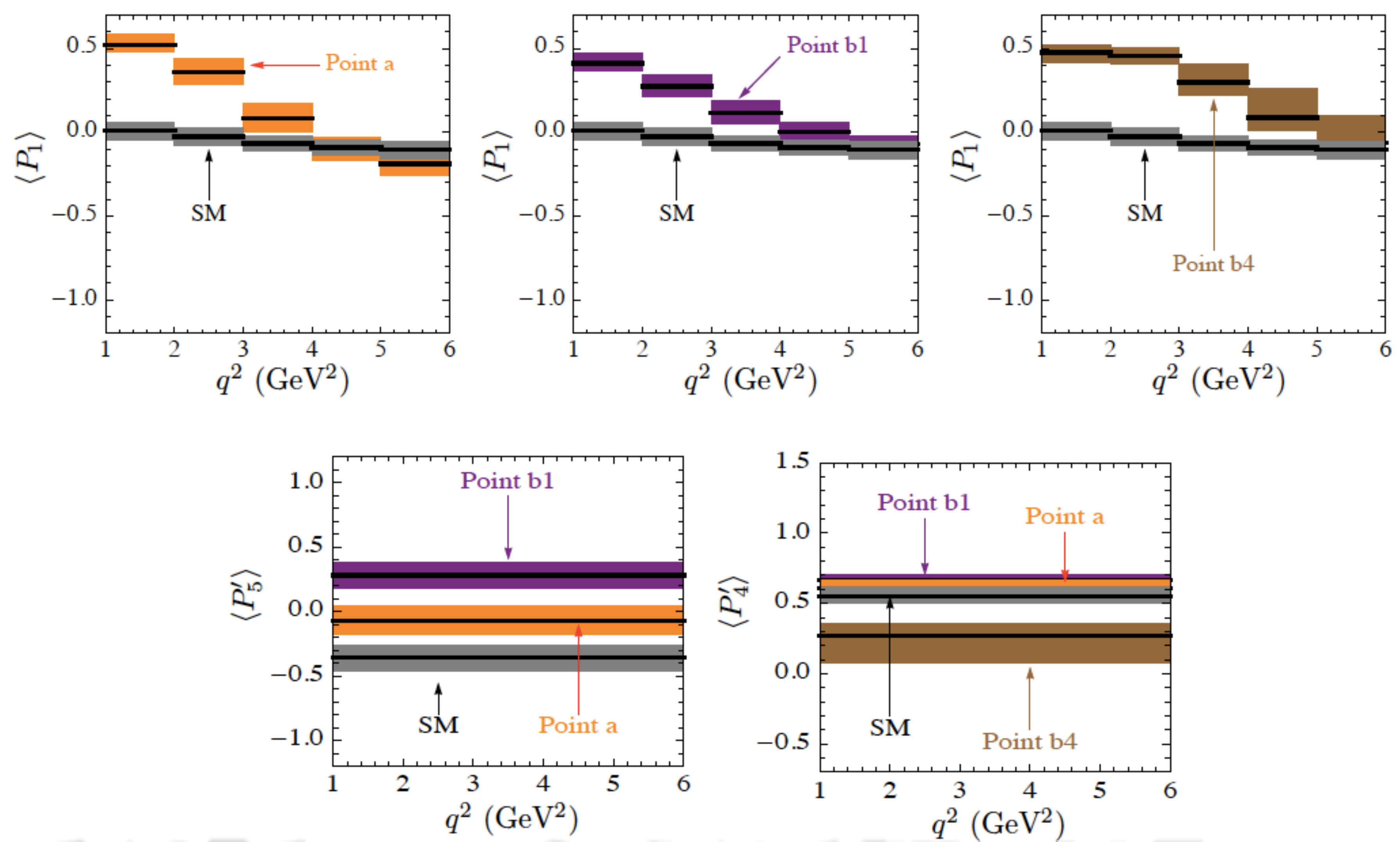}\\[-2mm]
\caption{Comparison of SM predictions for $\av{P_1}$, $\av{P'_4}_{[1,6]}$, $\av{P'_5}_{[1,6]}$ and the predictions within benchmark points $a$, $b1$, $b4$ given in Table \ref{TabPoints}.}
\label{SMNPplot1}
\end{figure}

\begin{figure}\centering
\includegraphics[height=9.4cm]{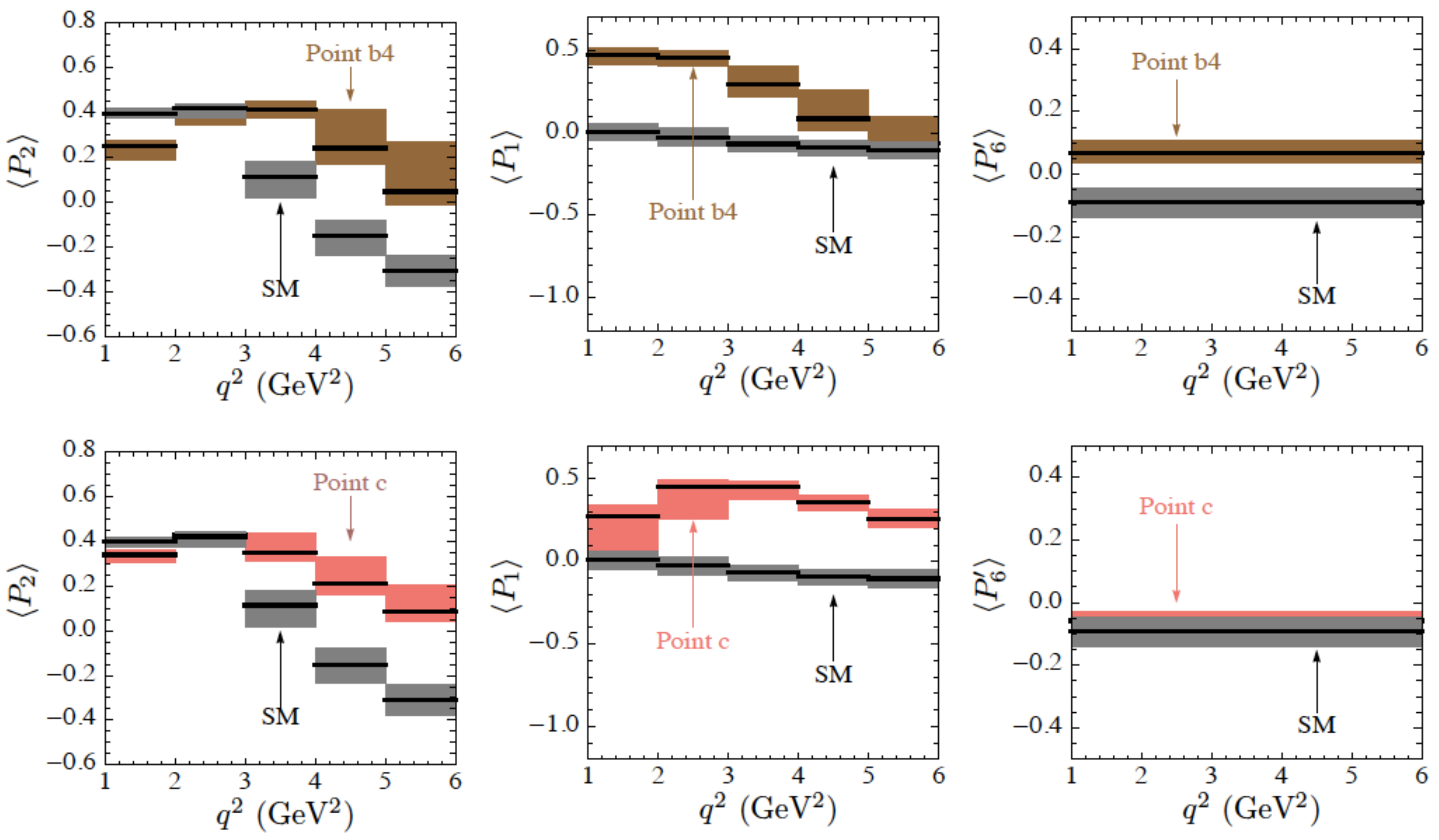}\\[-2mm]
\caption{Comparison of SM predictions for $\av{P_1}$, $\av{P_2}$, $\av{P'_6}$ and the predictions within benchmark points $b4$, $c$ given in Table \ref{TabPoints}.}
\label{SMNPplot4}
\end{figure}

These considerations apply equally well in terms of \emph{constraints} when no deviation from the SM is observed.
In Section~\ref{sec:const2} we will see explicitly how different observables constrain the NP contributions to the Wilson coefficients. It is already useful to build an intuition on how different observables in different $q^2$-bins are affected by shifts in the different Wilson coefficients, in order to have a better idea of the most promising observables in each case. Here we study briefly the effect of different New Physics scenarios on the integrated observables studied in Section \ref{sec:sm}. We focus on a set of ``benchmark points'' consistent with $BR(B\to X_s\gamma)$, $S_{K^*\gamma}$, $A_I(B\to K^*\gamma)$, $BR(B\to X_s\mu^+\mu^-)$, $\av{A_{\rm FB}}_{[1,6]}$ and $\av{F_L}_{[1,6]}$, according to the analysis of Section \ref{sec:const}. These points are specified in Table \ref{TabPoints}, and also indicated in Figs.~\ref{ConstraintsScA}-\ref{ConstraintsScC}.

\begin{figure}\centering
\includegraphics[height=9.5cm]{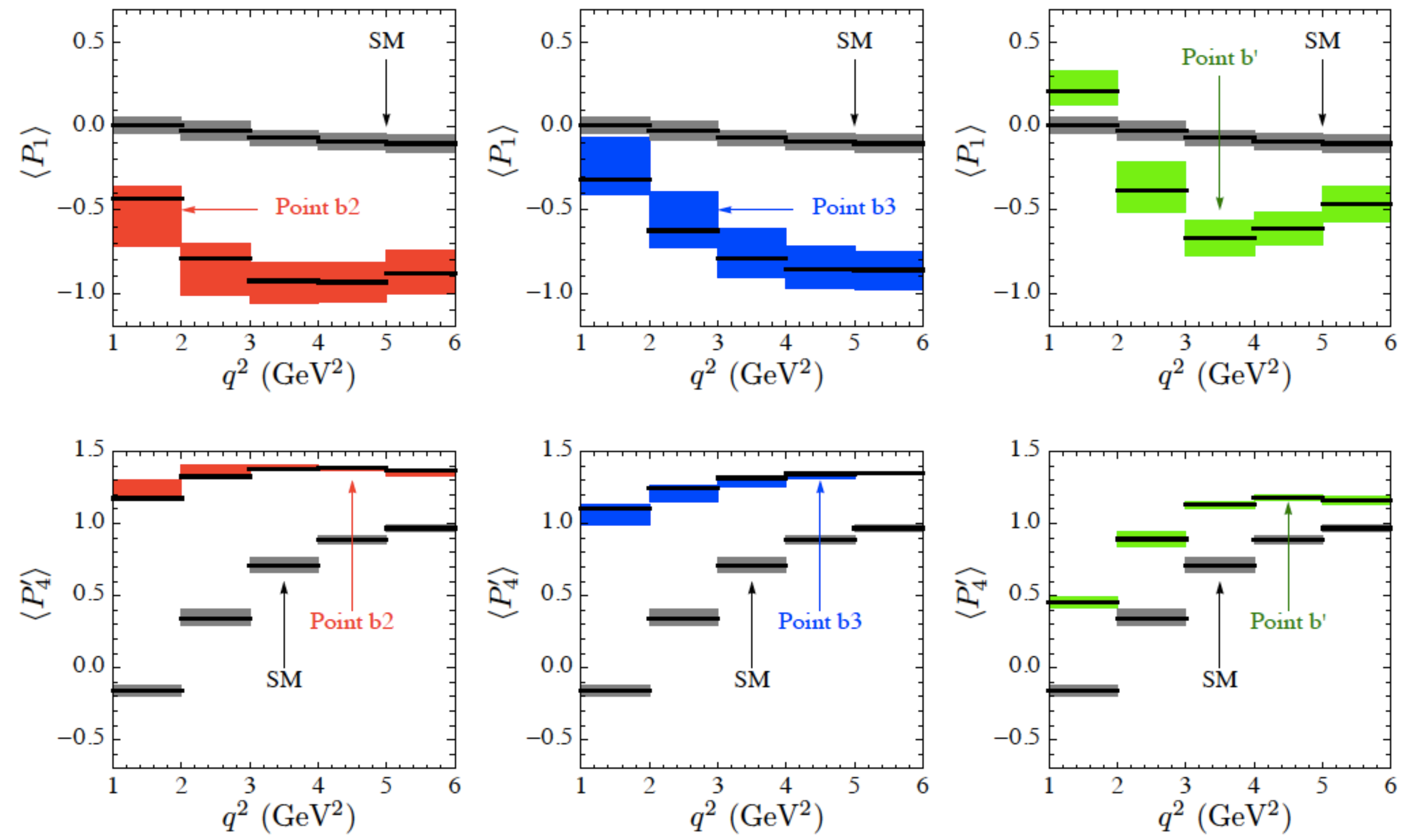}
\includegraphics[height=9.5cm]{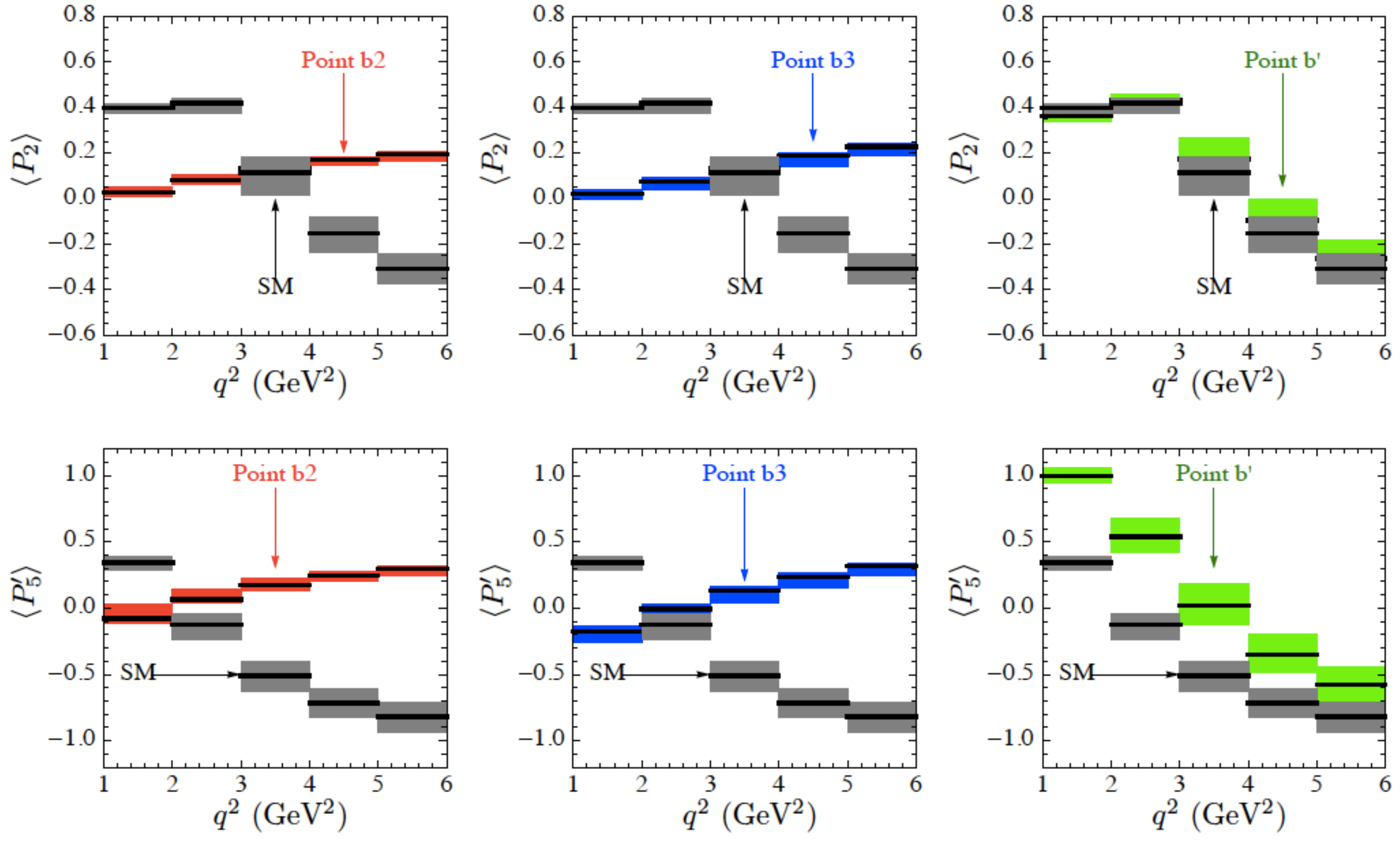}
\caption{Comparison of SM predictions for $\av{P_1}$, $\av{P_2}$, $\av{P'_4}$, $\av{P'_5}$ and the predictions within benchmark points $b2$, $b3$, $b'$ given in Table \ref{TabPoints}.}
\label{SMNPplot3}
\end{figure}

Each NP benchmark point gives a prediction for each observable in each $q^2$-bin. In  Figs.~\ref{SMNPplot1}, \ref{SMNPplot3} and \ref{SMNPplot4} we show the most significant cases, exemplifying the previous discussion.
In these figures, the gray rectangles correspond to the SM binned predictions, corresponding to Figs.~\ref{SMplotsPs1} and \ref{SMplotsPs2}. The colored rectangles correspond to the NP scenarios represented by each benchmark point collected in Table \ref{TabPoints}. We can draw a few conclusions:

\begin{itemize}
\item $\av{P_1}$ in the region $q^2\lesssim 3\,{\rm GeV}^2$ is able to discriminate the points $a$, $b1$ and $b4$ from the SM, but gives similar predictions for these scenarios. These scenarios could be resolved by $\av{P'_4}_{[1,6]}$ and $\av{P'_5}_{[1,6]}$ (see Fig.~\ref{SMNPplot1}). 
\item The points $b2$, $b3$ and $b'$ enhance the observables $\av{P_1}$ and $\av{P'_4}$ substantially, and the impact on $\av{P_1}$ ($\av{P'_4}$) is more important for $q^2>3\,{\rm GeV}^2$ ($q^2<3\,{\rm GeV}^2$) (see Fig.~\ref{SMNPplot3}). 
\item $\av{P_1}$ and $\av{P'_4}$ do not allow one to discriminate among $b2$, $b3$ and $b'$, but $\av{P_2}$ and $\av{P'_5}$ exhibit distinctive behaviours for $b2$
and $b3$ (Fig.~\ref{SMNPplot3}). For instance, a suppression of $\av{P_1}$ below $\sim -0.5$ together with an enhancement of  $\av{P_2}_{[4,6]}$ above $\sim 0$ would favour $b2$ and $b3$ with respect to $b'$. This conclusion could be verified by a suppression of $\av{P_2}_{[1,3]}$ below its SM value and an enhancement of $\av{P'_5}_{[4,6]}$.
\item A similar situation occurs with the points $b4$ and $c$. The observable $\av{P_2}_{[5,6]}$ could favour these scenarios, but cannot distinguish among them. However, $\av{P_1}_{[1,2]}$ and $\av{P_1}_{[5,6]}$ can discriminate these scenarios, as well as $\av{P'_6}_{[1,6]}$ if the experimental values are accurate enough (see Fig.~\ref{SMNPplot4}).
\end{itemize}

A full set of predictions for all benchmark points in comparison with the SM predictions can be found in Figs.~8 and 9 of Ref.~\cite{thisv1}.




\section{The benefit of using clean observables}
\label{sec:S3vsP1}

In this section we discuss the advantages of using clean observables in analyses of $B\to K^*\ell^+\ell^-$ as opposed to other observables such as $S_3$ or $\aim$. For definiteness we focus on $P_1$ and $S_3$, but it should be kept in mind that the conclusions are more general.

As discussed extensively in Refs.~\cite{kruger, matias1, matias2, primary}, clean observables are constructed in such a way that at LO and at large recoil, 
an exact cancellation of the form-factor dependence occurs. This indicates that clean observables should be stable under variation of hadronic uncertainties, as opposed to other observables, such as $F_L$, $A_{FB}$, $S_3$, etc. This is relevant because of the spread of published errors in the determination of form factors from light-cone sum rules (see Refs.~\cite{1006.4945,0412079} and the introduction).

If the form factors of Ref.~\cite{1006.4945} are used in the evaluation of $F_L$, for example, the error bars get enlarged by a factor of three. On the contrary, this enlargement does not happen in the case of $P_1$, which is practically insensitive to these uncertainties. In the case of $S_3$, an accidental circumstance makes its SM uncertainty smaller than what one would infer from the fact that $S_3\sim P_1 F_T$ (that is, a similar percentual enhancement of the errors as $F_L$). The fact that $P_1\sim 0$ in the SM, makes $S_3$ almost insensitive to the error in $F_L$ only near the SM point. This makes the situation with $S_3$ a bit more subtle. The important point here is that in the presence of New Physics, an enhancement of $P_1$ produces  an enlargement of the error bars in the theoretical prediction for $S_3$ automatically, which makes $S_3$ almost unable of discriminating between NP models where $P_1$ does not vanish.

\begin{figure}\centering
\includegraphics[height=6.3cm]{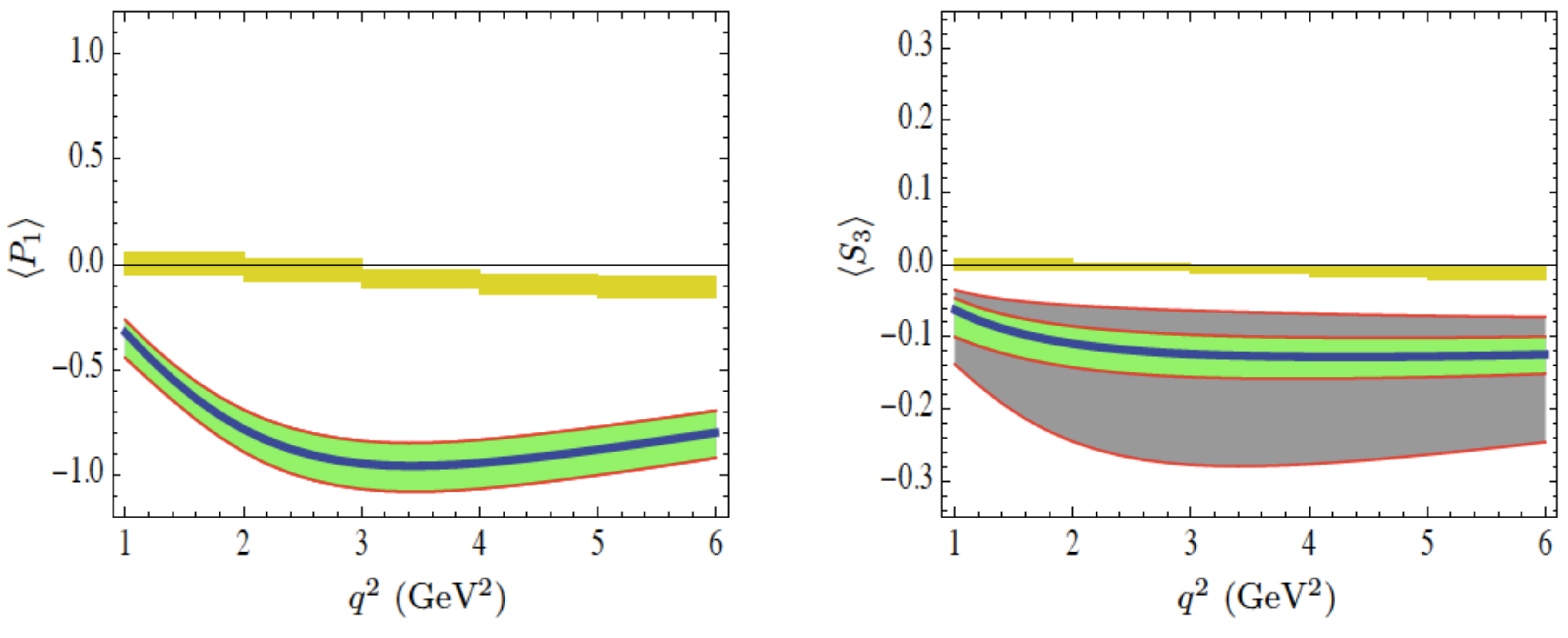}\\[-2mm]
\caption{Predictions in the SM and in the case of NP at the benchmark point $b2$ for $P_1$ (left) and $S_3$ (right). The yellow boxes are the SM predictions integrated in five 1 GeV$^2$ bins. The blue curve corresponds to the central values for the NP scenario. The green band is the total uncertainty considering the form factors of Refs.~\cite{0412079,buras}, while the gray band is the total uncertainty obtained using the form factors of Ref.~\cite{1006.4945}. In the case of $P_1$ the gray band is barely visible.
}
\label{P1vsS3fig}
\end{figure}

In Fig.~\ref{P1vsS3fig} we show the SM predictions and the predictions for benchmark point $b2$ (see Table \ref{TabPoints}) for $P_1$ and $S_3$ calculated with both choices of form factors (Refs.~\cite{1006.4945} and \cite{0412079}). We find  that:
\begin{itemize}
\item The SM prediction for $P_1$ is insensitive to the choice of form factors.
\item The SM prediction for $S_3$ shows a moderate dependence on the choice of form factors, and hadronic uncertainties are enlarged up to a 50\% when using the form factors of Ref.~\cite{1006.4945} compared to those in Ref.~\cite{0412079}.
\item The NP prediction for $P_1$ is insensitive to the choice of form factors.
\item The NP prediction for $S_3$ is \emph{very} sensitive to the choice of form factors. Indeed, the hadronic uncertainties increase from a factor 2 to a factor 3 when using the form factors of Ref.~\cite{1006.4945}.
\end{itemize}
The conclusion is that, in New Physics analyses of $B\to K^*\ell^+\ell^-$, when using the observable $S_3$, one is obliged to take into account hadronic uncertainties at each point in the NP parameter space, and these uncertainties can be substantial. On the other hand, the observable $P_1$ is not affected by this disease, and it is therefore a much more robust observable. Moreover, from Fig.~\ref{P1vsS3fig} we can see that $P_1$ has the potentiality to exclude completely a NP scenario given by benchmark point $b2$, while $S_3$ can barely distinguish this point from the SM case at more than 2$\sigma$. We expect the same results to hold in other regions of the NP parameter space. The conclusion does not change when we consider  the binned observables. For example, the corresponding predictions at benchmark point $b2$ for $\av{P_1}$ and $\av{S_3}$ in the bin $[1,6]$ GeV$^2$ are
\eq{\av{P_1}_{[1,6]}^{b2}=-0.82\pm 0.10 \ ,\quad \av{S_3}_{[1,6]}^{b2}=-0.16\pm 0.08\ .}

We would like to emphasise that similar conclusions are expected for other form factor dependent observables. For example, the observable $\aim \sim S_9  \sim P_3 F_T$ will also be protected from large hadronic uncertainties near the SM point, since $P_3\sim 0$ in the SM. However, complex NP can enhance considerably $P_3$ while being consistent with other data (see Ref.~\cite{primary}). In these NP scenarios, $\aim$ is expected to suffer from a similar problem as the one described for $S_3$ (a problem that does not affect $P_3$). Furthermore, such problems may also happen also in the case of CP conjugated observables such as $A_9$ and $A_3$. 

In view of this situation, one should be particularly careful with a rather usual hypothesis consisting in assigning the same hadronic uncertainty on observables in any NP scenario as in the SM case. This might be a good approximation in the case of clean observables (see for instance the left-hand side of Fig.~\ref{P1vsS3fig}, where the uncertainty on $P_1$ is indeed similar from the SM case to the NP scenario considered), but it can be very misleading for other quantities, sensitive to form factors (as illustrated on the right-hand side of Fig.~\ref{P1vsS3fig}, where the uncertainty on $S_3$ is significantly enlarged from the SM value to the NP scenario considered).


\section{Extracting clean observables from existing experimental measurements}
\label{sec:exp}

As pointed out in Section \ref{sec:S3vsP1}, there is a clear advantage in using the clean observables $P_{1,2,3}$ rather than $S_3$, $\aim$ and $\afb$ (or $F_L$) to put constraints on the Wilson coefficients. However, this can be achieved if the experimental fits are performed consistently considering these observables (see Ref.~\cite{primary}).
In its latest experimental analysis of $q^2$-dependent observables in $B\to K^*\ell^+\ell^-$, the LHCb collaboration \cite{LHCbinned} has preferred to fit directly the coefficients of the angular distributions, providing only observables proportional to the $J_i$ coefficients (in particular $\av{S_3}$ and $\av{\aim}$). We have collected for reference the relevant experimental results for these observables in Table \ref{TabLHCb}.

\begin{table}
\centering
\begin{tabular}{||c||r|r|r||}
\hline\hline
&&&\\[-4mm]
 & $[2,4.3]\qquad$ & $[4.3,8.68]\quad$ & $[1,6]\qquad$ \\[2mm]
\hline\hline
&&&\\[-4mm]
$\av{\afb}$ & $-0.20^{+0.08+0.01}_{-0.07-0.03}$ & $0.16^{+0.05+0.01}_{-0.05-0.01}$ & $-0.18^{+0.06+0.01}_{-0.06-0.02}$ \\[2mm]
\hline
&&&\\[-4mm]
$\av{F_L}$ & $0.74^{+0.09+0.02}_{-0.08-0.04}$ & $0.57^{+0.05+0.04}_{-0.05-0.03}$ & $0.66^{+0.06+0.04}_{-0.06-0.03}$ \\[2mm]
\hline
&&&\\[-4mm]
$\av{\aim}$ & $-0.02^{+0.10+0.05}_{-0.06-0.01}$ & $0.02^{+0.07+0.01}_{-0.07-0.01}$ & $0.07^{+0.07+0.02}_{-0.07-0.01}$  \\[2mm]
\hline
&&&\\[-4mm]
$\av{S_3}$ & $-0.05^{+0.18+0.05}_{-0.12-0.01}$ & $0.18^{+0.13+0.01}_{-0.13-0.01}$ & $0.10^{+0.15+0.02}_{-0.16-0.01}$  \\[2mm]
\hline\hline
\end{tabular}
\caption{LHCb experimental results for binned observables (from Ref.~\cite{LHCbinned}).}
\label{TabLHCb}
\end{table}

Of course, we can compute the clean observables from the measurements provided using the formulas of Sections 3 and 4 of Ref.~\cite{primary}, or equivalently from Eqs.~(\ref{Pis}). If this is done without knowing the correlation matrix (which is not provided by the experimental collaborations yet), one obtains errors much larger that the real uncertainties. It is still worth exploring the current situation based on these observables while waiting for correlated values. 
In addition, higher-statistics analyses from LHCb are expected to reduce the experimental errors on these observables considerably very soon.
 
Attending to these considerations, experimental values for $P_{1,2,3}$ can be derived from the measurements of $S_3$, $\aim$, $\afb$ and $F_L$ in Table \ref{TabLHCb} by means of Eq.~(\ref{Pis}):
\eq{
\av{P_1}_\bin = \frac{2 \av{S_3}_\bin}{ 1 - \av{F_L}_\bin}\ ,\quad
\av{P_2}_\bin =-\frac23 \frac{\av{\afb}_\bin}{(1 - \av{F_L}_\bin)}\ ,\quad
\av{P_3}_\bin =-\frac{\av{\aim}_\bin}{(1 - \av{F_L}_\bin)}\ .
\label{Pis2av}} 
In Table \ref{TabCleanObs} we present the resulting experimental values for $\av{P_1}$, $\av{P_2}$, $\av{P_3}$ in the different bins, together with their SM predictions. The experimental errors are calculated in the following way. We first add in quadrature both errors in Table \ref{TabLHCb}, and symmetrise upper and lower uncertainties. Assuming these errors are Gaussian, the errors for $P_{1,2,3}$ are obtained by the usual error propagation formula from Eqs.~(\ref{Pis2av}). 

\begin{table}
\centering
\begin{tabular}{||l|r|r||}
\hline\hline
Observable & Experiment & SM prediction \\
\hline\hline
$\av{P_1}_{[2,4.3]}$ & $-0.19 \pm 0.58$ & $-0.051 \pm 0.050$ \\
\hline
$\av{P_1}_{[4.3,8.68]}$ & $0.42 \pm 0.31$ & $-0.115\pm 0.060$ \\
\hline
$\av{P_1}_{[1,6]}$ & $0.29 \pm 0.47$ & $-0.055 \pm 0.051$\\
\hline\hline
$\av{P_2}_{[2,4.3]}$ & $0.51 \pm 0.27$ & $0.227 \pm 0.070$\\
\hline
$\av{P_2}_{[4.3,8.68]}$ & $-0.25 \pm 0.08$ & $-0.422\pm 0.074$ \\
\hline
$\av{P_2}_{[1,6]}$ & $0.35 \pm 0.14$ & $0.080 \pm 0.067$\\
\hline\hline
$\av{P_3}_{[2,4.3]}$ & $0.08 \pm 0.35$ & $-0.004 \pm 0.024$\\
\hline
$\av{P_3}_{[4.3,8.68]}$ & $-0.05 \pm 0.16$ & $-0.005\pm 0.027$ \\
\hline
$\av{P_3}_{[1,6]}$ & $-0.21 \pm 0.21$ & $-0.003\pm 0.024$\\
\hline\hline
\end{tabular}
\caption{Experimental values for the clean observables $P_1$, $P_2$ and $P_3$ within different $q^2$-bins, extracted from the measurements of $S_3$, $\aim$, $\afb$ and $F_L$, and their SM predictions.}
\label{TabCleanObs}
\end{table}


\section{Present and future constraints from $q^2$-dependent  $B\to K^*\ell^+\ell^-$ observables}
\label{sec:const2}

The recent LHCb measurements for $q^2$-dependent $B\to K^*\ell^+\ell^-$  observables \cite{LHCbinned} are divided into four bins --if we restrict ourselves to the low-$q^2$ region. These bins are $[0.05,2]$, $[2,4.3]$, $[4.3,8.68]$ and $[1,6]\,{\rm GeV}^2$. These results yield a first glimpse of the future, where 
precise measurements of the full angular distribution within fine $q^2$-bins will be available. The purpose of this section is to study the impact of the $q^2$-dependent observables provided in Ref.~\cite{LHCbinned} on the constraints on the Wilson coefficients, and to analyze what are the future expectations concerning the constraints from these observables.

A brief discussion is in order concerning the results in Ref.~\cite{LHCbinned}. From the theory point of view the first bin $[0.05,2]$ is very difficult to control, since the decay rate contains contributions from light resonances below $q^2\sim 1\,{\rm GeV}^2$. The third bin $[4.3,8.68]$ is also more difficult to handle theoretically, as it gets near to the $c\bar c$ resonance region. For this first attempt, we choose to drop the first bin and to consider constraints from the two others  $[2,4.3]$, $[4.3,8.68]$. We do not include $[1,6]$ as the results are likely to be strongly correlated with the two smaller bins considered for this study (we remind that the averaged experimental results in the bin $[1,6]$ have already been considered in Section \ref{sec:const} in the case of $\av{\afb}$ and $\av{F_L}$).

\subsection{Constraints from binned $\av{\afb}$ and $\av{F_L}$}

We first consider the observables $\av{\afb}_{[2,4.30]}$ , $\av{\afb}_{[4.30,8.68]}$, $\av{F_L}_{[2,4.30]}$ and $\av{F_L}_{[4.30,8.68]}$. The experimental numbers for these observables are given in Table~\ref{TabLHCb}. We study separately the constraints derived using the set of form factors of Ref.~\cite{buras} or the ones from Ref.~\cite{1006.4945}, the later being the choice throughout this article. In this case the constraints are implemented using Eq.~(\ref{Ok}) in App. \ref{appB} together with the coefficients in Tables~\ref{TableAFB} and~\ref{TableFL}. The constraints in the first case (form factors form Ref.~\cite{buras}) can be implemented from App. B in Ref.~\cite{thisv1}.

\begin{figure}\centering
\includegraphics[height=7.3cm]{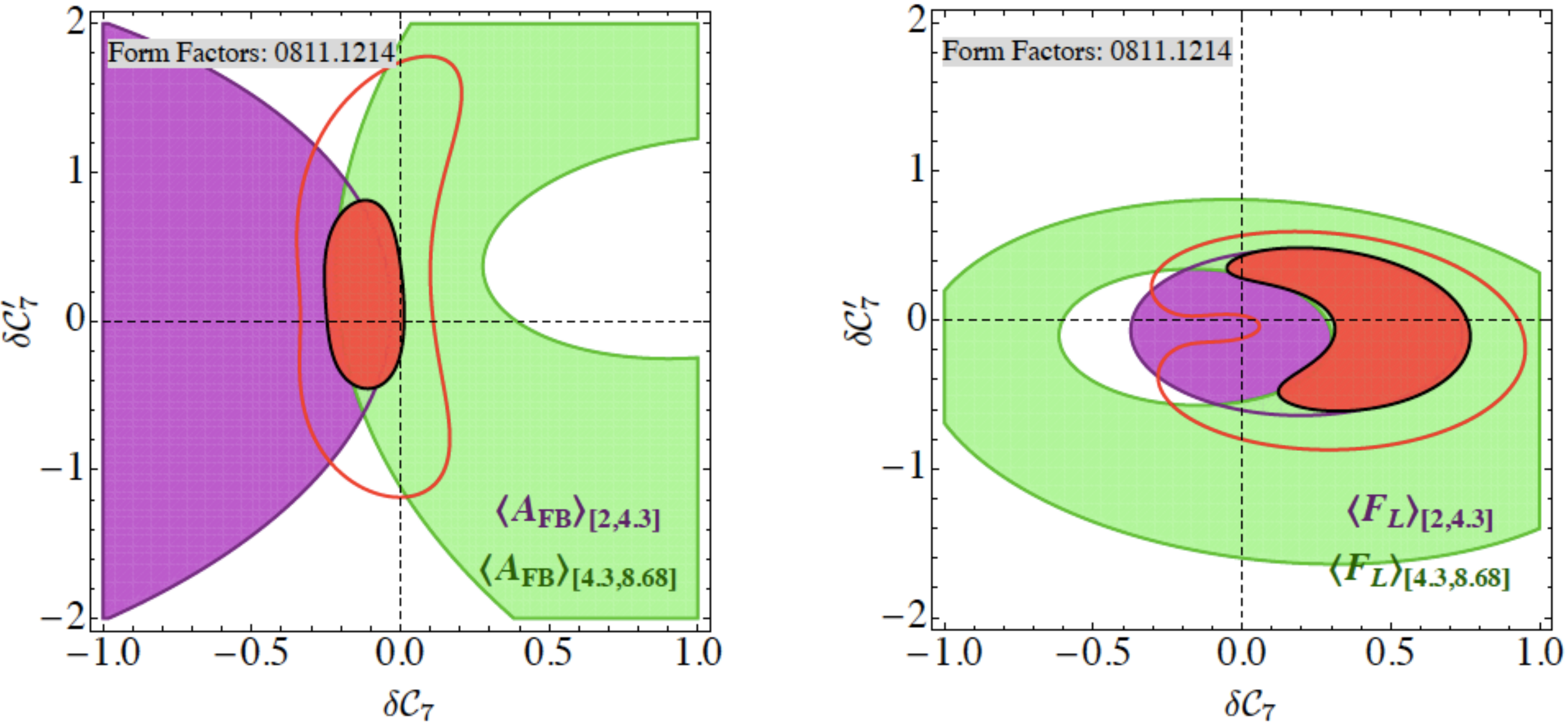}\\[4mm]
\includegraphics[height=7.3cm]{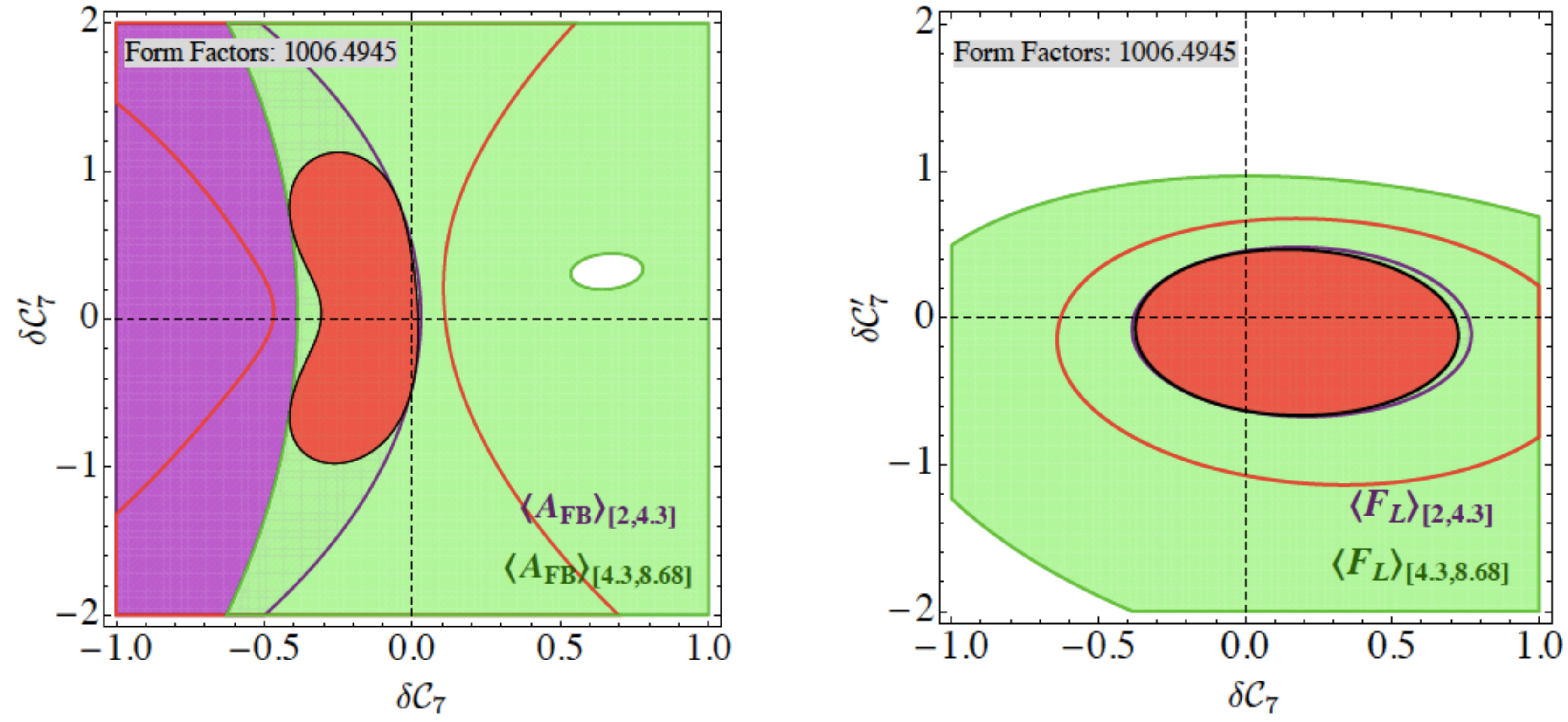}\\[-2mm]
\caption{Individual 68.3\% C.L. constraints in the $\delta\C7(\mu_b)$-$\delta \Cp7(\mu_b)$ plane from $\av{\afb}_{[2,4.30]}$ and $\av{\afb}_{[4.30,8.68]}$ (left), and  from $\av{F_L}_{[2,4.30]}$ and $\av{F_L}_{[4.30,8.68]}$ (right), taking form factors of Ref.~\cite{buras} (up) or Ref.~\cite{1006.4945} (down). The combined 68.3\% C.L. (red filled) and 95.5\% C.L. (red contour) regions are also shown. The origin corresponds to the SM value. }
\label{PlotAFBFLbinned}
\end{figure}

The individual 68.3\% C.L. and combined 68.3\% and 95.5\%C.L. contours for these observables in the $\C7$-$\Cp7$ plane (Scenario A) are shown in Fig.~\ref{PlotAFBFLbinned}. We see that:

\begin{itemize}
\item The constraints from $\av{\afb}$ are consistent with the SM at 95.5\%C.L. Using the form factors in Ref.~\cite{buras}, some tension is caused by $\av{\afb}_{[2,4.30]}$, while $\av{\afb}_{[4.30,8.68]}$ is compatible with the SM at 68.3\%C.L. This tension disappears if the form factors of Ref.~\cite{1006.4945} are used in the SM predictions. In this case the 95.5\%C.L. region widens considerably.
\item With form factors of Ref.~\cite{buras}, the constraints from $\av{F_L}$ show a discrepancy with the SM, which is just outside the 95.5\%C.L. region. Indeed $\av{F_L}_{[4.30,8.68]}$ has a clear tendency to avoid the SM point. This tension disappears completely if the form factors of Ref.~\cite{1006.4945} are used for the SM predictions.
\item Committing oneself to form factors in Ref.~\cite{buras} and taking seriously these tensions would require a NP that affects simultaneously $\afb$ around $q^2\sim 3\,{\rm GeV}^2$ and $F_L$ around $q^2\sim 6\,{\rm GeV}^2$. Predictions derived from form factors in Ref.~\cite{1006.4945} are perfectly consistent with the SM; this is the conservative conclusion.
\end{itemize}

\subsection{Constraints from binned $\av{P_1}$, $\av{P_2}$ and $\av{P_3}$}
\label{subsec:Ps}

We now consider the constraints from the observables $\av{P_i}_{[2,4.30]}$ , $\av{P_i}_{[4.30,8.68]}$, $\av{P_i}_{[1,6]}$ with $i=1,2,3$. The experimental values and SM predictions for these observables are collected in Table \ref{TabCleanObs}.
The constraints are implemented using Eq.~(\ref{Ok}) in Appendix \ref{appB} together with the coefficients in Tables \ref{TableP1}, \ref{TableP2} and \ref{TableP3}.

The individual constraints from these observables in the $\C7$-$\Cp7$ plane (Scenario A) are shown in Fig.~\ref{PlotPibinned}. We see that:

\begin{figure}\centering
\includegraphics[height=7cm]{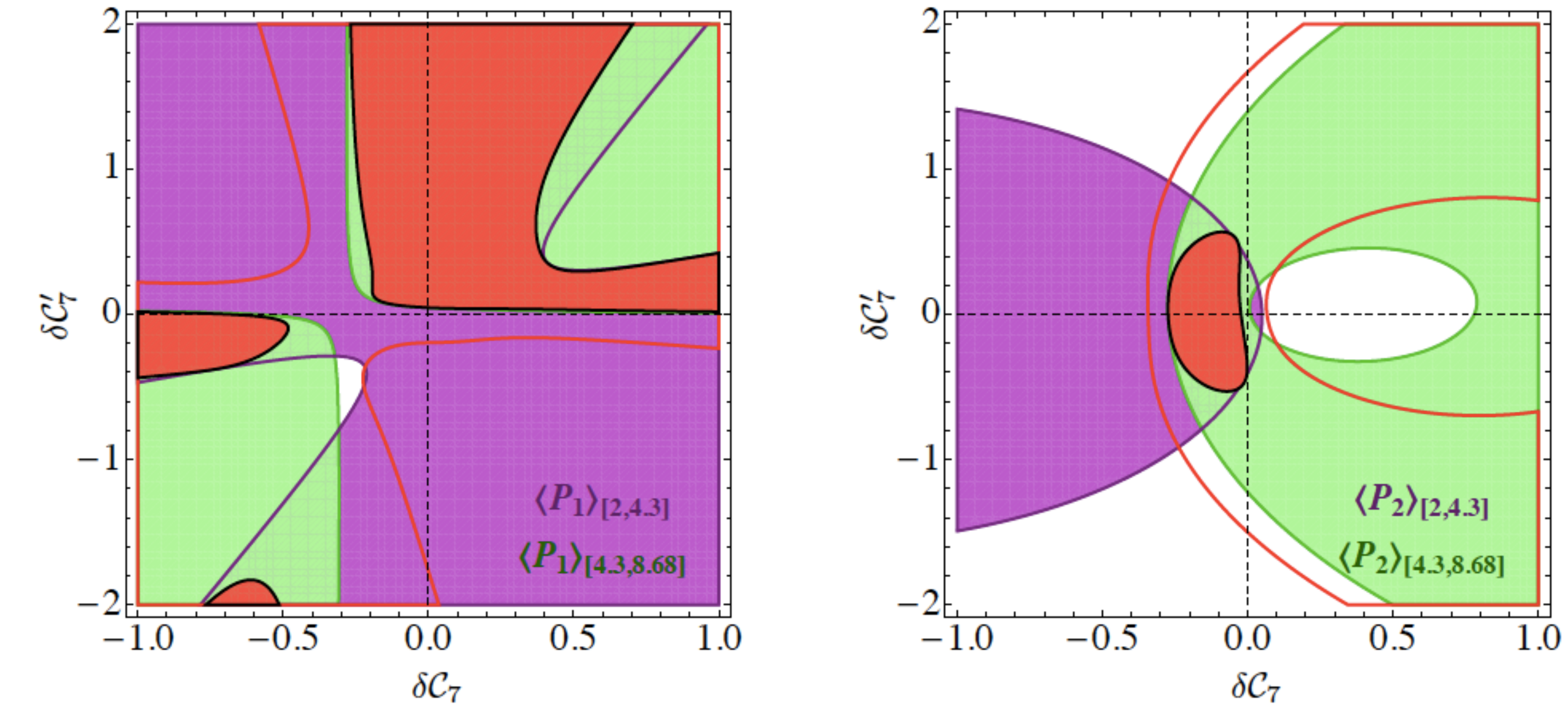}
\caption{Left: Individual 68.3\% CL constraints in the $\delta\C7(\mu_b)$-$\delta \Cp7(\mu_b)$ plane from the integrated clean observables $\av{P_1}_{[2,4.30]}$ and $\av{P_1}_{[4.30,8.68]}$, together with the combined result. The red region and contour correspond  to the combined 68.3\% and 95.5\% CL regions. Right: Same analysis for $\intbin{P_2}$. The origin corresponds to the SM point. Currently, $\intbin{P_3}$ does not provide any constraint on $\C7(\mu_b)$ and $\Cp7(\mu_b)$.}
\label{PlotPibinned}
\end{figure}

\begin{itemize}
\item The constraints from $\av{P_1}$ are not very stringent yet. However there is a very mild discrepancy of $\av{P_1}_{[4.30,8.68]}$ with respect to the SM, as well as the combined constraint from the two bins. This result is not affected by form factor uncertainties.
\item The constraints from $\av{P_2}$ are already quite interesting. The two bins point towards a negative value of $\delta \C7$, and the SM point is just outside the 68.3\% CL region. Again, this result is not affected by form factor uncertainties. While the theoretical prediction for $\av{P_2}_{[4.30,8.68]}$ can suffer from the proximity of the bin to the $c\bar c$ resonance, we point out that the same tendency to negative $\delta \C7$ is hinted at by the observable $\av{P_2}_{[1,6]}$, indicating that this is not a feature introduced by the data above 6 GeV$^2$. An enhancement of $\av{P_2}$ in the full low-$q^2$ region would be consistent with NP scenarios~$b4$ and~$c$ (see Fig.~\ref{SMNPplot4}).
\item The constraints from $\av{P_3}$ are inconclusive for the time being. This could be guessed already from the NP analysis of Section \ref{sec:comp} (see Fig.~8 in Ref.~\cite{thisv1}). It is well known that the CP-averaged version of $\intbin{P_3}$ (the one we are considering here) is not very sensitive to NP, and that the corresponding CP-asymmetry might be more interesting when constraining NP (see for example  Ref.~\cite{straub}). In this case we would suggest to focus, instead of $A_9$, on the corresponding clean CP-asymmetry, since $A_9$ can be affected by the problems discussed in Section \ref{sec:S3vsP1} concerning its sensitivity to form factors.
\end{itemize}

We stress again that these constraints should be considered as conservative, since they are based on the experimental numbers extracted in Section \ref{sec:exp} in absence of experimental correlations. Therefore, the uncertainties of $P_{1,2,3}$ that we quote in Table~\ref{TabCleanObs} are  most probably overestimated, and could be  reduced significantly once experimental correlations are available.

\subsection{Future prospects}

The experimental numbers in Table \ref{TabLHCb} for the various observables, as measured by the LHCb collaboration, contain uncertainties at the level of $\sim 0.10$, and up to $\sim 0.20$ for the observable $\av{S_3}$. The numbers for $\av{P_{1,2,3}}$ extracted in the previous section from the measurements in Table  \ref{TabLHCb} contain larger uncertainties, up to $\sim 0.5$ (see Table \ref{TabCleanObs}). As discussed above, these errors are probably overestimated since they do not take into account the relevant correlations among the observables. It is reasonable to expect that a direct extraction of $\av{P_{1,2,3}}$ from the data would give, with the present statistics, experimental uncertainties for these observables in the ballpark $\sim 0.10-0.20$, as is the case for the observables in Table \ref{TabLHCb}.

\begin{figure}\centering
\includegraphics[height=10cm]{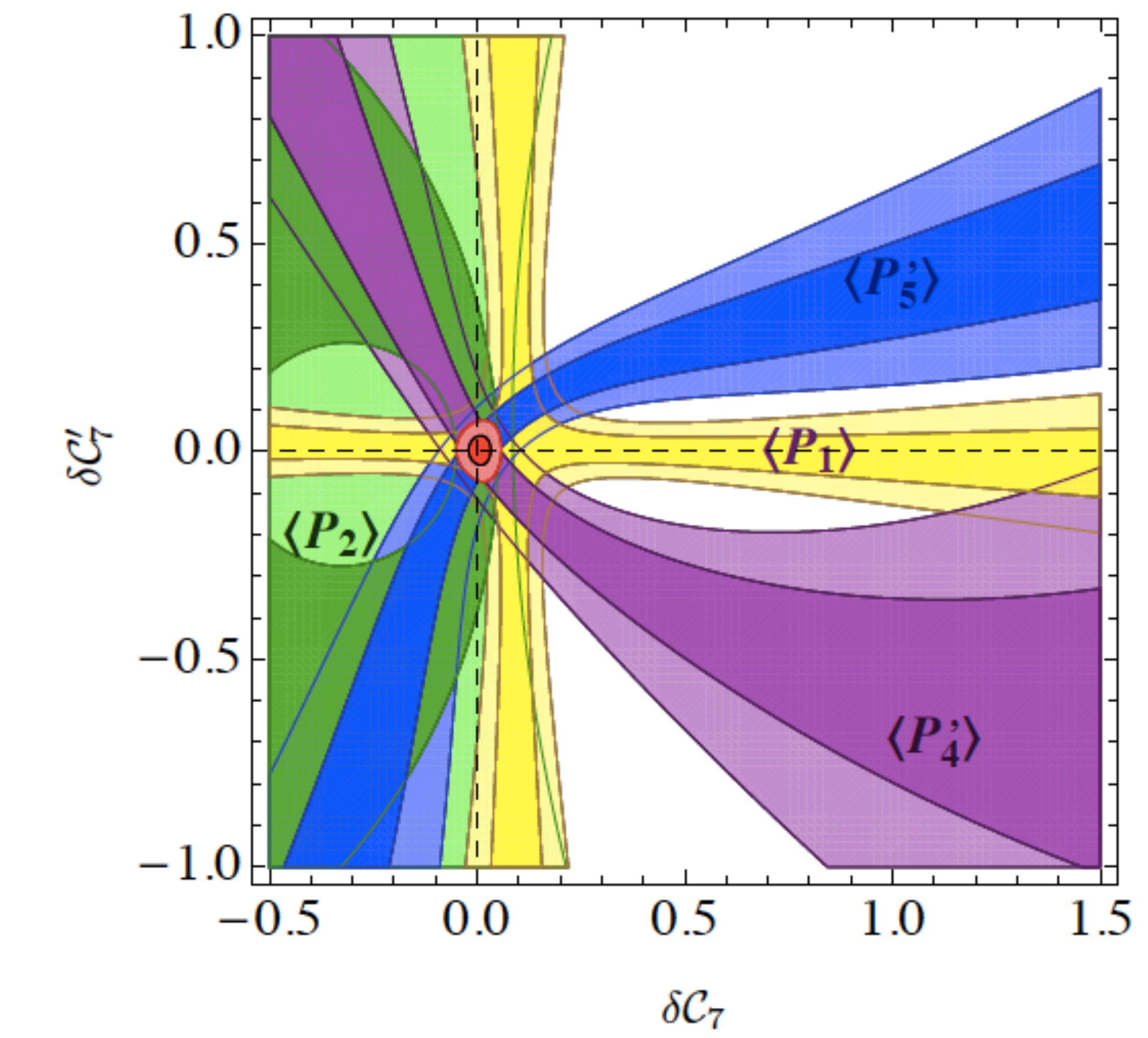}
\caption{Individual constraints in the $\delta\C7-\delta\Cp7$ plane from hypothetical measurements of the observables $\av{P_1}_{[2,4.3]}$, $\av{P_2}_{[2,4.3]}$, $\av{P'_4}_{[2,4.3]}$ and $\av{P'_5}_{[2,4.3]}$, corresponding to central values equal to the SM predictions and an experimental uncertainty $\sigma_{exp}=0.10$. The combined 68.3\% (dark red) and 95.5\% (light red) C.L. regions are also shown.}
\label{Future}
\end{figure}

We have also seen in Section \ref{subsec:Ps} that even with the enlarged uncertainties of Table \ref{TabCleanObs}, the constraints on $\C7$, $\Cp7$ from $\av{P_2}$ are nontrivial. These uncertainties will improve very significantly in the near future in forthcoming analyses of larger data sets by the LHCb collaboration.

In order to illustrate the very large impact that moderately precise measurements of the clean observables will have on New Physics, we consider the constraints on $\C7$, $\Cp7$ by hypothetical measurements of $\av{P_1}_{[2,4.3]}$, $\av{P_2}_{[2,4.3]}$, $\av{P'_4}_{[2,4.3]}$ and $\av{P'_5}_{[2,4.3]}$ with central values at their SM points, and experimental uncertainties of $\sigma_{exp}=0.10$. It is important to emphasise that these errors are not only realistic but also achievable with the current statistics. The result of this exercise is presented in Fig.~\ref{Future}, where the individual constraints in the  $\C7-\Cp7$ plane (corresponding to Scenario A) from these four observables are shown together with the combined 68.3\% and 95.5\% C.L. regions. Clearly these observables will play a very important role in the future, with the potentiality of providing the first unambiguous signal of New Physics in the flavour sector.

\section{Comparison with other works}
\label{sec:compar}

Several theoretical studies~\cite{DescotesGenon:2011yn,1202.2172,bobeth2,Altmannshofer:2011gn,bobeth3,bobeth4,straub,Damir3} have exploited the recent measurements of $b\to s\gamma$ and $b\to s \ell^+\ell^-$. We mention here briefly some differences of recent references with respect to the present work where only low-$q^2$ bins were considered. With that respect, one should notice that these references use at most the $B\to K^*\ell^+\ell^-$ observables integrated over [1,6] GeV$^2$, but not the values on finer bins included here.

In Ref.~\cite{Damir3}, a detailed discussion was provided concerning additional observables related to the photon polarisation in $b\to s\gamma$ transitions, allowing one to constrain $\C7$ and $\C7'$ more precisely. The three processes of interest were $B^0\to K^{*0} (\to K_S \pi^0) \gamma$, $B\to K_1(\to K\pi\pi)\gamma$ and $B^0\to K^{*0}(\to K^+\pi^-)\ell^+\ell^-$ at low $q^2$, showing the potential of an analysis combining all four processes even in the case of complex Wilson coefficients (assuming that there are only small contributions from the other operators for $B^0\to K^{*0}(\to K^+\pi^-)\ell^+\ell^-$ in this energy region). In particular, the current constraints on $\C7(')$ from $B(B\to X_s\gamma)$ and $S_{K^*\gamma}$ were studied, either with real or complex values for these Wilson coefficients (and SM values for the other operators). In the real case, a four-fold degeneracy is observed, corresponding essentially to the regions obtained in our Scenario A without $B\to K^*\gamma$ isospin asymmetry (see fig.~\ref{ConstraintsScA}). These ambiguities (which are even more numerous in the complex-valued cases) can in principle be lifted once more observables are measured from all three processes, with different strengths and weaknesses depending on the NP scenario considered.

In Ref.~\cite{straub} (which updates Ref.~\cite{Altmannshofer:2011gn}), branching ratios for the inclusive modes $B\to X_s \gamma$ and $B\to X_s\ell\ell$ (low and high-$q^2$) and the inclusive CP asymmetry for $b\to s\gamma$ are considered, as well as 
the branching ratio for $B\to K\ell\ell$ and $B_s\to\mu^+\mu^-$ and observables for $B\to K^*\mu^+\mu^-$ (branching ratio, longitudinal polarization fraction, forward-backward asymmetry, $A_9$ and $S_3$ both in low- and high-$q^2$ regions).
The constraints are put on $\C7('),\C9('),\C{10}('),\C{S}('),\C{P}(')$ with real or complex values, first in a frequentist framework, then in a Bayesian approach, with a focus on some specific NP scenarios. Once again, a good agreement with the SM is obtained. In addition, the possibility of a sign-flip in $\C7$ alone is disfavoured due to the branching ratio of $B\to X_s\gamma$ as well as $A_{FB}$. In the scenario without right-handed currents and with real values of the coefficients of the SM operators, the flipped sign solution $\C{7,9,10}\to - \C{7,9,10}$ remains allowed (since the $B\to K^*\gamma$ isospin asymmetry was not included in the analysis). In this scenario, the constraints obtained on the Wilson coefficients are similar to those that we obtain for scenario B, up to additional solutions with $\C7'\neq 0$. The very large parameter space and the choice of different scenarios prevents us from comparing our results in more detail, but we stress that the analysis in Ref.~\cite{straub} includes form-factor sensitive observables like $S_3$. 

In Ref.~\cite{bobeth4}, the authors did not consider the inclusive modes $B\to X_s \gamma$ and $B\to X_s\ell\ell$. On the other hand, they considered the branching ratio of the exclusive modes $B\to K^*\gamma$, $B\to K\ell\ell$, $B_s\to\mu^+\mu^-$ as well as the available observables for $B\to K^*\mu^+\mu^-$ (branching ratio, longitudinal fraction, forward-backward asymmetry, transverse asymmetry $A_T^{(2)}$ and $S_3$ both in low- and high-$q^2$ regions). The constraints were analysed in the SM basis, constraining only real values of $\C7,\C9,\C{10}$, in a Bayesian framework. The inputs for the hadronic form factors are taken from the light-cone sum rule analysis of Ref.~\cite{0412079}. They provide also predictions for the other transverse asymmetries $A_T^{(i)}$, as well as for the low-recoil observables $H_T^{(i)}$ introduced in Ref.~\cite{bobeth}. 
Obviously, as the list of inputs, scenarios and statistical frameworks are rather different, we can only perform a limited comparison with our benchmark points in different NP scenarios. In both analyses, the SM point is favoured.
A second solution, with flipped signs $\C{7,9,10}\to - \C{7,9,10}$ is allowed in Ref.~\cite{bobeth4} as in the previous references. The authors observe an update from prior to posterior p.d.f.'s of the form factors, which can be interpreted as the fact that the data themselves constrain partly the hadronic uncertainties, due to a slight tension between $B\to K\ell\ell$ and $B\to K^*\ell\ell$. In view
of our discussion concerning theoretically clean and form-factor sensitive observables, it would be interesting to perform a similar analysis separating the two sets of observables.

\section{Conclusions}
\label{sec:con}

The decay mode $B\to K^*\ell^+\ell^-$, with its large set of angular observables,  is becoming a more and more important process in constraining New Physics models. These constraints are complementary to those from inclusive and exclusive radiative decays. With increasingly precise experimental data on these modes and the prospects for the future from the LHC, theoretical uncertainties must be kept under control. In this context, the use of theoretically clean observables is not only convenient but also extremely recommendable, or even mandatory. The sensitivity of different observables to hadronic uncertainties  has been addressed in Section \ref{sec:S3vsP1}, and the conclusion is that clean observables such as $P_1$ are far more robust than other observables like $F_L$ or $S_3$, translating into a better performance in discriminating among different models.

A full description of the angular distribution of the $B\to K^*\ell^+\ell^-$ decay in terms of a maximum set of clean observables was presented in Ref.~\cite{primary}, where the observables $P_{1,2,3,4,5,6}$ where introduced. In this paper we have presented 
a simple and compact expression for the coefficients of the distribution in terms of these clean observables, and given
SM predictions for these observables integrated over a series of $q^2$-bins, that can be directly compared with experimental data.
These predictions are collected in Table \ref{TableSMPs} and in Figs.~\ref{SMplotsPs1} and \ref{SMplotsPs2}. As an important point, we have seen that the first three clean observables $P_{1,2,3}$ are \emph{already measured}, and can be extracted form the latest measurements by the LHCb collaboration \cite{LHCbinned}. The experimental numbers for these observables together with their SM predictions are given in Table \ref{TabCleanObs}. The uncertainties attached to these numbers are certainly overestimated, as we did not have the required correlations among experimental measurements (and thus treated them as uncorrelated).

We have also studied the model-independent constraints on the Wilson coefficients $\C7^{(')}$, $\C9^{(')}$, $\C{10}^{(')}$ from radiative decays $B\to X_s\gamma$ and $B\to K^*\gamma$, and semileptonic decays $B\to X_s\mu^+\mu^-$ and $B\to K^*\mu^+\mu^-$. Excluding all $B\to K^*\mu^+\mu^-$ observables except for the integrated observables $\av{\afb}_{[1,6]}$ and $\av{F_L}_{[1,6]}$ leads to the set of constraints shown in Figs.~\ref{ConstraintsScA}-\ref{ConstraintsScC}. Identifying a set of benchmark New Physics points compatible with those bounds, one can see that very large New Physics contributions to other observables in $B\to K^*\mu^+\mu^-$ are allowed, specially in $P_1$, $P_2$, $P_4'$ and $P_5'$ (see Fig.~\ref{SMNPplot3}). We have investigated the present constraints imposed by $P_1$, $P_2$ and $P_3$ on $\C7$, $\Cp7$, with already quite interesting constraints from $P_2$ suggesting a lower value of $\C7$ than the SM value.
We then showed the powerful prospects that the set of clean observables $P_{1,2,3}$ and $P_{4',5',6'}$ will have in the short term to discriminate possible New Physics contributions, illustrated in Fig.~\ref{Future}. 

Considering the advantages provided by the use of the $P_i$ observables at large recoil, we hope that the present study will be a strong incentive for experimentalists to rephrase their study of the low-$q^2$ region in terms of these observables.


\subsubsection*{Acknowledgments}

It is a pleasure to thank Nicola Serra and Thomas Blake for helpful discussions on experimental issues. We also acknowledge fruitful discussions with Rafel Escribano and Federico Mescia. 
J.M. enjoys financial support from FPA2011-25948, SGR2009-00894. J.V. is supported in part by ICREA-Academia funds and FPA2011-25948.

\appendix

\section{$J_8$ in terms of $P_i$ observables and the $Q$ observable}
\label{appA}

In Section~\ref{sec:2} we have provided the explicit expressions for the coefficients
of the distribution  in terms of the observables of the basis, neglecting mass terms
and scalar contributions.\footnote{The general case including lepton masses and scalars is
discussed in full detail in Ref.~\cite{primary}.} In this case, one finds very simple and
compact expressions for all these coefficients (see Eqs.(\ref{Jsintermsobs})) with the exception of $J_8$. The reason is that
$J_8$, in the absence of scalar contributions, is {\it not} an
independent quantity (exactly like $J_{1s}=3 J_{2s}$ and $J_{1c}=-J_{2c}$ in the massless case) and
deserves a separate discussion. The counting of degrees of freedom and continuous
symmetries in this case shows that there are only 8 degrees of freedom
parametrised by the observables $P_{i=1\ldots6}$, $F_L$ and $ {d\Gamma}/{dq^2}$. This
means that $J_8$ can be expressed in terms of these observables:
\eqa{
J_8 &=& - \sqrt{\dfrac{F_T F_L}{1-P_1}} \frac{d\Gamma}{dq^2} \Bigg\{ (P_2 P_6 -P_3
P_4) 
+\eta \left( (P_2 P_6-P_3 P_4)^2   +P_5 (P_2 P_4 + P_3 P_6) \sqrt{1-P_1^2} \right.
\nonumber \\ 
&+&\left. \frac{1}{4}(1-\sum_{i=4}^6 P_i^2) (1-P_1^2) -
P_2^2 - P_3^2\right)^{\frac{1}{2}} \Bigg\} 
 \label{j8}}
This expression is found by solving $J_8$ in terms of the other coefficients using
the relation in Eq.~(3.15) of Ref.~\cite{matias2} together with Eqs.~(\ref{Jsintermsobs}).
One can also replace $P_{4,5,6}$ by $P'_{4,5,6}$ using Eqs.~(\ref{P4'})-(\ref{P6'}).

We notice that in Eq.~(\ref{j8}), a discrete quantity $\eta$ is left as a free parameter
that  can take only two values $\pm 1$ in the massless case. This parameter is
indeed an observable, and its SM prediction is $\eta^{SM}=+1$  as
can be seen by substituting the SM values for the observables $P_i$.
This has the interesting consequence that a measurement of $\eta=-1$  would be an
unambiguous indication of New Physics (originating, for instance, from new weak phases
or sign flips in Wilson coefficients). Deviations from $|\eta|=1$  could be
expected, for instance, from  scalar contributions entering  $P_5$
and $P_6$, since such contributions would break the symmetry relation among the coefficients of the
angular distribution.

However, when one tries to write a similar relationship for ``binned" observables, it is clear that a naive substitution  $P_{i} \to \intbin{P_{i}}$ is not
possible  due to the highly non-linear form of Eq.~(\ref{j8}). One practical solution is to introduce an extra clean observable
$Q$ (or $Q'$):
\eqa{J_8=-\frac{1}{2} Q^\prime \sqrt{{F_T F_L}} \frac{d\Gamma}{dq^2}\ , \label{p8}}
with $Q^\prime=Q \sqrt{1+P_1}$ and whose definition, also valid in the massive case, is
%
\eqa{\label{defq}Q=\frac{{\rm Im}(n_0^\dagger n_\perp)}{\sqrt{|n_0|^2 |n_\perp|^2}}=
-\frac{\sqrt{2} J_8}{\sqrt{- J_{2c} (2 J_{2s}+J_3)}}\ .}
The vectors $n_i$ (with $i=0,\perp$) are defined in Ref.\cite{primary}.
This clean observable $Q$ (or $Q'$) is related to the form
factor-sensitive observable $S_8$ by means of
\eq{Q^\prime=-\frac{2 S_8}{\sqrt{F_T F_L}}\ .}
For completeness we also provide the expression of this observable in terms of our
basis of observables:
\eqa{
Q &=&  \frac{2}{\sqrt{1-P_1^2}}  \Bigg\{ (P_2 P_6 -P_3 P_4) 
+\eta \left( (P_2 P_6-P_3 P_4)^2   +P_5 (P_2 P_4 + P_3 P_6) \sqrt{1-P_1^2} \right.
\nonumber \\ 
&+&\left. \frac{1}{4}(1-\sum_{i=4}^6 P_i^2) (1-P_1^2) -
P_2^2 - P_3^2\right)^{\frac{1}{2}} \Bigg\} \ .
 \label{q}}
In principle, this implies that $Q$ (or $Q^\prime$) can be measured using either
Eq.~(\ref{defq}) or Eq.~(\ref{q}) (with $\eta=+1$ in the SM but free in general).
However, Eq.~(\ref{q}) would be of practical experimental use only in
the limit of the size of the binning going to zero.
Of course, Eq.~(\ref{q}) can be very easily turned into an equation
that relates $S_8$ with the other $S_i$ by using Eq.~(\ref{Pis}) and
Eq.~(\ref{P456}). This shows the redundancy of $S_8$ in the massless case, again up
to a single discrete parameter $\eta$.

\begin{figure}\centering
\includegraphics[height=7cm]{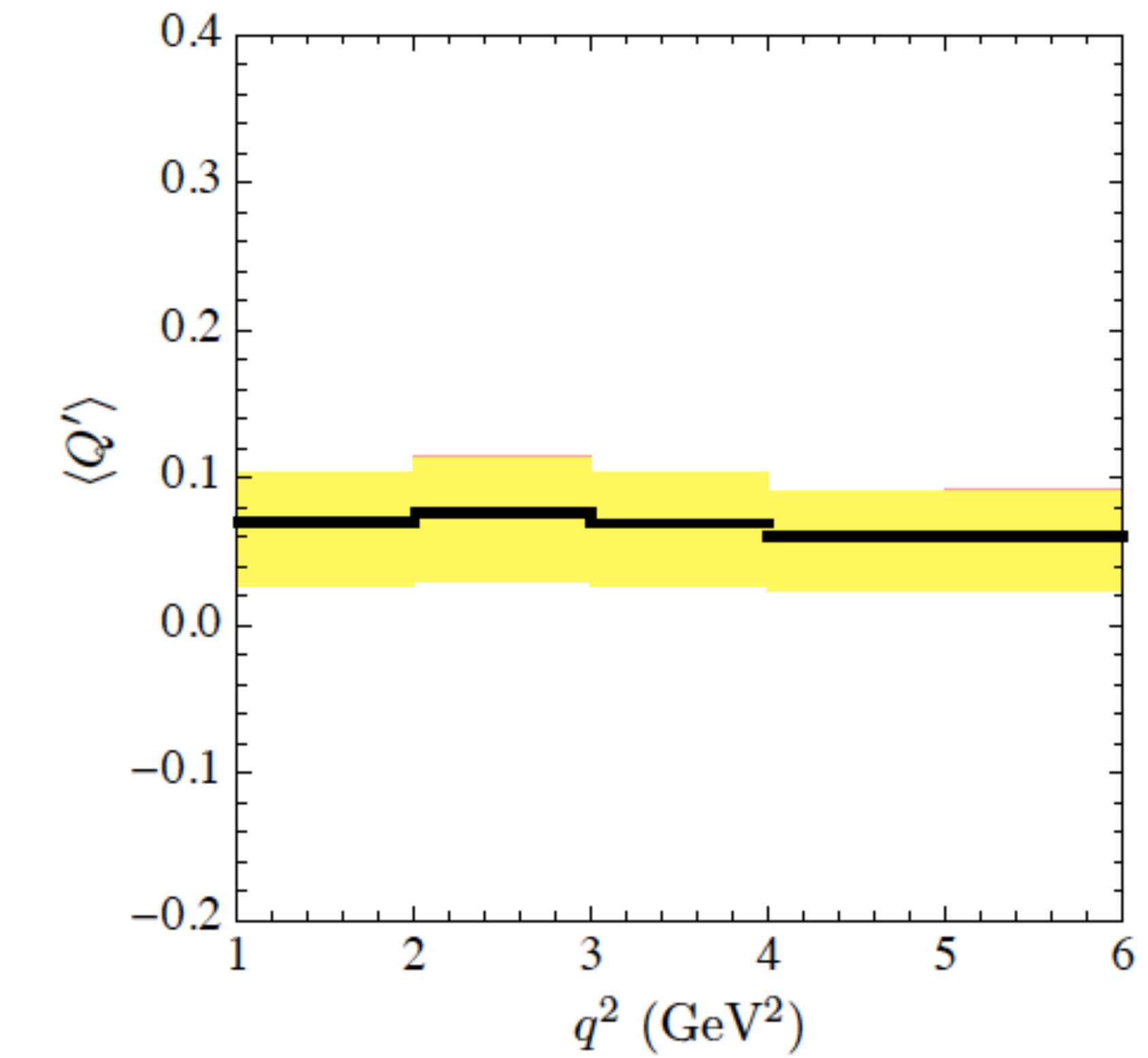}
\caption{SM prediction for the observable $Q'$, integrated in 5 bins of width 1 GeV$^2$.}
\label{PlotQ}
\end{figure}

At a more practical level, the integrated form of this observable from
Eq.~(\ref{defq}) is:
\eqa{
\av{Q'}_\bin &=& \frac{-2 \int_{{\rm bin}} dq^2 J_8(q^2)}{\sqrt{\int_{{\rm bin}}
dq^2 c_4(q^2)  \int_{{\rm bin}} dq^2 (c_0(q^2)-c_4(q^2))}} = \frac{-2
\intbin{J_8}}{\sqrt{\intbin{c_4}\big(\intbin{c_0}-\intbin{c_4}\big)}} \quad
}
and its SM prediction  is given in Fig.~\ref{PlotQ}. Notice that, as mentioned above, the
binning procedure breaks the relation Eq.~(3.15) of  Ref.~\cite{matias2} among the
coefficients of the distribution, which is only recovered in the limit of the size of
the binning going to zero, when the observable $\eta$ would become accessible.

In conclusion the set of Eqs.(\ref{Jsintermsobs}) and Eq.~(\ref{j8}) provides a
complete parametrisation of the distribution in the massless case with no scalars in
terms of only eight parameters (eight observables $P_{1,2,3,4,5,6}$ and $F_L,
d\Gamma/dq^2$) in agreement with the eight independent degrees of freedom, but there
is one extra redundant $Q$ (or $Q^\prime$) observable (up to a discrete parameter $\eta$)
which however can be fixed only once $\intbin{J_8}$ is measured, due to the binning procedure adopted 
by experimental analyses.

\section{New Physics expressions for binned observables}
\label{appB}

In this appendix we present the numerical expressions for the integrated observables $\av{\afb}$, $\av{F_L}$, $\av{P_{1,2,3}}$ and $\av{P'_{4,5,6}}$ as a function of the NP Wilson coefficients, for different choices of the $q^2$-binning.

The formula for any observable $\av{\op_k}$ has the general form:
\eq{
\av{\op_k} = \frac{\sum_{i,j =0,7,7',9,9',10,10'} N_{(i,j)}\, \delta \C{i}\, \delta \C{j}}{\sum_{i,j =0,7,7',9,9',10,10'} D_{(i,j)}\, \delta \C{i}\, \delta \C{j}} \pm \delta_k
\label{Ok}}
where $\delta C_0 \equiv 1$. The coefficients $\delta$, $N_{i,j}$ and $D_{i,j}$ corresponding to each observable are collected in Tables \ref{TableAFB}-\ref{TableP3}, where only nonzero coefficients are displayed. The coefficient $\delta\C{i}$ denotes the NP contribution to the Wilson coefficient $\C{i}$ at the hadronic scale $\mu_b$. The parameters $\delta_k$ correspond to the theoretical error assigned to each observable, and are also collected in Tables \ref{TableAFB}-\ref{TableP3}. We assume here that the uncertainty obtained within the SM can be considered as a good estimate of the uncertainty for arbitrary (NP) values of the Wilson coefficients. As discussed in detail in Section~\ref{sec:S3vsP1}, this assumption is appropriate for theoretically clean observables only.


\begin{table}
\scriptsize
\centering
\begin{tabular}{||c||r|r|r|r|r|r|r|r||}
\hline\hline
\multicolumn{9}{||c||}{$\av{\afb}$}\\
\hline\hline
  & [\,1\,,\,6\,] & [\,2\,,\,4.3\,] & [\,4.3\,,\,8.68\,] & [\,1\,,\,2\,] & [\,2\,,\,3\,] & [\,3\,,\,4\,] & [\,4\,,\,5\,] & [\,5\,,\,6\,] \\
\hline\hline

 $\delta$ & 0.035 & 0.059 & 0.126 & 0.127 & 0.106 & 0.04 & 0.051 & 0.089 \\
\hline

$N_{(0,0)}$ & $-132.63$ & $-129.81$ & $1075.21$ & $-169.19$ & $-103.04$ & $-31.81$ & $44.25$ & $127.17$ \\ 

$N_{(0,7)}$ & $3659.89$ & $1683.77$ & $3248.78$ & $709.73$ & $725.48$ & $736.$ & $742.78$ & $745.9$ \\ 

$N_{(0,7')}$ & $53.45$ & $24.59$ & $47.45$ & $10.37$ & $10.6$ & $10.75$ & $10.85$ & $10.89$ \\ 

$N_{(0,9)}$ & $273.22$ & $112.45$ & $446.79$ & $22.53$ & $38.38$ & $54.54$ & $70.81$ & $86.96$ \\ 


$N_{(0,10)}$ & $30.78$ & $30.13$ & $-249.55$ & $39.27$ & $23.92$ & $7.38$ & $-10.27$ & $-29.51$ \\ 

$N_{(0,10')}$ & $-9.55$ & $-4.44$ & $-7.25$ & $-1.92$ & $-1.93$ & $-1.93$ & $-1.91$ & $-1.85$ \\ 







$N_{(7,7')}$ & $-1210.54$ & $-544.54$ & $-1139.94$ & $-242.44$ & $-234.51$ & $-237.58$ & $-244.07$ & $-251.93$ \\ 

$N_{(7,9)}$ & $1502.03$ & $688.89$ & $1366.$ & $286.75$ & $295.32$ & $301.92$ & $307.12$ & $310.92$ \\ 

$N_{(7,9')}$ & $-415.72$ & $-187.6$ & $-424.87$ & $-73.44$ & $-78.45$ & $-83.25$ & $-87.97$ & $-92.62$ \\ 



$N_{(7',9)}$ & $-448.14$ & $-202.46$ & $-454.37$ & $-79.62$ & $-84.82$ & $-89.76$ & $-94.6$ & $-99.33$ \\ 

$N_{(7',9')}$ & $1511.64$ & $693.23$ & $1375.74$ & $288.45$ & $297.14$ & $303.84$ & $309.14$ & $313.05$ \\ 



$N_{(9,9')}$ & $-182.43$ & $-82.42$ & $-184.94$ & $-32.42$ & $-34.53$ & $-36.54$ & $-38.5$ & $-40.43$ \\ 





$N_{(10,10')}$ & $-182.43$ & $-82.42$ & $-184.94$ & $-32.42$ & $-34.53$ & $-36.54$ & $-38.5$ & $-40.43$ \\ 
\hline

$D_{(0,0)}$ & $3930.05$ & $1652.9$ & $4881.21$ & $788.91$ & $693.77$ & $727.04$ & $805.92$ & $914.41$ \\ 

$D_{(0,7)}$ & $-675.8$ & $-133.39$ & $3759.72$ & $-1510.4$ & $-367.15$ & $131.35$ & $427.89$ & $642.5$ \\ 

$D_{(0,7')}$ & $-1514.33$ & $-676.99$ & $-1832.12$ & $-243.59$ & $-277.12$ & $-303.71$ & $-329.95$ & $-359.96$ \\ 

$D_{(0,9)}$ & $607.74$ & $251.22$ & $1056.34$ & $62.61$ & $90.17$ & $119.3$ & $150.44$ & $185.22$ \\ 

$D_{(0,9')}$ & $-659.$ & $-294.16$ & $-766.64$ & $-114.1$ & $-122.32$ & $-130.82$ & $-140.19$ & $-151.57$ \\ 

$D_{(0,10)}$ & $-1121.92$ & $-494.21$ & $-1333.74$ & $-167.85$ & $-196.49$ & $-224.8$ & $-252.76$ & $-280.02$ \\ 

$D_{(0,10')}$ & $785.99$ & $355.1$ & $796.84$ & $139.7$ & $148.79$ & $157.42$ & $165.89$ & $174.19$ \\ 

$D_{(7,7)}$ & $19751.1$ & $8522.58$ & $8359.39$ & $7326.88$ & $4445.59$ & $3246.87$ & $2580.52$ & $2151.26$ \\ 

$D_{(7',7')}$ & $19812.1$ & $8549.86$ & $8412.14$ & $7340.65$ & $4457.7$ & $3258.55$ & $2592.18$ & $2163.05$ \\ 

$D_{(9,9)}$ & $260.39$ & $114.7$ & $309.56$ & $38.96$ & $45.6$ & $52.18$ & $58.66$ & $64.99$ \\ 

$D_{(9',9')}$ & $260.39$ & $114.7$ & $309.56$ & $38.96$ & $45.6$ & $52.18$ & $58.66$ & $64.99$ \\ 

$D_{(10,10)}$ & $260.39$ & $114.7$ & $309.56$ & $38.96$ & $45.6$ & $52.18$ & $58.66$ & $64.99$ \\ 

$D_{(10',10')}$ & $260.39$ & $114.7$ & $309.56$ & $38.96$ & $45.6$ & $52.18$ & $58.66$ & $64.99$ \\ 

$D_{(7,7')}$ & $-1210.54$ & $-544.54$ & $-1139.94$ & $-242.44$ & $-234.51$ & $-237.58$ & $-244.07$ & $-251.93$ \\ 

$D_{(7,9)}$ & $1502.03$ & $688.89$ & $1366.$ & $286.75$ & $295.32$ & $301.92$ & $307.12$ & $310.92$ \\ 

$D_{(7,9')}$ & $-415.72$ & $-187.6$ & $-424.87$ & $-73.44$ & $-78.45$ & $-83.25$ & $-87.97$ & $-92.62$ \\ 



$D_{(7',9)}$ & $-448.14$ & $-202.46$ & $-454.37$ & $-79.62$ & $-84.82$ & $-89.76$ & $-94.6$ & $-99.33$ \\ 

$D_{(7',9')}$ & $1511.64$ & $693.23$ & $1375.74$ & $288.45$ & $297.14$ & $303.84$ & $309.14$ & $313.05$ \\ 



$D_{(9,9')}$ & $-182.43$ & $-82.42$ & $-184.94$ & $-32.42$ & $-34.53$ & $-36.54$ & $-38.5$ & $-40.43$ \\ 





$D_{(10,10')}$ & $-182.43$ & $-82.42$ & $-184.94$ & $-32.42$ & $-34.53$ & $-36.54$ & $-38.5$ & $-40.43$ \\ 
\hline\hline
\end{tabular}
\caption{Coefficients for the New Physics formula of $\av{\afb}$.}
\label{TableAFB}
\end{table}


\begin{table}
\scriptsize
\centering
\begin{tabular}{||c||r|r|r|r|r|r|r|r||}
\hline\hline
\multicolumn{9}{||c||}{$\av{F_L}$}\\
\hline\hline
  & [\,1\,,\,6\,] & [\,2\,,\,4.3\,] & [\,4.3\,,\,8.68\,] & [\,1\,,\,2\,] & [\,2\,,\,3\,] & [\,3\,,\,4\,] & [\,4\,,\,5\,] & [\,5\,,\,6\,] \\
\hline\hline

 $\delta$ & 0.178 & 0.162 & 0.18 & 0.211 & 0.164 & 0.158 & 0.169 & 0.179 \\
\hline

$N_{(0,0)}$ & $1413.04$ & $635.91$ & $1559.39$ & $251.77$ & $266.81$ & $281.63$ & $297.22$ & $315.6$ \\ 

$N_{(0,7)}$ & $681.08$ & $305.96$ & $758.66$ & $120.71$ & $128.16$ & $135.6$ & $143.55$ & $153.06$ \\ 

$N_{(0,7')}$ & $-696.01$ & $-312.66$ & $-775.28$ & $-123.35$ & $-130.97$ & $-138.58$ & $-146.69$ & $-156.42$ \\ 

$N_{(0,9)}$ & $323.9$ & $145.53$ & $360.32$ & $57.46$ & $60.98$ & $64.49$ & $68.24$ & $72.74$ \\ 

$N_{(0,9')}$ & $-323.9$ & $-145.53$ & $-360.32$ & $-57.46$ & $-60.98$ & $-64.49$ & $-68.24$ & $-72.74$ \\ 

$N_{(0,10)}$ & $-378.21$ & $-171.92$ & $-367.37$ & $-68.88$ & $-72.6$ & $-75.94$ & $-79.01$ & $-81.79$ \\ 

$N_{(0,10')}$ & $378.21$ & $171.92$ & $367.37$ & $68.88$ & $72.6$ & $75.94$ & $79.01$ & $81.79$ \\ 

$N_{(7,7)}$ & $388.11$ & $176.38$ & $377.96$ & $70.55$ & $74.44$ & $77.92$ & $81.14$ & $84.06$ \\ 

$N_{(7',7')}$ & $405.3$ & $184.19$ & $394.71$ & $73.68$ & $77.73$ & $81.38$ & $84.73$ & $87.79$ \\ 

$N_{(9,9)}$ & $87.78$ & $39.9$ & $85.27$ & $15.99$ & $16.85$ & $17.62$ & $18.34$ & $18.98$ \\ 

$N_{(9',9')}$ & $87.78$ & $39.9$ & $85.27$ & $15.99$ & $16.85$ & $17.62$ & $18.34$ & $18.98$ \\ 

$N_{(10,10)}$ & $87.78$ & $39.9$ & $85.27$ & $15.99$ & $16.85$ & $17.62$ & $18.34$ & $18.98$ \\ 

$N_{(10',10')}$ & $87.78$ & $39.9$ & $85.27$ & $15.99$ & $16.85$ & $17.62$ & $18.34$ & $18.98$ \\ 

$N_{(7,7')}$ & $-605.27$ & $-272.27$ & $-569.97$ & $-121.22$ & $-117.25$ & $-118.79$ & $-122.04$ & $-125.97$ \\ 

$N_{(7,9)}$ & $751.01$ & $344.45$ & $683.$ & $143.37$ & $147.66$ & $150.96$ & $153.56$ & $155.46$ \\ 

$N_{(7,9')}$ & $-207.86$ & $-93.8$ & $-212.44$ & $-36.72$ & $-39.23$ & $-41.62$ & $-43.98$ & $-46.31$ \\ 



$N_{(7',9)}$ & $-224.07$ & $-101.23$ & $-227.18$ & $-39.81$ & $-42.41$ & $-44.88$ & $-47.3$ & $-49.67$ \\ 

$N_{(7',9')}$ & $755.82$ & $346.62$ & $687.87$ & $144.23$ & $148.57$ & $151.92$ & $154.57$ & $156.53$ \\ 



$N_{(9,9')}$ & $-91.21$ & $-41.21$ & $-92.47$ & $-16.21$ & $-17.27$ & $-18.27$ & $-19.25$ & $-20.21$ \\ 





$N_{(10,10')}$ & $-91.21$ & $-41.21$ & $-92.47$ & $-16.21$ & $-17.27$ & $-18.27$ & $-19.25$ & $-20.21$ \\ 
\hline

$D_{(0,0)}$ & $1965.02$ & $826.45$ & $2440.6$ & $394.45$ & $346.88$ & $363.52$ & $402.96$ & $457.21$ \\ 

$D_{(0,7)}$ & $-337.9$ & $-66.69$ & $1879.86$ & $-755.2$ & $-183.57$ & $65.68$ & $213.94$ & $321.25$ \\ 

$D_{(0,7')}$ & $-757.17$ & $-338.49$ & $-916.06$ & $-121.8$ & $-138.56$ & $-151.85$ & $-164.98$ & $-179.98$ \\ 

$D_{(0,9)}$ & $303.87$ & $125.61$ & $528.17$ & $31.31$ & $45.09$ & $59.65$ & $75.22$ & $92.61$ \\ 

$D_{(0,9')}$ & $-329.5$ & $-147.08$ & $-383.32$ & $-57.05$ & $-61.16$ & $-65.41$ & $-70.1$ & $-75.79$ \\ 

$D_{(0,10)}$ & $-560.96$ & $-247.1$ & $-666.87$ & $-83.93$ & $-98.25$ & $-112.4$ & $-126.38$ & $-140.01$ \\ 

$D_{(0,10')}$ & $393.$ & $177.55$ & $398.42$ & $69.85$ & $74.4$ & $78.71$ & $82.94$ & $87.09$ \\ 

$D_{(7,7)}$ & $9875.55$ & $4261.29$ & $4179.7$ & $3663.44$ & $2222.8$ & $1623.43$ & $1290.26$ & $1075.63$ \\ 

$D_{(7',7')}$ & $9906.06$ & $4274.93$ & $4206.07$ & $3670.32$ & $2228.85$ & $1629.28$ & $1296.09$ & $1081.53$ \\ 

$D_{(9,9)}$ & $130.2$ & $57.35$ & $154.78$ & $19.48$ & $22.8$ & $26.09$ & $29.33$ & $32.5$ \\ 

$D_{(9',9')}$ & $130.2$ & $57.35$ & $154.78$ & $19.48$ & $22.8$ & $26.09$ & $29.33$ & $32.5$ \\ 

$D_{(10,10)}$ & $130.2$ & $57.35$ & $154.78$ & $19.48$ & $22.8$ & $26.09$ & $29.33$ & $32.5$ \\ 

$D_{(10',10')}$ & $130.2$ & $57.35$ & $154.78$ & $19.48$ & $22.8$ & $26.09$ & $29.33$ & $32.5$ \\ 

$D_{(7,7')}$ & $-605.27$ & $-272.27$ & $-569.97$ & $-121.22$ & $-117.25$ & $-118.79$ & $-122.04$ & $-125.97$ \\ 

$D_{(7,9)}$ & $751.01$ & $344.45$ & $683.$ & $143.37$ & $147.66$ & $150.96$ & $153.56$ & $155.46$ \\ 

$D_{(7,9')}$ & $-207.86$ & $-93.8$ & $-212.44$ & $-36.72$ & $-39.23$ & $-41.62$ & $-43.98$ & $-46.31$ \\ 



$D_{(7',9)}$ & $-224.07$ & $-101.23$ & $-227.18$ & $-39.81$ & $-42.41$ & $-44.88$ & $-47.3$ & $-49.67$ \\ 

$D_{(7',9')}$ & $755.82$ & $346.62$ & $687.87$ & $144.23$ & $148.57$ & $151.92$ & $154.57$ & $156.53$ \\ 



$D_{(9,9')}$ & $-91.21$ & $-41.21$ & $-92.47$ & $-16.21$ & $-17.27$ & $-18.27$ & $-19.25$ & $-20.21$ \\ 





$D_{(10,10')}$ & $-91.21$ & $-41.21$ & $-92.47$ & $-16.21$ & $-17.27$ & $-18.27$ & $-19.25$ & $-20.21$ \\ 

\hline\hline
\end{tabular}
\caption{Coefficients for the New Physics formula of $\av{F_L}$.}
\label{TableFL}
\end{table}


\begin{table}
\scriptsize
\centering
\begin{tabular}{||c||r|r|r|r|r|r|r|r||}
\hline\hline
\multicolumn{9}{||c||}{$\av{P_1}$}\\
\hline\hline
  & [\,1\,,\,6\,] & [\,2\,,\,4.3\,] & [\,4.3\,,\,8.68\,] & [\,1\,,\,2\,] & [\,2\,,\,3\,] & [\,3\,,\,4\,] & [\,4\,,\,5\,] & [\,5\,,\,6\,] \\
\hline\hline

 $\delta$ & 0.051 & 0.05 & 0.06 & 0.052 & 0.056 & 0.046 & 0.046 & 0.053 \\
\hline

$N_{(0,0)}$ & $-30.44$ & $-9.75$ & $-103.1$ & $1.11$ & $-1.96$ & $-5.4$ & $-9.53$ & $-14.65$ \\ 

$N_{(0,7)}$ & $-82.98$ & $-33.81$ & $-116.67$ & $-17.23$ & $-14.27$ & $-14.77$ & $-16.77$ & $-19.95$ \\ 

$N_{(0,7')}$ & $-1017.67$ & $-372.1$ & $1124.22$ & $-875.94$ & $-311.58$ & $-69.64$ & $70.79$ & $168.7$ \\ 

$N_{(0,9)}$ & $-5.6$ & $-1.55$ & $-23.$ & $0.41$ & $-0.18$ & $-0.92$ & $-1.85$ & $-3.05$ \\ 

$N_{(0,9')}$ & $-20.03$ & $-19.92$ & $167.85$ & $-26.15$ & $-15.89$ & $-4.84$ & $6.98$ & $19.87$ \\ 

$N_{(0,10)}$ & $14.79$ & $5.63$ & $31.05$ & $0.98$ & $1.8$ & $2.78$ & $3.94$ & $5.3$ \\ 

$N_{(0,10')}$ & $-182.75$ & $-75.18$ & $-299.5$ & $-15.05$ & $-25.65$ & $-36.46$ & $-47.37$ & $-58.22$ \\ 

$N_{(7,7)}$ & $-5.14$ & $-4.4$ & $-102.13$ & $27.93$ & $4.9$ & $-6.01$ & $-13.18$ & $-18.78$ \\ 

$N_{(7',7')}$ & $-412.36$ & $-179.74$ & $-265.45$ & $-126.24$ & $-87.31$ & $-72.35$ & $-65.09$ & $-61.37$ \\ 

$N_{(9,9)}$ & $-3.43$ & $-1.31$ & $-7.21$ & $-0.23$ & $-0.42$ & $-0.64$ & $-0.91$ & $-1.23$ \\ 

$N_{(9',9')}$ & $-3.43$ & $-1.31$ & $-7.21$ & $-0.23$ & $-0.42$ & $-0.64$ & $-0.91$ & $-1.23$ \\ 

$N_{(10,10)}$ & $-3.43$ & $-1.31$ & $-7.21$ & $-0.23$ & $-0.42$ & $-0.64$ & $-0.91$ & $-1.23$ \\ 

$N_{(10',10')}$ & $-3.43$ & $-1.31$ & $-7.21$ & $-0.23$ & $-0.42$ & $-0.64$ & $-0.91$ & $-1.23$ \\ 

$N_{(7,7')}$ & $-208.65$ & $-92.03$ & $-183.73$ & $-49.13$ & $-41.19$ & $-39.16$ & $-39.12$ & $-40.06$ \\ 

$N_{(7,9)}$ & $566.44$ & $260.55$ & $503.48$ & $109.79$ & $112.25$ & $113.9$ & $114.99$ & $115.51$ \\ 

$N_{(7,9')}$ & $-23.29$ & $-9.91$ & $-32.92$ & $-3.14$ & $-3.81$ & $-4.56$ & $-5.41$ & $-6.36$ \\ 



$N_{(7',9)}$ & $-35.45$ & $-15.5$ & $-43.73$ & $-5.49$ & $-6.22$ & $-7.01$ & $-7.88$ & $-8.84$ \\ 

$N_{(7',9')}$ & $567.2$ & $260.89$ & $504.42$ & $109.91$ & $112.38$ & $114.05$ & $115.15$ & $115.7$ \\ 



$N_{(9,9')}$ & $-3.43$ & $-1.31$ & $-7.21$ & $-0.23$ & $-0.42$ & $-0.64$ & $-0.91$ & $-1.23$ \\ 





$N_{(10,10')}$ & $-3.43$ & $-1.31$ & $-7.21$ & $-0.23$ & $-0.42$ & $-0.64$ & $-0.91$ & $-1.23$ \\ 
\hline

$D_{(0,0)}$ & $551.99$ & $190.54$ & $881.21$ & $142.68$ & $80.07$ & $81.89$ & $105.74$ & $141.61$ \\ 

$D_{(0,7)}$ & $-1018.98$ & $-372.65$ & $1121.2$ & $-875.91$ & $-311.74$ & $-69.93$ & $70.4$ & $168.19$ \\ 

$D_{(0,7')}$ & $-61.16$ & $-25.83$ & $-140.78$ & $1.56$ & $-7.59$ & $-13.28$ & $-18.28$ & $-23.56$ \\ 

$D_{(0,9)}$ & $-20.03$ & $-19.92$ & $167.85$ & $-26.15$ & $-15.89$ & $-4.84$ & $6.98$ & $19.87$ \\ 

$D_{(0,9')}$ & $-5.6$ & $-1.55$ & $-23.$ & $0.41$ & $-0.18$ & $-0.92$ & $-1.85$ & $-3.05$ \\ 

$D_{(0,10)}$ & $-182.75$ & $-75.18$ & $-299.5$ & $-15.05$ & $-25.65$ & $-36.46$ & $-47.37$ & $-58.22$ \\ 

$D_{(0,10')}$ & $14.79$ & $5.63$ & $31.05$ & $0.98$ & $1.8$ & $2.78$ & $3.94$ & $5.3$ \\ 

$D_{(7,7)}$ & $9487.44$ & $4084.91$ & $3801.73$ & $3592.89$ & $2148.36$ & $1545.51$ & $1209.12$ & $991.57$ \\ 

$D_{(7',7')}$ & $9500.76$ & $4090.74$ & $3811.36$ & $3596.65$ & $2151.12$ & $1547.9$ & $1211.35$ & $993.74$ \\ 

$D_{(9,9)}$ & $42.42$ & $17.45$ & $69.51$ & $3.49$ & $5.95$ & $8.46$ & $10.99$ & $13.51$ \\ 

$D_{(9',9')}$ & $42.42$ & $17.45$ & $69.51$ & $3.49$ & $5.95$ & $8.46$ & $10.99$ & $13.51$ \\ 

$D_{(10,10)}$ & $42.42$ & $17.45$ & $69.51$ & $3.49$ & $5.95$ & $8.46$ & $10.99$ & $13.51$ \\ 

$D_{(10',10')}$ & $42.42$ & $17.45$ & $69.51$ & $3.49$ & $5.95$ & $8.46$ & $10.99$ & $13.51$ \\ 

$D_{(7,7')}$ & $-208.65$ & $-92.03$ & $-183.73$ & $-49.13$ & $-41.19$ & $-39.16$ & $-39.12$ & $-40.06$ \\ 

$D_{(7,9)}$ & $566.44$ & $260.55$ & $503.48$ & $109.79$ & $112.25$ & $113.9$ & $114.99$ & $115.51$ \\ 

$D_{(7,9')}$ & $-23.29$ & $-9.91$ & $-32.92$ & $-3.14$ & $-3.81$ & $-4.56$ & $-5.41$ & $-6.36$ \\ 



$D_{(7',9)}$ & $-35.45$ & $-15.5$ & $-43.73$ & $-5.49$ & $-6.22$ & $-7.01$ & $-7.88$ & $-8.84$ \\ 

$D_{(7',9')}$ & $567.2$ & $260.89$ & $504.42$ & $109.91$ & $112.38$ & $114.05$ & $115.15$ & $115.7$ \\ 



$D_{(9,9')}$ & $-3.43$ & $-1.31$ & $-7.21$ & $-0.23$ & $-0.42$ & $-0.64$ & $-0.91$ & $-1.23$ \\ 





$D_{(10,10')}$ & $-3.43$ & $-1.31$ & $-7.21$ & $-0.23$ & $-0.42$ & $-0.64$ & $-0.91$ & $-1.23$ \\

\hline\hline
\end{tabular}
\caption{Coefficients for the New Physics formula of $\av{P_1}$.}
\label{TableP1}
\end{table}


\begin{table}
\scriptsize
\centering
\begin{tabular}{||c||r|r|r|r|r|r|r|r||}
\hline\hline
\multicolumn{9}{||c||}{$\av{P_2}$}\\
\hline\hline
  & [\,1\,,\,6\,] & [\,2\,,\,4.3\,] & [\,4.3\,,\,8.68\,] & [\,1\,,\,2\,] & [\,2\,,\,3\,] & [\,3\,,\,4\,] & [\,4\,,\,5\,] & [\,5\,,\,6\,] \\
\hline\hline

 $\delta$ & 0.067 & 0.07 & 0.074 & 0.023 & 0.032 & 0.085 & 0.082 & 0.071 \\
\hline

$N_{(0,0)}$ & $44.21$ & $43.27$ & $-358.4$ & $56.4$ & $34.35$ & $10.6$ & $-14.75$ & $-42.39$ \\ 

$N_{(0,7)}$ & $-1219.96$ & $-561.26$ & $-1082.93$ & $-236.58$ & $-241.83$ & $-245.33$ & $-247.59$ & $-248.63$ \\ 

$N_{(0,7')}$ & $-17.82$ & $-8.2$ & $-15.82$ & $-3.46$ & $-3.53$ & $-3.58$ & $-3.62$ & $-3.63$ \\ 

$N_{(0,9)}$ & $-91.07$ & $-37.48$ & $-148.93$ & $-7.51$ & $-12.79$ & $-18.18$ & $-23.6$ & $-28.99$ \\ 


$N_{(0,10)}$ & $-10.26$ & $-10.04$ & $83.18$ & $-13.09$ & $-7.97$ & $-2.46$ & $3.42$ & $9.84$ \\ 

$N_{(0,10')}$ & $3.18$ & $1.48$ & $2.42$ & $0.64$ & $0.64$ & $0.64$ & $0.64$ & $0.62$ \\ 







$N_{(7,7')}$ & $-208.65$ & $-92.03$ & $-183.73$ & $-49.13$ & $-41.19$ & $-39.16$ & $-39.12$ & $-40.06$ \\ 

$N_{(7,9)}$ & $566.44$ & $260.55$ & $503.48$ & $109.79$ & $112.25$ & $113.9$ & $114.99$ & $115.51$ \\ 

$N_{(7,9')}$ & $-23.29$ & $-9.91$ & $-32.92$ & $-3.14$ & $-3.81$ & $-4.56$ & $-5.41$ & $-6.36$ \\ 



$N_{(7',9)}$ & $-35.45$ & $-15.5$ & $-43.73$ & $-5.49$ & $-6.22$ & $-7.01$ & $-7.88$ & $-8.84$ \\ 

$N_{(7',9')}$ & $567.2$ & $260.89$ & $504.42$ & $109.91$ & $112.38$ & $114.05$ & $115.15$ & $115.7$ \\ 



$N_{(9,9')}$ & $-3.43$ & $-1.31$ & $-7.21$ & $-0.23$ & $-0.42$ & $-0.64$ & $-0.91$ & $-1.23$ \\ 





$N_{(10,10')}$ & $-3.43$ & $-1.31$ & $-7.21$ & $-0.23$ & $-0.42$ & $-0.64$ & $-0.91$ & $-1.23$ \\ 
\hline

$D_{(0,0)}$ & $551.99$ & $190.54$ & $881.21$ & $142.68$ & $80.07$ & $81.89$ & $105.74$ & $141.61$ \\ 

$D_{(0,7)}$ & $-1018.98$ & $-372.65$ & $1121.2$ & $-875.91$ & $-311.74$ & $-69.93$ & $70.4$ & $168.19$ \\ 

$D_{(0,7')}$ & $-61.16$ & $-25.83$ & $-140.78$ & $1.56$ & $-7.59$ & $-13.28$ & $-18.28$ & $-23.56$ \\ 

$D_{(0,9)}$ & $-20.03$ & $-19.92$ & $167.85$ & $-26.15$ & $-15.89$ & $-4.84$ & $6.98$ & $19.87$ \\ 

$D_{(0,9')}$ & $-5.6$ & $-1.55$ & $-23.$ & $0.41$ & $-0.18$ & $-0.92$ & $-1.85$ & $-3.05$ \\ 

$D_{(0,10)}$ & $-182.75$ & $-75.18$ & $-299.5$ & $-15.05$ & $-25.65$ & $-36.46$ & $-47.37$ & $-58.22$ \\ 

$D_{(0,10')}$ & $14.79$ & $5.63$ & $31.05$ & $0.98$ & $1.8$ & $2.78$ & $3.94$ & $5.3$ \\ 

$D_{(7,7)}$ & $9487.44$ & $4084.91$ & $3801.73$ & $3592.89$ & $2148.36$ & $1545.51$ & $1209.12$ & $991.57$ \\ 

$D_{(7',7')}$ & $9500.76$ & $4090.74$ & $3811.36$ & $3596.65$ & $2151.12$ & $1547.9$ & $1211.35$ & $993.74$ \\ 

$D_{(9,9)}$ & $42.42$ & $17.45$ & $69.51$ & $3.49$ & $5.95$ & $8.46$ & $10.99$ & $13.51$ \\ 

$D_{(9',9')}$ & $42.42$ & $17.45$ & $69.51$ & $3.49$ & $5.95$ & $8.46$ & $10.99$ & $13.51$ \\ 

$D_{(10,10)}$ & $42.42$ & $17.45$ & $69.51$ & $3.49$ & $5.95$ & $8.46$ & $10.99$ & $13.51$ \\ 

$D_{(10',10')}$ & $42.42$ & $17.45$ & $69.51$ & $3.49$ & $5.95$ & $8.46$ & $10.99$ & $13.51$ \\ 

$D_{(7,7')}$ & $-208.65$ & $-92.03$ & $-183.73$ & $-49.13$ & $-41.19$ & $-39.16$ & $-39.12$ & $-40.06$ \\ 

$D_{(7,9)}$ & $566.44$ & $260.55$ & $503.48$ & $109.79$ & $112.25$ & $113.9$ & $114.99$ & $115.51$ \\ 

$D_{(7,9')}$ & $-23.29$ & $-9.91$ & $-32.92$ & $-3.14$ & $-3.81$ & $-4.56$ & $-5.41$ & $-6.36$ \\ 



$D_{(7',9)}$ & $-35.45$ & $-15.5$ & $-43.73$ & $-5.49$ & $-6.22$ & $-7.01$ & $-7.88$ & $-8.84$ \\ 

$D_{(7',9')}$ & $567.2$ & $260.89$ & $504.42$ & $109.91$ & $112.38$ & $114.05$ & $115.15$ & $115.7$ \\ 



$D_{(9,9')}$ & $-3.43$ & $-1.31$ & $-7.21$ & $-0.23$ & $-0.42$ & $-0.64$ & $-0.91$ & $-1.23$ \\ 





$D_{(10,10')}$ & $-3.43$ & $-1.31$ & $-7.21$ & $-0.23$ & $-0.42$ & $-0.64$ & $-0.91$ & $-1.23$ \\

\hline\hline
\end{tabular}
\caption{Coefficients for the New Physics formula of $\av{P_2}$.}
\label{TableP2}
\end{table}


\begin{table}
\scriptsize
\centering
\begin{tabular}{||c||r|r|r|r|r|r|r|r||}
\hline\hline
\multicolumn{9}{||c||}{$\av{P_3}$}\\
\hline\hline
  & [\,1\,,\,6\,] & [\,2\,,\,4.3\,] & [\,4.3\,,\,8.68\,] & [\,1\,,\,2\,] & [\,2\,,\,3\,] & [\,3\,,\,4\,] & [\,4\,,\,5\,] & [\,5\,,\,6\,] \\
\hline\hline

 $\delta$ & 0.024 & 0.024 & 0.027 & 0.027 & 0.027 & 0.022 & 0.022 & 0.024 \\
\hline

$N_{(0,0)}$ & $-1.73$ & $-0.72$ & $-0.57$ & $-0.49$ & $-0.34$ & $-0.3$ & $-0.28$ & $-0.32$ \\ 

$N_{(0,7)}$ & $3.7$ & $1.54$ & $0.98$ & $1.29$ & $0.79$ & $0.59$ & $0.51$ & $0.52$ \\ 

$N_{(0,7')}$ & $172.63$ & $71.81$ & $45.75$ & $60.34$ & $36.74$ & $27.69$ & $23.8$ & $24.07$ \\ 

$N_{(0,9)}$ & $-0.16$ & $-0.07$ & $-0.07$ & $-0.03$ & $-0.03$ & $-0.03$ & $-0.03$ & $-0.04$ \\ 

$N_{(0,9')}$ & $10.89$ & $4.61$ & $4.74$ & $1.84$ & $1.92$ & $2.04$ & $2.27$ & $2.81$ \\ 









$N_{(7,7')}$ & $-208.65$ & $-92.03$ & $-183.73$ & $-49.13$ & $-41.19$ & $-39.16$ & $-39.12$ & $-40.06$ \\ 

$N_{(7,9)}$ & $566.44$ & $260.55$ & $503.48$ & $109.79$ & $112.25$ & $113.9$ & $114.99$ & $115.51$ \\ 

$N_{(7,9')}$ & $-23.29$ & $-9.91$ & $-32.92$ & $-3.14$ & $-3.81$ & $-4.56$ & $-5.41$ & $-6.36$ \\ 



$N_{(7',9)}$ & $-35.45$ & $-15.5$ & $-43.73$ & $-5.49$ & $-6.22$ & $-7.01$ & $-7.88$ & $-8.84$ \\ 

$N_{(7',9')}$ & $567.2$ & $260.89$ & $504.42$ & $109.91$ & $112.38$ & $114.05$ & $115.15$ & $115.7$ \\ 



$N_{(9,9')}$ & $-3.43$ & $-1.31$ & $-7.21$ & $-0.23$ & $-0.42$ & $-0.64$ & $-0.91$ & $-1.23$ \\ 





$N_{(10,10')}$ & $-3.43$ & $-1.31$ & $-7.21$ & $-0.23$ & $-0.42$ & $-0.64$ & $-0.91$ & $-1.23$ \\ 
\hline

$D_{(0,0)}$ & $551.99$ & $190.54$ & $881.21$ & $142.68$ & $80.07$ & $81.89$ & $105.74$ & $141.61$ \\ 

$D_{(0,7)}$ & $-1018.98$ & $-372.65$ & $1121.2$ & $-875.91$ & $-311.74$ & $-69.93$ & $70.4$ & $168.19$ \\ 

$D_{(0,7')}$ & $-61.16$ & $-25.83$ & $-140.78$ & $1.56$ & $-7.59$ & $-13.28$ & $-18.28$ & $-23.56$ \\ 

$D_{(0,9)}$ & $-20.03$ & $-19.92$ & $167.85$ & $-26.15$ & $-15.89$ & $-4.84$ & $6.98$ & $19.87$ \\ 

$D_{(0,9')}$ & $-5.6$ & $-1.55$ & $-23.$ & $0.41$ & $-0.18$ & $-0.92$ & $-1.85$ & $-3.05$ \\ 

$D_{(0,10)}$ & $-182.75$ & $-75.18$ & $-299.5$ & $-15.05$ & $-25.65$ & $-36.46$ & $-47.37$ & $-58.22$ \\ 

$D_{(0,10')}$ & $14.79$ & $5.63$ & $31.05$ & $0.98$ & $1.8$ & $2.78$ & $3.94$ & $5.3$ \\ 

$D_{(7,7)}$ & $9487.44$ & $4084.91$ & $3801.73$ & $3592.89$ & $2148.36$ & $1545.51$ & $1209.12$ & $991.57$ \\ 

$D_{(7',7')}$ & $9500.76$ & $4090.74$ & $3811.36$ & $3596.65$ & $2151.12$ & $1547.9$ & $1211.35$ & $993.74$ \\ 

$D_{(9,9)}$ & $42.42$ & $17.45$ & $69.51$ & $3.49$ & $5.95$ & $8.46$ & $10.99$ & $13.51$ \\ 

$D_{(9',9')}$ & $42.42$ & $17.45$ & $69.51$ & $3.49$ & $5.95$ & $8.46$ & $10.99$ & $13.51$ \\ 

$D_{(10,10)}$ & $42.42$ & $17.45$ & $69.51$ & $3.49$ & $5.95$ & $8.46$ & $10.99$ & $13.51$ \\ 

$D_{(10',10')}$ & $42.42$ & $17.45$ & $69.51$ & $3.49$ & $5.95$ & $8.46$ & $10.99$ & $13.51$ \\ 

$D_{(7,7')}$ & $-208.65$ & $-92.03$ & $-183.73$ & $-49.13$ & $-41.19$ & $-39.16$ & $-39.12$ & $-40.06$ \\ 

$D_{(7,9)}$ & $566.44$ & $260.55$ & $503.48$ & $109.79$ & $112.25$ & $113.9$ & $114.99$ & $115.51$ \\ 

$D_{(7,9')}$ & $-23.29$ & $-9.91$ & $-32.92$ & $-3.14$ & $-3.81$ & $-4.56$ & $-5.41$ & $-6.36$ \\ 



$D_{(7',9)}$ & $-35.45$ & $-15.5$ & $-43.73$ & $-5.49$ & $-6.22$ & $-7.01$ & $-7.88$ & $-8.84$ \\ 

$D_{(7',9')}$ & $567.2$ & $260.89$ & $504.42$ & $109.91$ & $112.38$ & $114.05$ & $115.15$ & $115.7$ \\ 



$D_{(9,9')}$ & $-3.43$ & $-1.31$ & $-7.21$ & $-0.23$ & $-0.42$ & $-0.64$ & $-0.91$ & $-1.23$ \\ 





$D_{(10,10')}$ & $-3.43$ & $-1.31$ & $-7.21$ & $-0.23$ & $-0.42$ & $-0.64$ & $-0.91$ & $-1.23$ \\ 

\hline\hline
\end{tabular}
\caption{Coefficients for the New Physics formula of $\av{P_3}$.}
\label{TableP3}
\end{table}

\section{Statistical approach}
\label{appC}

We determine and combine our constraints in a frequentist framework, treating theoretical and experimental uncertainties on the same footing (i.e., taking them as normally distributed random variables).
The model independent bounds are obtained in the following way. A chi-square function is constructed, according to:
\eq{
\chi^2(p) = \sum _k \frac{(\op^{th}_k(p) - \op^{exp}_k)^2}{\delta_k^2 + \sigma_k^{2}}
\label{chi2}}
where $p$ are the theoretical parameters constrained by the analysis (Wilson coefficients, hadronic quantities, CKM matrix elements\ldots), $\op^{th}_k$, $\delta_k$ are the central value for theoretical prediction and error for the observable $\op_k$, and $\op^{exp}_k\pm \sigma_k$ is the experimental average. In case of several uncertainties (statistical and systematic ones, for instance), we combine them in quadrature.
In the case of the observables $\av{\afb}$, $\av{F_L}$, $\av{P_{1,2,3}}$ and $\av{P'_{4,5,6}}$, the numbers $\op^{th}_k$ and $\delta_k$ are given by Eq.~(\ref{Ok}).
We add to the $\chi^2$ similar quadratic terms for the theoretical quantities involved (decay constants, form factors, quark masses, CKM matrix elements).

We want to obtain the constraints in the two-dimensional plane corresponding to two (shifts of) Wilson coefficients $(\delta\C{a},\delta\C{b})$ among all the parameters $p$ (we denote the remaining theoretical parameters $q$). We obtain these regions by drawing the region of $(\delta\C{a},\delta\C{b})$ where:
\begin{equation}
\min_q\chi^2(\delta\C{a},\delta\C{b})-\min_p\chi^2<\delta
\end{equation}
where $\delta$ depends on the dimension of the parameter space where the region is drawn (here, a two-dimensional plane), and the required confidence level (here, 2.3 for 68.3\% CL and 6.18 for 95.5\% CL, see the review of statistics and Monte Carlo techniques in  Ref.~\cite{pdg})

At the practical level, one could in principle compute the $\chi^2$ on a grid for the Wilson coefficients, performing the minimisation over all the nuisance parameters. One would obtain the two-dimensional contours for a given pair of Wilson coefficients $(\delta\C{a},\delta\C{b})$ by performing a further minimisation on the other Wilson coefficients. It turns out that one can improve the accuracy of the method by sampling the parameter space through a Metropolis-Hastings Markov-Chain Monte Carlo algorithm with a weight that favours the minima of the $\chi^2$, for instance 
$\exp(-\chi^2/2)$. Once this sampling has been performed, one creates the grid of points (and compute the value of the $\chi^2$ points) by considering all the points sampled by the algorithm. The points of grid which have not been sampled at all are given a very large $\chi^2$. One then proceeds to a smearing procedure to get a smooth dependence of the reconstructed $\chi^2$ grid.



\end{document}